\documentclass[aps,prd,amsmath,amssymb,amsfonts,floats,floatfix,twocolumn,superscriptaddress,nofootinbib,notitlepage]{revtex4-2}
\usepackage{graphicx}
\usepackage{graphics}
\usepackage[retainorgcmds]{IEEEtrantools}
\usepackage{color}
\usepackage[dvipsnames]{xcolor}
\usepackage{multirow}
\usepackage{xspace} 
\usepackage{slashed}
\usepackage[colorlinks=true, citecolor=blue,urlcolor=blue]{hyperref}
\usepackage{graphicx,marvosym,amssymb,amsmath}
\usepackage{comment}
\usepackage{graphicx}
\usepackage{soul}
\usepackage[caption = false]{subfig}

\newcommand{\sgGTret}[0]{{}_{-2}{G}_{\nt{ret}}^T}

\newcommand{\RR}[0]{{\textrm{R}}}
\newcommand{\sGTret}{{}_{s}G_\nt{ret}^{T}}

\newcommand{\gRIn}{{}_{-2}{R}^{\nt{in}}_{\ell\omega}}

\newcommand{\gRInUps}{{}_{\s}{R}^{\nt{in/up}}_{\ell\omega}}

\newcommand{\gXIn}{\sg{X}^\nt{in}_{\ell\omega}}
\newcommand{\gXUp}{\sg{X}^\nt{up}_{\ell\omega}}
\newcommand{\gXInUp}{\sg{X}^\nt{in/up}_{\ell\omega}}

\newcommand{\sGTwl}[0]{\ensuremath{{}_{s}G_{\ell\omega}^{T}}}
\newcommand{\sgGTwl}[0]{\ensuremath{{}_{-2}G_{\ell\omega}^{T}}}

\newcommand{\sGTl}[0]{\ensuremath{{}_{s}{G}_{\ell}^T}}
\newcommand{\sgGTl}[0]{\ensuremath{{}_{-2}{G}_{\ell}^T}}

\newcommand{\ii}{i}
\newcommand{\Hsd}[1]{\ensuremath{\partial_{#1}}}
\newcommand{\Ylm}[0]{\ensuremath{Y_{\ell m}}}
\newcommand{\sYlm}[0]{\ensuremath{{}_{s}Y_{\ell m}}}
\newcommand{\algn}[1]
{
    \begin{align}#1\end{align}
}

\newcommand{\comm}[1]{}

\newcommand{\nt}[1]{\textrm{#1}}
\newcommand{\s}{{}_{s}}
\newcommand{\sz}{{}_{0}}
\newcommand{\sg}{{}_{2}}

\newcommand{\df}{\textrm{d}}

\newcommand{\pd}[2]{\frac{\partial#1}{\partial#2}}
\newcommand{\dd}[2]{\frac{\textrm{d}#1}{\textrm{d}#2}}

\newcommand{\Gret}{G_\nt{ret}}

\newcommand{\lC}{\left[}
\newcommand{\rC}{\right]}
\newcommand{\lP}{\left(}
\newcommand{\rP}{\right)}
\newcommand{\order}[1]{\mathcal{O}\lP#1\rP}

\newcommand{\sGwl}[0]{\s G_{\ell\omega}}

\newcommand{\nn}{\nonumber\\}

\newcommand{\eqn}[1]
{
    \begin{equation}#1\end{equation}
}

\newcommand{\eqnalgn}[1]
{
    \begin{align}#1\end{align}
}

\newcommand{\GlCID}{{}_{s}g_\ell}

\newcommand{\sgGwl}[0]{\sg G_{\ell\omega}}
\newcommand{\sGl}[0]{\s{G}_{\ell}}
\newcommand{\sgGl}[0]{\sg{G}_{\ell}}
\newcommand{\Dt}[0]{\ensuremath{\Delta t}}
\newcommand{\rL}{r_{\scriptscriptstyle <}}
\newcommand{\rG}{r_{\scriptscriptstyle >}}

\newcommand{\RWFSln}{\s{X}^\nt}
\newcommand{\sXIn}{\s{X}^\nt{in}_{\ell\omega}}
\newcommand{\sXUp}{\s{X}^\nt{up}_{\ell\omega}}

\newcommand{\sgXInUp}{{}_{2}{X}^\nt{in/up}_{\ell\omega}}

\newcommand{\OpCh}[3]{%
  {}_{#1}\hat{\mathcal{C}}_{#3}(#2)
}
\newcommand{\OpChR}[3]{%
  {}_{#1}\hat{\mathcal{C}}^{\RR}_{#3}(#2)
}

\newcommand{\OpORW}[2]{%
  {}_{#1}\hat{\mathcal{O}}_{#2}^{RW}
}
\newcommand{\OpOTw}[2]{%
  {}_{#1}\hat{\mathcal{O}}_{#2}^{T}
}

\newcommand{\TOp}{\hat{\mathcal{T}}}
\newcommand{\RWOp}{\hat{\mathcal{R}}}
\newcommand{\ChOp}{\hat{\mathcal{C}}}

\begin{document}
\global\parskip 6pt

\author{David Q. Aruquipa}
\email{david.q.aruquipa@gmail.com}
\noaffiliation

\author{Marc Casals}
\email{marc.casals@uni-leipzig.de}
\affiliation{
Institut f\"ur Theoretische Physik, Universit\"at Leipzig,\\ Br\"uderstra{\ss}e 16, 04103 Leipzig, Germany}
\affiliation{School of Mathematics and Statistics, University College Dublin, Belfield, Dublin 4, D04 V1W8, Ireland}
\affiliation{Centro Brasileiro de Pesquisas F\'isicas (CBPF), Rio de Janeiro, CEP 22290-180, Brazil}

\title{Green functions of the Regge-Wheeler and Teukolsky equations in Schwarzschild spacetime}

\begin{abstract} 
We present a  calculation of the {\it full} retarded Green functions of the Regge-Wheeler and Teukolsky equations obeyed by gravitational field perturbations of Schwarzschild spacetime. 
We perform the calculations for spacetime points along: (i) a timelike circular geodesic (where null-separated points are not at caustics); and (ii) a static worldline (where null-separated points are at caustics).
These Green functions show a $4$-fold singularity structure away from caustics, and $2$-fold at caustics (similarly to the case of scalar field perturbations, which we also reproduce). Physical oscillations near the singularities appear in the gravitational case, which were not present in the scalar case.
We obtain our results by developing various numerical and analytical methods. 
\end{abstract}

\date{\today}
\maketitle


\section{Introduction}

The time evolution of field perturbations of  a black hole background spacetime can be obtained via the retarded Green function (GF)
of the wave equation obeyed by the perturbations.
The GF is thus a very important object for understanding  black hole properties and 
field propagation.
Classically,
 the GF is 
 valuable, for example,  for analysing the stability properties of  black hole spacetimes (e.g.,~\cite{casals2016horizon});
calculating the self-force (e.g., \cite{lrr-2011-7} for a review) on a point particle moving on a background spacetime via the so-called MiSaTaQuWa equation~\cite{PhysRevD.56.3381,PhysRevD.62.064029};
understanding the merger-ringdown stage of a gravitational waveform (e.g.,~\cite{berti2025blackholespectroscopytheory,2026CQGra..43a5018K,2026arXiv260122015S}).
Also, quantum communication between particle detectors on a background spacetime is determined, to leading order in the coupling with the field, 
by worldline integrals of the GF~\cite{jonsson2015information,blasco2015violation}.

Let us henceforth focus on the case that the background spacetime is that of a spherically-symmetric,  Schwarzschild black hole.
In the particular case of scalar field perturbations of  Schwarzschild spacetime, the GF has already been used for the analysis
of the time evolution of initial data (e.g.,~\cite{Leaver:1986,PhysRevD.92.124055,PhysRevLett.109.111101}), the calculation of the (scalar) self-force~\cite{CDOW13,CDGOWZ} and for the study of  quantum communication between particle detectors~\cite{Jonsson:2020npo}.
The calculation (and application) of the GF in the case of {\it gravitational} perturbations of Schwarzschild spacetime, which are the relevant ones for modelling black hole inspirals, is, however,  significantly more scarce.

In 
Schwarzschild spacetime, the components of the metric perturbation obey {\it coupled} equations.
Unfortunately, it is the GF of these coupled equations which appears in the MiSaTaQuWa equation for the gravitational self-force.
However, there exist two types of quantities,  involving the metric perturbation and its derivatives,  which  do obey decoupled and separable equations.
One type are the Regge-Wheeler (RW)~\cite{PhysRev.108.1063} (and Zerilli~\cite{PhysRevD.2.2141,PhysRevLett.24.737}) quantities.
The second type are the Weyl scalars, which are projections of the Weyl tensor
onto a null tetrad and obey the so-called (Bardeen-Press-)Teukolsky equation (BPT)~\cite{barden-doi:10.1063/1.1666175}.
To the best of our knowledge, the calculation of the GF of 
the {\it four}-dimensional 
RW
or BPT
equations for gravitational perturbations has not previously been achieved;
only  the specific $\ell=2$ multipolar mode of the GF of the RW equation
has been presented 
in~\cite{otoole2020characteristic,2026arXiv260122015S}.
In this paper, we present, for the first time,
the calculation of the {\it full} (i.e., after summing over ``all" $\ell$-modes, if performing a multipolar decomposition) GF of both the RW (for RW spin $s=2$) and BPT (for BPT spin $s=-2$) equations.

The calculation of
GFs is
plagued
with a myriad of technical difficulties, mainly due to their distributional character and divergences.
GFs are functions of two spacetime points and 
it is well-known that
they diverge 
when the two 
points 
coincide~\cite{hadamard2014lectures}.
Importantly, GFs also 
diverge when the two points are  separated but connected by a null geodesic~\cite{Garabedian, Ikawa}.
It has been shown~\cite{Ori1short,CDOWa,Dolan:2011fh,Casals:2012px,Zenginoglu:2012xe,harte2012caustics,Yang:2013shb,casals2020global,CDOW13} in the case of scalar field perturbations 
that this yields 
an interesting cycle in the singularity structure of the GF, which is
$2$-fold at caustics and $4$-fold away from caustics.
Our results in this paper show that these $2$-fold and $4$-fold singularity structures of the scalar GF are both also displayed by the RW and BPT GFs, which also exhibit some extra, physical oscillations near the singularities.
We show this by explicitly calculating these GFs for the two spacetime points on the following two settings: (i) along a timelike circular geodesic (where divergences of the GF are not at caustics); and (ii) along a static worldline (where divergences of the GF are at caustics).

In order to achieve our calculation of the RW and BPT GFs, we use and develop a combination of analytical and numerical methods.
As for the numerical methods, we use 
numerical evolution of characteristic initial data in the time domain as well as
(inverse-)Fourier integrals.
As for the analytical methods, we obtain the equations satisfied by the biscalars in the so-called Hadamard form~\cite{hadamard2014lectures,friedlander} for the RW and BPT GFs and provide a code for calculating the coefficients in the coordinate distance expansion of the Hadamard tail term for the RW GF.
We do not derive such an expansion for the BPT GF and, because of that,
while we achieve the calculation of the RW GF throughout the whole time domain,
we do not achieve the calculation of the BPT GF for points very near coincidence -- we leave that for future work. 
Another analytical method that we use is 
the calculation of BPT $\ell$-modes
via a spectroscopical decomposition on the complex-frequency plane from the 
poles  (quasinormal modes, QNMs)  and
a branch cut of the Fourier modes; we only do it for the dominant contributions from the QNM and branch cut, and as a check for one $\ell$-mode of  our Fourier-integral results. 

For the reader only interested in the final GF results, we here point to where to find them: 
the spin-0 GF is presented in Fig.~\ref{fig:PlotScalarGretCircularAndStatic}; 
the spin-2 RW GF, in Fig.~\ref{fig:spin2MatchingRegion}, and, minus its unphysical $\ell=0,1$ modes and including also its radial derivative, in Figs.~\ref{fig:LogPlotRWGretCircular} (circular geodesic setting) and \ref{fig:PlotsRWGretStatic} (static setting);
the spin-2 BPT GF and its radial derivative, in Figs.~\ref{fig:BPTGretCircular}  and \ref{fig:PlotBPTCircularGretNearSings} (circular geodesic setting) and Figs.~\ref{fig:BPTGretStatic} and \ref{fig:PlotBPTStaticGretNearSings} (static setting).

The layout of the rest of this paper is as follows.
In Sec.~\ref{sec:Schw} we 
present the global singularity structures of the GF (known in the scalar case and shown in this paper in the gravitational case) in  Schwarzschild spacetime.
In Secs.~\ref{sec:RW} and~\ref{sec:BPT} we present the calculation of the GF for the RW and BPT equations, respectively.
We finish in Sec.~\ref{sec:discussion} with a general discussion.
We relegate some technical details to the appendices: In App.~\ref{app:sVijkCoefficients}, we give the coefficients in a small coordinate distance expansion of the RW Hadamard tail bitensor;
In App.~\ref{app:Jaffe} we detail the method of Jaff\'e series  for calculating the ingoing radial RW  solution;
In App.~\ref{app:LightCrossingsAnalysis} we show how the singularities of the GFs arise as resonances between $\ell$-modes; We analyze the  asymptotics for large real frequency of the Fourier modes of the GFs in App.~\ref{app:FourierModesAsymptotics};
In  App.~\ref{sec:late-time} we obtain the spectroscopical decomposition (dominant quasinormal mode plus branch cut) 
for the $\ell$-modes of the BPT GF.

Throughout this paper we use geometric units $c=G=1$.


\section{Global Singularity structure of the GF in Schwarzschild spacetime}
\label{sec:Schw}

The metric of the exterior of Schwarzschild black hole spacetime is given, in Schwarzschild coordinates $(t,r,\theta,\varphi)\in \mathbb{R}\times (2M,\infty)\times \mathbb{S}^2$,  by
\eqn{
    \df s^2=-f\df t^2+f^{-1}\df r^2+r^2\lP\df\theta^2+\sin^2\theta\df\varphi^2\rP,
}
where $f=f(r)\equiv 1-2M/r$, $M$ is the mass of the black hole and $r=2M$ is the radius of the event horizon.
It is convenient to define $\Delta_S(r)\equiv r(r-2M)$
and to denote by  $\gamma$  the angular separation between two spacetime points $x=(t,r,\theta,\varphi)$ and $x'=(t',r',\theta',\varphi')$.

Null geodesics in Schwarzschild spacetime may orbit around the black hole an arbitrary number of times. These orbiting null geodesics asymptote the  circular null geodesic at $r=3M$ as the number of orbits goes to infinity.
As null geodesics orbit around the black hole, they cross caustics 
(i.e., spacetime points where neighboring null geodesics are
focused). Given the spherical symmetry of Schwarzschild, caustics happen at $\gamma=0$ and $\pi$, with $\gamma$ corresponding to the angular separation between the point of emission of a null geodesic and an arbitrary point along that null geodesic.

These orbiting null geodesics have important consequences for the 
GFs.
A GF is a function of two spacetime points: a field point $x$ and a base point $x'$.
As mentioned in the introduction, the GFs diverge when $x$ and $x'$ are connected via a null geodesic, which we generically denote by $\xi$.
The type of singularity of the GFs 
in the case that $\xi$ is
a {\it direct} null geodesic (i.e., a null geodesic joining $x$ and $x'$ which has not crossed any caustic) is well-known since a long time ago: it is $\delta(\sigma)$~\cite{hadamard2014lectures,friedlander}
(see also the later Eqs.~\eqref{eqn:GretHadamarForm} and \eqref{eqn:hadamardTGret}), where 
 $\delta$ is the Dirac delta distribution and
 $\sigma=\sigma(x,x')$ is Synge's world function 
(i.e., $\sigma$ is equal to half of the squared distance along the geodesic connecting $x$ and $x'$, which is unique for points `near enough' the direct null geodesic). 

In  recent years, the singularity structure of the GF  has been shown~\cite{Ori1short,CDOW13,CDOWa,Dolan:2011fh,harte2012caustics,PhysRevD.86.024038,Zenginoglu:2012xe,Yang:2013shb,2023PhRvD.108d4033C} for the case of the scalar field and when 
$\xi$
has  crossed at least one caustic. In particular, the singularity structure, in the case that the point $x$ which is connected to $x'$ via $\xi$ is not a caustic of $x'$, has a  $4$-fold cycle. This $4$-fold singularity structure, starting after $\xi$ has crossed one caustic, is\footnote{The extension of the global singularity structure to the case of the Feynman Green function has been found in Ref.~\cite{Buss-2016} and has found applications to entanglement harvesting in Ref.~\cite{Caribe:2023fhr}.}, to leading order:
\begin{equation}\label{eq:4-fold}
\text{PV}\left(\frac{1}{\sigma}\right)\to -\delta(\sigma) \to 
-\text{PV}\left(\frac{1}{\sigma}\right)\to \delta(\sigma)\to \text{PV}\left(\frac{1}{\sigma}\right) \dots
\end{equation}
where $\text{PV}$ denotes the Cauchy principal value distribution.
In the case that $x$ is a caustic of $x'$, then the cycle is instead $2$-fold  and the singularity `strength' is enhanced~\cite{Zenginoglu:2012xe,2023PhRvD.108d4033C}. To leading order and away from coincidence (i.e., for $x\neq x'$), this $2$-fold singularity structure is~\cite{2023PhRvD.108d4033C}: 
\begin{align}\label{eq:2-fold,g=0}
&
-\frac{d}{d\sigma}\left(|\sigma|^{-1/2}\theta(-\sigma)\right) \to \frac{d}{d\sigma}\left(|\sigma|^{-1/2}\theta(-\sigma)\right) \to
\nonumber \\ &
-\frac{d}{d\sigma}\left(|\sigma|^{-1/2}\theta(-\sigma)\right) 
\dots
\end{align}
if the spacetime points $x$ and $x'$ are at the same spatial position ($\gamma=0$) and
\begin{align}\label{eq:2-fold,g=pi}
&
\frac{d}{d\sigma}\left(|\sigma|^{-1/2}\theta(\sigma)\right) \to -\frac{d}{d\sigma}\left(|\sigma|^{-1/2}\theta(\sigma)\right) \to
\nonumber \\ &
\frac{d}{d\sigma}\left(|\sigma|^{-1/2}\theta(\sigma)\right) 
\dots
\end{align}
if they are antipodal points ($\gamma=\pi$).
Note that \eqref{eq:2-fold,g=0} and \eqref{eq:2-fold,g=pi} possess not only a Dirac-$\delta$ singularity at $\sigma=0$ but also a $\sigma^{-3/2}$ singularity (peaking in the opposite direction of the Dirac-$\delta$) when approaching $\sigma=0$ only from, respectively, $\sigma<0$ and $\sigma>0$. Thus, these $2$-fold singularities are asymmetric about $\sigma=0$.

The following remarks about Eqs.~\eqref{eq:4-fold}, \eqref{eq:2-fold,g=0} and \eqref{eq:2-fold,g=pi} are in order.~(i) 
Like in the earlier case  of the direct null geodesic, we have omitted the positive-valued coefficients of the singularity factors.
(ii) Every change in the singularity type 
takes place after 
$\xi$
crosses a caustic.
(iii) The 
subleading
discontinuity
has also been obtained in~\cite{2023PhRvD.108d4033C}. 
(iv) Strictly speaking, the world function $\sigma(x,x')$ is only defined when $x'$ is in a local (normal) neighbourhood of $x$  (see the later Sec.~\ref{subsec:quasilocalRegion}); In the global structure in Eqs.~\eqref{eq:4-fold}, \eqref{eq:2-fold,g=0} and \eqref{eq:2-fold,g=pi}, by `$\sigma$' we really mean a well-defined
extension of $\sigma$ outside such local neighborhood (see Ref.~\cite{2023PhRvD.108d4033C} for details).

\begin{figure}
    \centering
    \includegraphics[scale=0.65]{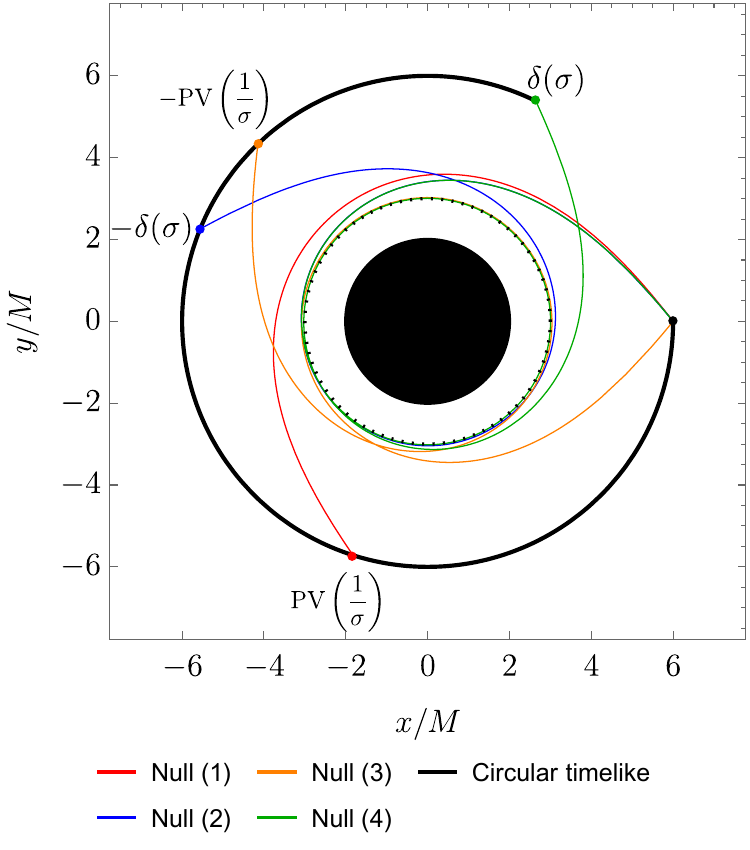}
        \includegraphics[width=.45\textwidth]{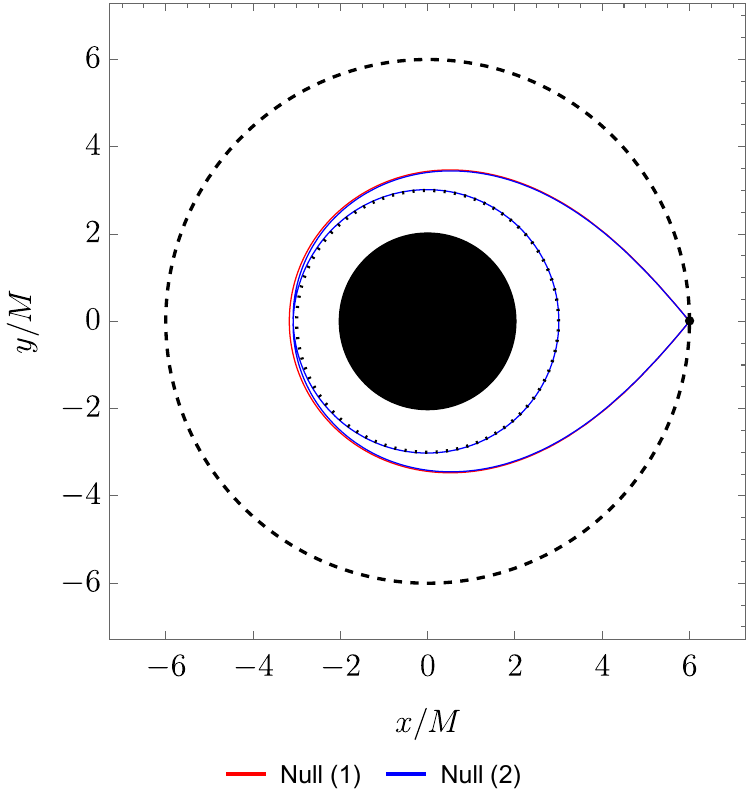}
    \caption{Past-directed null geodesics emanating from a field point $x$ on a timelike worldline, orbiting around the equator of a Schwarzschild black hole and re-intersecting the worldline at various base points $x'$ (light crossings). Top: first four orbiting null geodesics in the setting of a timelike circular geodesic at 
    $r_0=6M$; the type of singularity in the GF at each intersection is indicated and follows the $4$-fold cycle in Eq.~\eqref{eq:4-fold}. Bottom: first four (two if taking the symmetry into account) orbiting null geodesics in the setting of a static worldline  at
    $r_0=6M$; the GF singularity structure  follows the $2$-fold cycle in Eq.~\eqref{eq:2-fold,g=0}.    
     The dotted, inner circular curve corresponds to the photon orbit at $r=3M$; a circle of radius $r=6M$ is the solid, black curve on the top plot and the dashed, black curve on the bottom plot.
    The number in parenthesis in the labels indicates the number of caustic crossings by the null geodesics before re-intersecting the timelike worldline.}
    \label{fig:lightCrossings}
\end{figure}

In this paper we are interested in the GF at points $x$ and $x'$ along some timelike wordline. 
Such points with $x\neq x'$ may also be connected by null geodesics
orbiting around the black hole (and thus crossing caustics).
Given a point $x$ on the timelike worldline,
we refer to the points $x'$ on the worldline 
which are connected to 
$x$
via a past-directed null geodesic as {\it light crossings}.
The GF between $x'$ and a light-crossing $x$ will diverge and, as described above, the type of divergence will depend on how many caustics the null geodesic $\xi$ joining 
$x$ and $x'$
has crossed.
Let us illustrate 
this 
in 
the settings of the timelike worldlines on which we will focus in this paper.

The first  worldline setting is that of a timelike circular geodesic with radius $r=r_0$, whose angular velocity is $\Omega=\sqrt{M}/r_0^{3/2}$.
Given an arbitrary point $x$ on that timelike geodesic,
there will be an arbitrarily large number of points $x'$ lying on that geodesic which are also connected to $x'$ by a past-directed, orbiting null geodesic which has crossed an arbitrarily large number of caustics.
For the specific scase of
$r_0=6M$, it may be easily calculated that the crossings between the timelike geodesic and null geodesics occur at the following times: $\Delta t /M \approx 27.62, 51.84, 58.05, 75.96, 100.09, 108.55, 124.21,\dots$ where $\Delta t\equiv t-t'$. 
Generically, the light-crossings in this setting do not take place at caustics and so the singularity structure of the GF is given by Eq.~\eqref{eq:4-fold}.
  See the top plot in Fig.~\ref{fig:lightCrossings} for an illustration.

The second  worldline setting that we consider is a static worldline. In this case, clearly, the light-crossings will take place at 
caustics with
angular separation $\gamma=0$,
and so the singularity structure of the GF 
is given by Eq.~\eqref{eq:2-fold,g=0}
instead. For a static worldline at the specific radius $r_0=6M$, the first caustic happens at $\Delta t/M\approx37.50$ and the second one at $\Delta t/M\approx70.17$. See the bottom plot in Fig.~\ref{fig:lightCrossings} for an illustration.

The numerical results in this paper show that the $4$-fold singularity structure of Eq.~\eqref{eq:4-fold} for $x'$ not at a caustic of $x$ (as in our setting of the timelike circular geodesic) and the $2$-fold singularity structure of Eq.~\eqref{eq:2-fold,g=0} for $x'$ at a caustic of $x$ with $\gamma=0$ (as in our static setting), seen in the case of the  scalar GF, also hold for the gravitational RW and BPT GFs.


\section{Regge-Wheeler Green function}\label{sec:RW}

 First we shall solve the GF of the RW equation \cite{10.1143/PTP.105.197, PhysRevD.92.124055}
\begin{align}\label{eqn:RWEGret}
    \RWOp\,
    \s\Gret(x,x')=\,&-4\pi\delta_4(x,x'),
\end{align}
where 
\eqn{\label{eqn:RW op}\RWOp\equiv \Box+s^2\frac{2M}{r^3},}
\eqn{\Box=\nabla_\mu\nabla^\mu=-\frac{1}{f}\frac{\partial^2}{\partial t^2}+\frac{1}{r^2}\frac{\partial}{\partial r}\lP r^2f\frac{\partial}{\partial r}\rP+\frac{1}{r^2}\Delta_{\mathbb{S}^2}
}
is the D'Alembertian operator in Schwarzschild spacetime,
\eqn{\Delta_{\mathbb{S}^2}\equiv \frac{1}{\sin\theta}\frac{\partial}{\partial\theta}\lP\sin\theta\frac{\partial}{\partial\theta}\rP+\frac{1}{\sin^2\theta}\frac{\partial^2}{\partial{\varphi^2}}}
 is the Laplacian on the 2-sphere $\mathbb{S}^2$,
and 
\eqn{\delta_4(x,x')\equiv \frac{\delta^{(4)}(x-x')}
{\lP g(x)g(x')\rP^{1/4}}=\frac{\delta(t-t')\delta(r-r')\delta(\Omega-\Omega')}{r \cdot r'}} is an invariant Dirac distribution, with $\Omega$ being the solid angle of the 2-sphere and $g(x)$ the metric determinant at $x$.
Here, RW spin $s=0$, $1$ and $2$ corresponds to, respectively, scalar~\cite{PhysRevD.5.2419,PhysRevD.5.2439}, electromagnetic~\cite{Wheeler:1955zz,Ruffini1972} and gravitational~\cite{Regge:1957td} perturbations.
In particular, metric perturbations in the odd sector can be readily obtained from solutions of the  spin-2 RW equation, which are gauge-invariant quantities~\cite{moncrief1974gravitational}.
The causality condition on the RW GF requires that $\s\Gret(x,x')$ is the solution of Eq.~\eqref{eqn:RWEGret} such that it is equal to zero if the field point $x$ is not in the causal future of the base point $x'$.

In the scalar case (i.e., $s=0$), a direct application of the GF is the obtention of the scalar self-force \footnote{As mentioned, the gravitational self-force can be obtained via a worldline integral similar to that in Eq.~\eqref{eqn:scalarChargeTailIntegral} but involving the GF of the metric perturbation equations (in the Lorenz gauge) rather than the RW or BPT GFs.} via \cite{lrr-2011-7}
\eqn{\label{eqn:scalarChargeTailIntegral}
    F_\mu(\tau)=q\nabla_\mu\,\int_{-\infty}^{\tau^-}{}_0\Gret(z(\tau),z(\tau'))\,\df\tau',
}
where $q$ is the charge of the scalar particle, $\tau$ is its proper time and the integral is performed along the particle's entire past described by its worldline $z(\tau)$.
It is worth noting that, due to a physically-motivated  regularization procedure, the upper limit in the integral in Eq.~\eqref{eqn:scalarChargeTailIntegral} excludes $\tau'=\tau$, i.e., coincidence  points $z(\tau)=z(\tau')$.

Ref.~\cite{Anderson_2005} proposed the so-called method of matched expansions for evaluating the integrand and integral in Eq.~\eqref{eqn:scalarChargeTailIntegral}. Although this method was originally derived for the spin-0 case and for calculating Eq.~\eqref{eqn:scalarChargeTailIntegral}, it is  also useful for applying to calculating the GF for general (RW or BPT) spin, as we do here, regardless of whether it goes into a worldline integral like the one in Eq.~\eqref{eqn:scalarChargeTailIntegral}. The method of matched expansions essentially consists of calculating $\s\Gret(x,x')$ differently in two regions. The first region, called the quasi-local (QL) region, corresponds to points $x$ and $x'$ that are ``close''. The second region,  called distant past (DP), corresponds to points $x$ and $x'$ that are ``not close''. 
The GF is evaluated using different expansions in these two regions and one would hope that there is a matching region where both expansions agree within sufficient accuracy. 
Such matching was achieved in the scalar case in~\cite{CDOW13} and in this paper we achieve it in the gravitational RW case.

In the next two subsections we describe the expansions in the QL region and DP.
In the third, and last, subsection, we present the calculation of the full RW GF for $s=0$ and $2$, as well as its radial derivative for $s=2$, throughout the whole time domain, in the two worldline settings indicated in Sec.~\ref{sec:Schw} (namely, timelike circular geodesic and static wordline).


\subsection{Calculation of RW GF in a Quasi-Local region}\label{subsec:quasilocalRegion}

The calculation of the GF in a QL region may be naturally achieved via the Hadamard form, which
is an analytic expression for the GF that makes explicit its singularity for points in local, normal neighbourhoods.
A normal neighbourhood $\mathcal{N}(x)$ of a point $x$ is a neighbourhood of $x$ such that
every $x'\in \mathcal{N}(x)$ is connected to $x$ by a unique geodesic which
lies in $\mathcal{N}(x)$. Thus, given a point $x$, the QL region would in principle lie inside the maximal normal neighbourhood of $x$.

For $x'\in \mathcal{N}(x)$, the Hadamard form of the RW GF $\s\Gret$ is given by \cite{hadamard2014lectures,friedlander}
\eqn{\label{eqn:GretHadamarForm}
    \s\Gret(x,x')=U(x,x')\delta_+(\sigma)-V_s(x,x')\theta_+(-\sigma),
}
where $U(x,x')$ and $V_s(x,x')$ are two regular biscalars
\footnote{A biscalar is a scalar function that depends on two spacetime points, $x$ and $x'$.}.
We have defined the distributions
\begin{align}\label{eq:delta+}
    \delta_+(\sigma)\equiv \,&\delta(\sigma)\theta_+(x,x'),\nonumber\\
    \theta_+(-\sigma)\equiv\,&\theta(-\sigma)\theta_+(x,x'),
\end{align}
where 
$\theta$ is the Heaviside distribution and $\theta_+(x,x')$ equals 1 if $x$ lies in the future of $x'$ and equals 0 otherwise.
The first term on the right hand side of Eq.~\eqref{eqn:GretHadamarForm} (which involves $U$) is usually referred to as the direct term,
and the second term (which involves $V_s$) as the tail term.

In order to determine $U$ and $V_s$, we take a similar approach to that in Refs.~\cite{friedlander,lrr-2011-7} in the case of $s=0$.
In order to deal with the distributional nature of Eq.~\eqref{eqn:GretHadamarForm}, we note that it can be obtained as the limit $\epsilon\to0^+$  of the following functions:
\eqn{
    \label{eqn:HadamardFormLimit}
    \s\Gret^\epsilon(x,x')\equiv U(x,x')\delta_+(\sigma+\epsilon)-V_s(x,x')\theta_+(-\sigma-\epsilon).
}
By substituting Eq.~(\ref{eqn:HadamardFormLimit}) into Eq.~(\ref{eqn:RWEGret}) and applying the identities \cite{lrr-2011-7}
\begin{align}\label{eq:ids}
    \sigma_\mu\sigma^\mu=\,&2\sigma\\
    \dd{}{\sigma}\theta_+(-\sigma-\epsilon)=\,&-\delta_+(\sigma+\epsilon),\nonumber\\
    \sigma\delta_+(\sigma+\epsilon)=\,&-\epsilon\delta_+(\sigma+\epsilon),\nonumber
\end{align}
where $\sigma_\mu \equiv \nabla_\mu\sigma$,
and
\eqn{
    \label{eqn:diracDeltaDerivativesIdentity}
    \sigma\dd{^n}{\sigma^n}\delta_+(\sigma+\epsilon)=-n\dd{^{n-1}}{\sigma^{n-1}}\delta_+(\sigma+\epsilon)-\epsilon\dd{^n}{\sigma^n}\delta_+(\sigma+\epsilon),
}
for $n\in \mathbb{Z}_{>0}$, we find that
\begin{widetext}
\begin{align}
\label{eq:GF eq-eps}
\RWOp\,
\s\Gret^\epsilon=\,&-2U\epsilon\delta''(\sigma+\epsilon)+\lP2\sigma^\mu\nabla_\mu U+({\sigma^\mu}_\mu-4)U\rP\delta'_+(\sigma+\epsilon)-2V_s\epsilon\,\delta'_+(\sigma+\epsilon)+\\\notag
&\left(2\sigma^\mu\nabla_\mu V_s+({\sigma^\mu}_\mu-2)V_s+
\RWOp\,
U\right)\delta_+(\sigma+\epsilon)+
\theta_+(-\sigma-\epsilon)\RWOp\,V_s,
\end{align}
\end{widetext}
where ${\sigma^\mu}_\mu\equiv \nabla_\mu\sigma^\mu$. Since Eq.~\eqref{eq:GF eq-eps} should reduce to Eq.~\eqref{eqn:RWEGret} in the limit $\epsilon\to0^+$, it follows that
\begin{align}
    \label{eqn:UTranspEq}
    \begin{aligned}
        2\sigma^\mu\nabla_\mu U+({\sigma^\mu}_\mu-4)U=\,&0,\\
    \left[U\right]
    =\,&1,
    \end{aligned}
\end{align}
where $\left[U\right] \equiv U|_{x'=x}$ (henceforth, square brackets around a bitensor will mean evaluation of that bitensor at coincidence: $\left[...\right]\equiv \left. ...\right|_{x=x'}$) and
\begin{align}
    \label{eq:PDE-V-RW}
    &
    \RWOp\,
    V_s = 0,\\\label{eqn:TranspEqVHat}
    &2\sigma^\mu\nabla_\mu \check{V}_s+({\sigma^\mu}_\mu-2)\check{V}_s=-
    \left.
    \lP\RWOp\,
    U\rP\right|_{\sigma=0},\\
    &\left[V_s\right]
    = 
    -s^2\frac{M}{r^3},\label{eq:Vs x=x'}
\end{align}
where $\hat{V_s}\equiv V_s|_{\sigma=0}$ and we have used the distributional identities~\cite{lrr-2011-7}
\begin{align}\label{eq:ids-eps}
    \begin{aligned}
        \lim\limits_{\epsilon\to0^+}\epsilon\, \delta'_+(\sigma+\epsilon)=\,&0,\\
        \lim\limits_{\epsilon\to0^+}\epsilon\,\delta''_+(\sigma+\epsilon)=\,&2\pi\delta_4(x,x').
    \end{aligned}
\end{align}
In order to obtain Eq.~\eqref{eq:Vs x=x'} from Eq.~\eqref{eqn:TranspEqVHat}, we have also used that $[\Box U]=\frac{1}{6}R=0$ \cite{lrr-2011-7}, where $R$ is the Ricci scalar.

The solution to Eq.~\eqref{eqn:UTranspEq} is  \cite{lrr-2011-7}
\eqn{\label{eq:U-VV}
U=\Delta^{1/2},
}
where 
\eqn{
    \Delta\equiv -\frac{\det\lC-\nabla_\alpha\nabla_{\beta'}\sigma\rC}{\sqrt{-g(x)}\sqrt{-g(x')}},
}
is the van Vleck determinant

We note that the world function $\sigma$ and the direct Hadamard coefficient $U$ are independent of the field spin $s$, and so the direct RW Hadamard term is the same as for the scalar (wave) operator. This is a consequence of the fact that the RW operator in Eq.~\eqref{eqn:RW op} is the same as the wave operator but just modified by a spin-dependent potential.
On the other hand, the tail Hadamard coefficient $V_s$ depends on the spin $s$.
In 
particular, Eq.~\eqref{eq:Vs x=x'}
 shows that, at coincidence (i.e, $x'=x$), $V_s$  is zero for $s=0$ and nonzero for $s\neq 0$.

The Hadamard form may be used as the integrand in Eq.~\eqref{eqn:scalarChargeTailIntegral} for proper time $\tau'$ sufficiently close to $\tau$ so that  $z(\tau')\in \mathcal{N}(z(\tau))$. 
Within the context of  the worldline integral in Eq.~\eqref{eqn:scalarChargeTailIntegral} and the black hole-orbiting null geodesics described in Sec.~\ref{sec:Schw}, the point $z(\tau')$ corresponding  to the end of the normal neighbourhood $\mathcal{N}(z(\tau))$ is the first light-crossing.
It is clear from the explicit form of Eq.~\eqref{eqn:GretHadamarForm} that the regularization whereby $\tau'=\tau$ is excluded at the upper integral limit directly amounts to replacing ${}_0\Gret(z(\tau),z(\tau'))$ by $-V_s(z(\tau),z(\tau'))$ in the integrand where $z(\tau')\in \mathcal{N}(z(\tau))$, thus rendering the regularization trivial. 
This way, the integrand in the {\it whole} worldline integration range $\tau':-\infty\to \tau^-$ would satisfy the {\it homogeneous} RW equation (Eq.~\eqref{eq:PDE-V-RW} for $V_s$ and Eq.~\eqref{eqn:RWEGret} away from coincidence for $\s \Gret$).

We next proceed to describe methods for calculating the direct and tail terms in the RW Hadamard form, Eq.~\eqref{eqn:GretHadamarForm}.

\subsubsection{Direct term in the Hadamard form for RW GF}
\label{sec:direct}

Eqs.~\eqref{eqn:GretHadamarForm} and \eqref{eq:delta+} show that the direct term $U\delta_+$ has support only on the lightcone, where it is singular. Despite this singularity, it is useful in certain situations to be able to calculate the Hadamard coefficient $U$ or to provide a multipolar decomposition of the whole direct term. For example, $U\delta_+$ was used in Ref.~\cite{Yang:Casals:2014} to model the direct image of a source near a (rotating) black hole.
Also, it was shown in Ref.~\cite{jonsson2020communication} that $U$ is a key component in determining the signal strength of a quantum communication channel.
In its turn, Ref.~\cite{PhysRevD.100.104037} showed that subtracting the multipoles of the direct term from the multipoles of the full GF, greatly improves the $\ell$-sum convergence for calculating the GF from its multipoles for spacetime points near --but not exactly at-- coincidence (which, the integral in Eq.~\eqref{eqn:scalarChargeTailIntegral}, corresponds to near its upper limit).
The $\ell$-multipoles of the direct term may also be used for helping to model more accurately the early stage of the ringdown of a gravitational waveform (e.g.,~\cite{oshita2025probingdirectwavesblack, deamicis2026postminkowskianexpansionpromptresponse,2026arXiv260122015S}, where methods very different  from the Hadamard form were used to calculate the direct propagation of the field).
The direct term being spin-independent, methods for calculating it and its $\ell$-modes have already been developed for the spin-0 field and are readily valid for general RW spin $s$.
In this subsection, we  briefly describe already-existing methods for calculating the
direct term and its $\ell$-modes, which we later use.

An approach proposed in~\cite{avramidi1995covariant,PhysRevD.84.104039} for calculating the van Vleck determinant $\Delta$ consists of solving the following set of transport equations along the (unique) geodesic joining $x$ and $x'\in\mathcal{N}(x)$:
\begin{align}\label{eqn:VVDTransportEqn}
    \dd{\Delta^{1/2}}{\tau} =& -\frac{1}{2\tau}{Q^\mu}_\mu\Delta^{1/2},\\\notag
    \dd{{Q^\mu}_\nu}{\tau} =&\, u^\alpha\lP{Q^\mu}_\beta{\Gamma^\beta}_{\alpha\nu}-{\Gamma^\mu}_{\alpha\beta}{Q^\beta}_\nu\rP\\\label{eqn:QTransportEqn}
    &-\frac{1}{\tau}\lP{Q^\mu}_\alpha{Q^\alpha}_\nu+{Q^\mu}_\nu\rP-\tau{R^\mu}_{\alpha\nu\beta}u^\alpha u^\beta.
\end{align}
Here, $R^{\mu}{}_{\alpha\nu\beta}$  is the Riemann tensor, ${\Gamma^\mu}_{\alpha\beta}$ are the Christoffel symbols, $\tau$ is an affine parameter along the geodesic joining $x$ and $x'$, $u^{\mu}=\frac{dx^{\mu}}{d\tau}$ is a vector tangent to the geodesic and  ${Q^\mu}_\nu\equiv {\delta^\mu}_\nu-{\sigma^\mu}_\nu$. The initial conditions to solve these equations are, respectively, 
$\left[\Delta^{1/2}\right]=1$ (see Eq.~\eqref{eqn:UTranspEq}) and $\left[Q^{\mu}{}_\nu\right]=0$.

Eqs.~\eqref{eqn:VVDTransportEqn} and \eqref{eqn:QTransportEqn} are valid for spacetimes in an arbitrary number of dimensions. Instead of using them by directly solving for the van Vleck determinant $\Delta$ and the full direct term in Eq.~\eqref{eqn:HadamardFormLimit} in 4-dimensional Schwarzschild, we will use them for solving for   the van Vleck determinant in a 2-dimensional spacetime, as we next explain.

Ref.~\cite{PhysRevD.100.104037} showed how  the multipolar $\ell$-modes of the direct part of Eq.~\eqref{eqn:GretHadamarForm} can be calculated. In this approach, the Schwarzschild line element is decomposed as $\df s^2=r^{2}(\df s_2^2+\df\Omega_2^2)$, where $\df s_2^2$ is the line element of a 2-dimensional spacetime $\mathcal{M}_2$, and the direct term is decomposed as
\eqnalgn{\notag
    &U(x,x')\delta_+(x,x')=\,U(r,r';\Delta t)\delta(\sigma)\theta(\Delta t)\\\label{eqn:GdirModes}
    &=\,\frac{1}{rr'}\sum\limits_{\ell=0}^{\infty}(2\ell+1)G_\ell^{\textrm{dir}}(r,r';\Delta t)P_\ell(\cos\gamma),
}
where $G_\ell^\nt{dir}$
are the multipolar $\ell$-modes of the direct part.
The $\ell$-modes of the direct part are calculated as \cite{PhysRevD.100.104037}
\eqn{\label{eqn:directModesum}
    G^\nt{dir}_\ell(r,r';\Delta t)=\frac{\theta(\Delta t)}{2}\theta(\pi-\eta) \Delta^{1/2}_\nt{2d}\,P_\ell(\cos\eta)\sqrt{\frac{\sin\eta}{\eta}},
}
where 
$\Delta_\nt{2d}=\Delta_\nt{2d}(r,r';\Delta t)$ is the van Vleck determinant  in $\mathcal{M}_2$ and
$\eta$ is the proper time/zero/proper distance between the points $(t,r)$ and $(t',r')$ in $\mathcal{M}_2$
 along the, respectively, 
timelike/null/spacelike geodesic joining them.
Thus, the world function in $\mathcal{M}_2$ is given by $\sigma_2\equiv \varepsilon\, \eta^2/2$, where $\varepsilon=-1,0$ and $+1$ for, respectively, timelike, null, and spacelike
separations in  $\mathcal{M}_2$. It is worth noting that $\mathcal{M}_2$ is a causal domain~\cite{PhysRevD.92.104030}, implying that $\sigma_2$ and $\Delta_\nt{2d}$ are {\it globally} defined in $\mathcal{M}_2$.
We thus have that, for obtaining $G^\nt{dir}_\ell$, we need to calculate two quantities: $\Delta_\nt{2d}$ and $\eta$, which
we can calculate via two different methods.

In the first, numerical method, we obtain $\eta$ by  solving the  equation for the geodesic joining the points  $(t,r)$ and $(t',r')$ in $\mathcal{M}_2$ (see~\cite{PhysRevD.92.104030} for details).
We then calculate $\Delta_\nt{2d}$ by numerically solving Eqs.~\eqref{eqn:VVDTransportEqn}--\eqref{eqn:QTransportEqn} (which are valid in arbitrary dimensions) in $\mathcal{M}_2$, and so with  $\sigma$ replaced by $\sigma_2$ and $\Delta$ by $\Delta_\nt{2d}$.

In the second method, we use the expansions in the  coordinate separations $\Delta x^{A}\equiv \{\Delta t,\Delta r\equiv r-r'\}$ for $\Delta_\nt{2d}$ and $\eta$ that are provided in~\cite{PhysRevD.100.104037}.
Here we merely note the leading order:
\eqn{\label{eq:Gldir-LO}G^\nt{dir}_\ell(r,r';\Delta t)=\frac{1}{2}+o(1),}
using the `Little-o' asymptotic notation (see, e.g.,~\cite{Olver:1974}).

We find it better to calculate $\Delta_\nt{2d}$ and $\eta$ via the second method (i.e., expansions in the  coordinate separations) rather than via the first method (i.e., numerical integration) for the following reason.
In principle, the coordinate expansion should do particularly well near coincidence but less so further away, whereas the numerical integration could potentially do well throughout. However, we found that, in practise, it is hard to obtain precise values for the  parameters of the geodesic joining the points in $\mathcal{M}_2$ when these points are near coincidence\footnote{Coincidence in $\mathcal{M}_2$ is of course equivalent to coincidence in Schwarzschild spacetime if the points are causally separated in the latter spacetime.}. This then leads to a numerical solution for $\Delta_\nt{2d}$ of the transport equations \eqref{eqn:VVDTransportEqn}--\eqref{eqn:QTransportEqn} which is not very precise near coincidence. Since we will be particularly interested in $G^\nt{dir}_\ell$ near coincidence, we will use  the coordinate expansion method instead and we will do so for an expansion up to order $20$ in $\Delta x^{A}$.
In one particular setting --namely, in the later Fig.~\ref{fig:spin2MatchingRegion}-- we will make a comparison between these two methods.

\subsubsection{Tail term  in the Hadamard form for RW GF}\label{Sec:RW-QL-tail}

In order to calculate the Hadamard tail term $V_s$,
we carry out an expansion in the  coordinate distances. By considering the symmetries of Schwarzschild spacetime, we may write the expansion \cite{PhysRevD.77.104002}
\eqn{\label{eqn:VCoordExpAnzats}
    V_s=V_s(x,x')=\sum\limits_{i,j,k=0}^{\infty}\s v_{ijk}(r)\lP\Delta t\rP^{2i}(1-\cos\gamma)^j\lP\Delta r\rP^k,
}
where $\s v_{ijk}$ are some coefficients to be determined. The expansion in Eq.~\eqref{eqn:VCoordExpAnzats}  is not guaranteed to converge for all points $x$ and $x'$ even inside a normal neighbourhood~\cite{CDOWb}. 
In practise, of course, the infinite sums in \eqref{eqn:VCoordExpAnzats} are truncated at some upper values for $i$, $j$ and $k$.
Thus, within this approach, the QL region is given by  the region where the truncated expansion in Eq.~\eqref{eqn:VCoordExpAnzats} is sufficiently accurate.  

In the spin-0 case, the so-called Hadamard-WKB method for calculating ${}_0v_{ijk}$
was proposed in Ref.~\cite{PhysRevD.69.064039},  corrected and developed in Ref.~\cite{PhysRevD.77.104002} and implemented in Ref.~\cite{CDOWb}.
Within this method, one obtains the coefficients ${}_0v_{ijk}$ by first Euclideanizing the metric, then decomposing the Euclidean Green function into Fourier- and $\ell$-modes, and finally  carrying out WKB asymptotics for large-$\omega$ and/or large-$\ell$. 
We availed ourselves of the publicly available code provided in \cite{CDOWb} for spin-0 and generalized it to arbitrary RW spin $s$. With this generalized code, which we make publicly available in \cite{SchwHadamardV}, we calculated the  coefficients $\s v_{ijk}$ necessary to obtain Eq.~\eqref{eqn:VCoordExpAnzats} to order 26, which is the truncation order we use in this paper. We list in Appendix~\ref{app:sVijkCoefficients} the first few coefficients.

Additionally, assuming that the points $x$ and $x'$ lie on a worldline $z(\tau)$ with light-crossings as described in Sec.~\ref{sec:Schw}, we construct a Padé approximant for Eq.~\eqref{eqn:VCoordExpAnzats} in the following manner. 
We first isolate  the singularity of the GF, and of $V_s$, 
at the first light-crossing (given $x$) by re-writing Eq.~\eqref{eqn:VCoordExpAnzats} as
\eqn{\label{eq:Pade}
    V_s= \frac{\mathcal{V}_s}{\lP\Delta t-t_\textrm{NN}\rP^N},
}
for some $N\in \mathbb{Z}_{>0}$, 
where $t_\textrm{NN}$ is the value of $\Delta t$ corresponding to the first light-crossing, which marks the end of the normal neighborhood of $x$ along the worldline $z(\tau)$, and
\eqn{\label{eqn:noSingVs}
    \mathcal{V}_s\equiv (\Delta t-t_\textrm{NN})^N\sum\limits_{i,j,k=0}^{\infty}\s v_{ijk}(r)\Delta t^{2i}(1-\cos\gamma)^j(r-r')^k.
}
The form of \eqref{eq:Pade} is motivated by the form $\text{PV}\left(1/\sigma\right)$ of the singularity at the first light-crossing given in Eq.~\eqref{eq:4-fold} -- we shall address later the value of $N$.
Next, given some fixed point $x$ (and so fixed radius $r$), we replace $r'=r'(\Delta t)$ and $\gamma=\gamma(\Delta t)$ as corresponding to the worldline $z(\tau)$ of interest
on the right hand side of Eq.~\eqref{eqn:noSingVs}, and then re-express it
as a Padé approximant about $\Delta t=0$. This Padé approximant is a  fraction of two polynomials in $\Delta t$ (with $r$-dependent coefficients) of orders $m\geq0$ for the numerator and $n\geq1$  for the denominator.
The result of using such Padé approximant of $\mathcal{V}_s$ in Eq.~\eqref{eq:Pade} for $V_s$
 may extend the region where  $V_s$ is a good approximation to the GF
beyond the radius of convergence of the series in Eq.~\eqref{eqn:VCoordExpAnzats}.
The choices of $m$ and $n$ depend on the truncation order of Eq.~\eqref{eqn:VCoordExpAnzats}, and we find that $m=n$ equal to the truncation order (=26) yields more accurate approximants. 
We note that this Padé approximant that we use is slightly different from that used in previous literature like~\cite{CDOWb,PhysRevD.100.104037,PhysRevD.89.084021}, where, the  Padé approximant was used directly on $V_s$ (for $s=0$) instead of $\mathcal{V}_s$.
Our  Padé approximant requires prior knowledge of the value of $t_\textrm{NN}$ but we prefer to use it as a new alternative which, as we shall see, does as well as, or even better than, those previously-used Padé approximants of $V_s$.

\subsection{Calculation of RW GF in the Distant Past}\label{subsec:distantPastRegion}

In the DP, we  exploit the spherical symmetry of Schwarzschild spacetime in order to perform a multipolar decomposition of the RW GF as
\begin{align}\label{eqn:GretModeSum}
 &   \s\Gret(x,x')=\frac{1}{r\, r'}
    \sum_{\ell=0}^{\infty}
    \s\mathcal{G}_\ell(x,x'),
    \\ \nonumber
& \s\mathcal{G}_\ell(x,x')\equiv 
 4\pi {}_s G_\ell(r,r';\Delta t)
\sum\limits_{m=-\ell}^{\ell}Y_{\ell m}(\theta,\varphi)Y_{\ell m}^*(\theta',\varphi')
\\&
= 
 (2\ell+1){}_s G_\ell(r,r';\Delta t) P_\ell(\cos\gamma),
 \nonumber 
\end{align}
where $\s G_{\ell}\in \mathbb{R}$ are the $\ell$-modes of $\s\Gret$ and
\eqn{\label{eq:r*}
    r_*\equiv r+2M \ln\lP\frac{r}{2M}-1\rP \in \mathbb{R},
}
is the so-called tortoise coordinate.
The spherical harmonics $Y_{\ell m}$ are normalized so that
\eqn{
\int_{\mathbb{S}^2}Y_{\ell m}(\theta,\varphi)\, Y^*_{\ell',m'}(\theta,\varphi)\df\Omega=\delta_{\ell\ell'}\delta_{mm'}.
}

By substituting Eq.~\eqref{eqn:GretModeSum} into Eq.~\eqref{eqn:RWEGret}, 
and using 
\eqn{
\Delta_{\mathbb{S}^2}P_{\ell}(\cos\gamma)=-\ell(\ell+1)P_{\ell}(\cos\gamma),
}
it can be seen that the $\ell$-modes of the GF satisfy the following (1+1)-dimensional partial differential equation (PDE):
\eqn{\label{eqn:1plus1GModes}
    \lP\pd{^2}{r_*^2}-\pd{^2}{t^2}-4\mathcal{Q}_s(r)\rP \s G_\ell=-\delta(\Delta t)\delta(\Delta r_*),
}
where $\Delta r_*\equiv r_*-r_*'$ and 
\eqn{\label{eqn:RW potential}
\mathcal{Q}_s(r)\equiv \frac{f(r)}{4r^2}\lP\ell(\ell+1)+\frac{2M(1-s^2)}{r}\rP.
}

The boundary conditions for Eq.~\eqref{eqn:1plus1GModes} are determined by the {\it retarded} boundary conditions that $\s\Gret$ obeys. 

We describe two methods for calculating the $\ell$-modes $\s G_{\ell}\in \mathbb{R}$:
the first one is in the time domain and the second one in the frequency domain.

\subsubsection{RW $\ell$-modes in the time domain}\label{subsubsec:time}

 A  time-domain method for calculating the modes $\s G_\ell$ was  introduced and implemented first in the spin-0 case in~\cite{PhysRevD.96.084002} to order 2 and later extended to order 4 
in \cite{jonsson2020communication}.
In this subsection we present the generic RW-spin extension of the method in \cite{jonsson2020communication}.
We note that~\cite{otoole2020characteristic} also extended the method in~\cite{PhysRevD.96.084002} to order 4 for generic RW-spin, and they calculated the single $\ell=2$ mode ${}_{2}G_{2}$ but not the full GF ${}_{2}\Gret$.
However, the extension in~\cite{PhysRevD.96.084002} is slightly different from the one that is presented here for generic spin and in~\cite{jonsson2020communication} for spin-0.

From the Hadamard form in (1+1)-dimensions, and taking into account the retarded boundary conditions,
$\s G_\ell$ can be expressed as
\eqn{\label{eqn:sGlAnsatz}
\s G_\ell(v,u)=\s g_\ell(v,u)\theta(u-u')\theta(v-v'),
} 
where
$u\equiv t-r_*$ and $v\equiv t+r_*$ are null coordinates and
 $\s g_\ell(v,u)$ satisfies the {\it homogeneous} PDE
\eqn{\label{eqn:sglCIDEqn}
\frac{\partial^2}{\partial v\partial u}\s g_\ell(v,u)+\mathcal{Q}_s(r)\s g_\ell(v,u)=0.
}
By inserting Eq.~\eqref{eqn:sGlAnsatz} into Eq.~\eqref{eqn:1plus1GModes} we obtain the following  initial data for $\s g_\ell(v,u)$:
\eqnalgn{\label{eq:CID}
\begin{cases}
\GlCID(v=v',u)=\frac{1}{2},\\
\GlCID(v,u=u')=\frac{1}{2}.
\end{cases}
}
This is {\it characteristic} initial data (CID) since it is imposed on null hypersurfaces.
\begin{figure}
    \centering
    \includegraphics[scale=0.75]{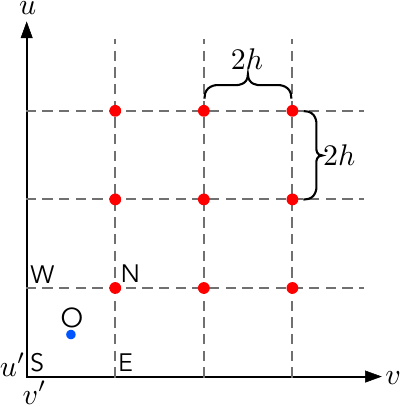}
    \caption{Uniform grid distribution used to solve the characteristic initial value problem \eqref{eqn:sglCIDEqn}--\eqref{eq:CID} for $\s g_\ell(v,u)$. For the cell $SENW$ on the grid, the value of $\s g_\ell(v,u)$ at $N$ is determined from its previously calculated values at $S$, $E$ and $W$, together with the value of $\mathcal{Q}_s(r)$ at  $O$.}
    \label{fig:CIDGrid}
\end{figure}
The CID in Eq.~\eqref{eq:CID} together with the PDE~\eqref{eqn:sglCIDEqn} define a characteristic initial value problem for $\s g_\ell$, which is a smooth solution.
We presented our method for solving this CID problem in detail in Ref.~\cite{jonsson2020communication} for $s=0$. Its generalization to arbitrary spin is straightforward: We simply add the spin term ``$-s^2\frac{2M}{r^3}$" into the $\mathcal{Q}(r)$ in App.~C in~\cite{jonsson2020communication} so that it becomes the $\mathcal{Q}_s(r)$ in Eq.~\eqref{eqn:RW potential} here. Once a uniform grid on the  $(u,v)$-plane  with a stepsize of $2h$ is constructed, as illustrated in Fig.~\ref{fig:CIDGrid}, Eq.~\eqref{eqn:sglCIDEqn} is integrated over each cell of the grid. The integration starts at coincidence and proceeds to the adjacent cells. We carried out this scheme to $\order{h^4}$ 
and chose a stepsize of $2h=2\frac{5}{12}\cdot 10^{-2}$.

We also note that it follows from Eq.~\eqref{eq:CID} that
\eqn{\label{eq:gsl-LO}\GlCID(v,u)=\frac{1}{2}+o(1),} 
when carrying out a small coordinate expansion to leading order, similarly to $G^\nt{dir}_\ell$ in Eq.~\eqref{eq:Gldir-LO}.
Comparison between these two equations shows that the $\ell$-modes of the RW GF and its direct part agree at coincidence: $[\GlCID]=[G^\nt{dir}_\ell]$. We shall comment further on this in Sec.~\ref{sec:RW GF spin-2}.

\begin{figure}
    \centering
        \includegraphics[width=0.48\textwidth]{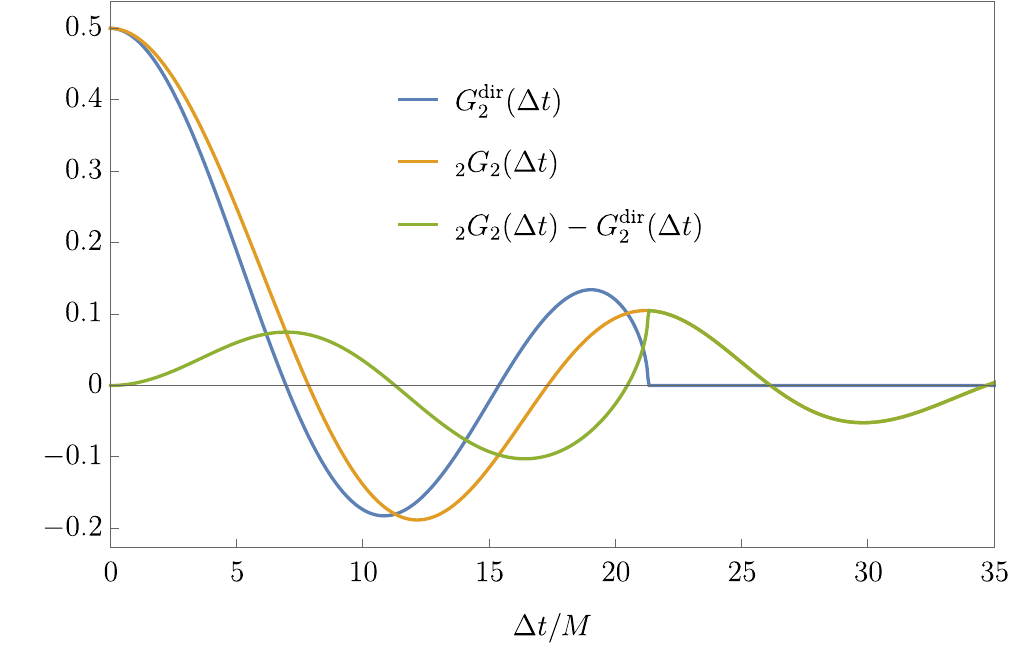}
        \includegraphics[width=0.48\textwidth]{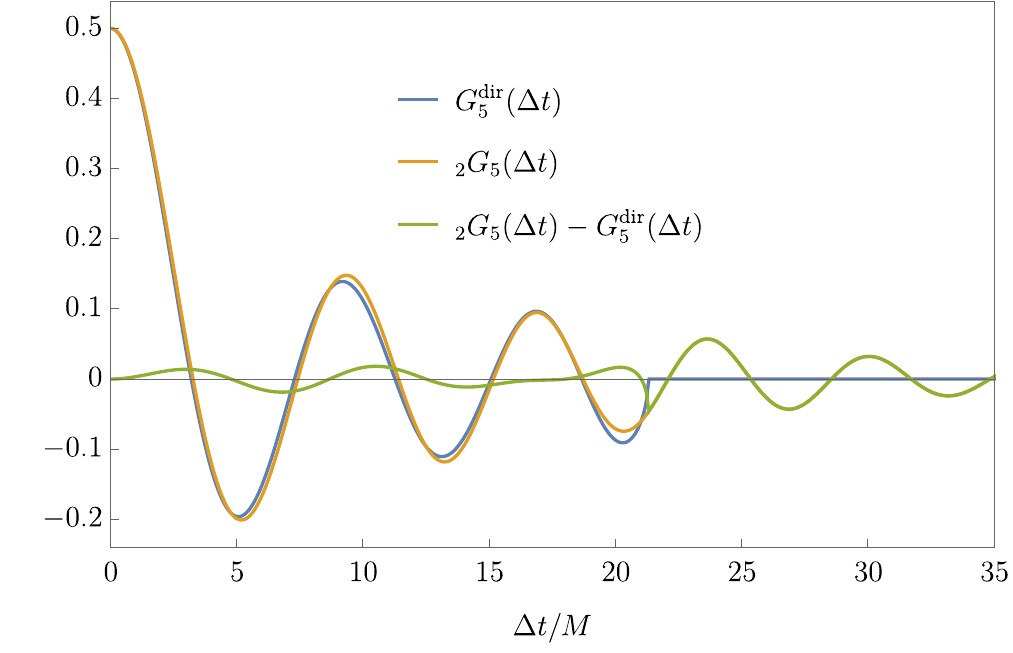}
    \caption{Modes ${}_{2}G_{\ell}$ and $G^{\textrm{dir}}_\ell$  of, respectively, the spin-2 RW GF  (orange curves) and
    its direct Hadamard term
    (blue curves), as well as their difference (green curves),
as functions of $\Delta t$. The points are on $r=r'=6M$ and the modes are for $\ell=2$ (top) and 5 (bottom).}
    \label{fig:ellModes}
\end{figure}

In Fig.~\ref{fig:ellModes} we plot the $\ell$-modes ${}_{2}G_{\ell}$ and   $G^{\textrm{dir}}_\ell$  of, respectively,   the spin-2 RW GF and its direct Hadamard term, 
for $r=r'=6M$ and $\ell=2$  and $5$. 
Here, $G^{\textrm{dir}}_\ell$ was computed using Eq.~\eqref{eqn:directModesum}  and $\s G_\ell$ using Eqs.~\eqref{eqn:sGlAnsatz}--\eqref{eq:CID}. 
The GF modes ${}_{2}G_{\ell}$ are, of course, smooth, whereas the support of $G^{\textrm{dir}}_\ell$ has one-sided compact support due to the stepfunctions in Eq.~\eqref{eqn:sGlAnsatz}.
There is numerical agreement between the  modes ${}_{2}G_{\ell}$ and   $G^{\textrm{dir}}_\ell$ at $\Delta t=0$, as predicted analytically below Eq.~\eqref{eq:gsl-LO}.
Also, as $\ell$ increases, the proximity between ${}_{2}G_{\ell}$ and   $G^{\textrm{dir}}_\ell$ reaches out to larger values of $\Delta t$. 
For comparison with similar plots in the spin-0 case, see  Fig.~1 in \cite{PhysRevD.100.104037}.

The radial derivative of the GF is also an important quantity -- for example, it may be used for calculating the radial component of the self-force, as per Eq.~\eqref{eqn:scalarChargeTailIntegral} for the scalar case.
Differentiating Eq.~\eqref{eqn:sGlAnsatz}, we obtain:
\begin{align}\label{eq:dr-CID}
\partial_r\lP\s G_\ell\rP
&=
\theta(\Delta t-|\Delta r_*|)\lP\partial_r\s g_\ell\rP+
\\&
\left(\theta(\Delta t-\Delta r_*)\delta(\Delta t+\Delta r_*)-
\right.\nonumber \\&\left.
\delta(\Delta t-\Delta r_*)\theta(\Delta t+\Delta r_*)\right)f^{-1}\s g_\ell.\nonumber
\end{align}
This equation shows  that the radial derivative of the GF calculated via \eqref{eqn:sGlAnsatz} has the same radial-coincidence limit, whether it is taken in the direction $r'\to r^-$ or $r'\to r^+$, at least leaving aside  the case $\Dt=\Delta r_*=0$. Thus, $\lim_{r'\to r}\lP\partial_r\, \s G_\ell(r,r';\Delta t)\rP=\lP\partial_r\,\s g_\ell(r,r';\Delta t)\rP_{r=r'}$ for $\Delta t\neq 0$.
In fact, from Eq.~\eqref{eq:dr-CID} and since $\s g_\ell$ is smooth, $\s G_\ell$ is also smooth other than along the characteristic lines $\Dt=\Delta r_*$ and $\Dt=-\Delta r_*$.


\subsubsection{RW $\ell$-modes in the  frequency domain}\label{sec:RWEFreq}

An alternative to the time-domain approach of the preceding subsection for finding the $\ell$-modes of the RW GF, is to instead carry out a(n inverse) Fourier transform as
\eqn{\label{eqn:sGlFourierDecomposition}
    \sGl(r,r';\Delta t)=\frac{1}{2\pi}\int_{-\infty}^{\infty}\sGwl(r,r')e^{-i\omega\Dt}\df\omega,
}
where $\sGwl$ are the Fourier modes of $\s G_\ell$. Inserting this decomposition into Eq.~\eqref{eqn:1plus1GModes} yields the radial RW GF equation:
\eqn{\label{eqn:radiialRWEsGwl}
   \OpORW{s}{\ell\omega}
    \sGwl=-\delta(\Delta r_*),
}
where
\eqn{\label{eq:radial RW op}
\OpORW{s}{\ell\omega}\equiv
\dd{^2}{r_*^2}+\omega^2-4\mathcal{Q}_s(r).
}
We calculate $\sGwl$ from two linearly independent  solutions $\RWFSln{in}_{\ell\omega}$ and $\RWFSln{up}_{\ell\omega}$  of the homogeneous version of Eq.~\eqref{eqn:radiialRWEsGwl}:  

\eqn{\label{eqn:RWEFreqDomain}
 \OpORW{s}{\ell\omega}
\RWFSln{in/up}_{\ell\omega}=0,
}
satisfying the  boundary conditions

\algn{\label{eqn:sXInConditions}
    \sXIn\sim\left\lbrace
        \begin{tabular}{l l}
             $
             e^{-i\omega r_*},$& $r_*\to-\infty$,\\
             $\s{X}_{\ell\omega}^{\nt{in,inc}}e^{-i\omega r_*}+\s{X}_{\ell\omega}^{\nt{in,ref}}e^{i\omega r_*},$& $r_*\to\infty$
        \end{tabular}
    \right.,
}
and
\algn{\label{eqn:sXUpConditions}
    \sXUp\sim\left\lbrace
        \begin{tabular}{l l}
            $\s{X}_{\ell\omega}^{\nt{up,inc}}e^{i\omega r_*}+\s{X}_{\ell\omega}^{\nt{up,ref}}e^{-i\omega r_*},$& $r_*\to-\infty$,\\
             $
             e^{i\omega r_*},$& $r_*\to\infty$
        \end{tabular}
    \right.,
}
where $\s{X}_{\ell\omega}^{\nt{in,inc/ref}}$  are the incidence/reflection coefficients of the ingoing solution and  $\s{X}_{\ell\omega}^{\nt{up,inc/ref}}$ those of the upgoing solution.

Then, taking into account the retarded boundary conditions of the RW GF, the Fourier modes of $\s G_\ell$ are given by
\eqn{\label{eqn:RWGellModes}
    \begin{aligned}
         \sGwl (r,r')
        =&-\frac{\RWFSln{in}_{\ell\omega}(r_<)\RWFSln{up}_{\ell\omega}(r_>)}{W(\RWFSln{in}_{\ell\omega},\RWFSln{up}_{\ell\omega})},
    \end{aligned}
}
where $r_<\equiv \min(r,r')$, $r_>\equiv\max(r,r')$ and 
\eqnalgn{
W(\RWFSln{in}_{\ell\omega},\RWFSln{up}_{\ell\omega}) &= 
\RWFSln{in}_{\ell\omega}\dd{}{r_*}\RWFSln{up}_{\ell\omega}-\RWFSln{up}_{\ell\omega}\dd{}{r_*}\RWFSln{in}_{\ell\omega}\nonumber\\
&=2i\,\omega \s X^\textrm{in,inc}_{\ell\omega}\label{eqn:RWEWronskian}
}
is the RW Wronskian.

There exist various methods for computing the homogeneous solutions  $\RWFSln{in/up}_{\ell\omega}$. For example, the so-called MST method~\cite{10.1143/PTP.96.549,Sasaki:2003xr,PhysRevD.92.124055} 
provides infinite-series representations of the solutions in terms of hypergeometric functions.
Alternatively, 
one can  numerically  integrate the ordinary differential equation \eqref{eqn:RWEFreqDomain} together with the appropriate boundary conditions \eqref{eqn:sXInConditions} and \eqref{eqn:sXUpConditions}. 
Depending on the parameter regime we are interested in, we  select the most suitable method.
For example, the MST method is very efficient in providing  solutions to high precision for $M\omega$  small. In other regimes, the numerical integration might be more practical. 

We chose to use the following methods. We calculate $\sXUp$ and $d\sXUp/dr$ by
using the method for numerical integration of Eq.~\eqref{eqn:RWEFreqDomain}  implemented in the Black Hole Perturbation Toolkit (BHPT) \cite{BHPToolkit}. The boundary conditions for the numerical integration  are placed at some large but finite value of $r$, where the values of $\sXUp$ and $d\sXUp/dr$ are calculated via asymptotic large-$r$ expansions. As for $\sXIn$ and $d\sXIn/dr$, we use instead the so-called Jaff\'e series, introduced in~\cite{Leaver:1986a},
and which we detail in App.~\ref{app:Jaffe}.
We prefer to employ the Jaffé series to compute $\sXIn$ over using the numerical integration in the BHPT since
the Jaff\'e series yields precision comparable to the numerical integration while offering significantly reduced computational time up to $\ell=90$ (for higher values of $\ell$, though, the Jaff\'e series for $\sXIn$  starts becoming slower than using the BHPT, whether its numerical method or MST).

Once we obtain $\RWFSln{in/up}_{\ell\omega}$ and their derivatives,  we calculate the Wronskian $W$ via the top line in \eqref{eqn:RWEWronskian}.
We calculated $\sgXInUp$ and their derivatives
at $r=6M$ at precisions varying between 39 and 28, depending on the values of $\omega$ and $\ell$. We made sure that, with these precisions, the resulting value  of $W$ is such that $\sgGl$ as obtained from Eq.~\eqref{eqn:RWGellModes} has at least 16 digits of precision across $M\omega: 0\to 10$ and all $\ell: 0\to 90$.
We checked this by comparing our numerical values of $\sgGwl$ with those obtained by using the MST method within the BHPT to calculate $\sgXInUp$ and $W$ via the bottom line in \eqref{eqn:RWEWronskian}, at $M\omega=10$ (the precision worsens as $M\omega$ increases) and for all $\ell: 0\to 90$.

Once the Fourier modes $\sGwl$ are obtained, we calculate $\s G_\ell$, not
directly from Eq.~\eqref{eqn:sGlFourierDecomposition}, but from  the following  expressions.
By applying the relation ${\s G^*_{\ell,-\omega}}=\sGwl$, which follows directly from $\s G_\ell\in\mathbb{R}$ together with the causality condition $\s\Gret=0$ for $\Delta t<0$ (implying $\s G_\ell=0$ for $\Delta t<0$), we find that
\begin{align}
    \label{eqn:sGwlFourierIntegral}
    \s G_\ell(r,r';\Delta t)=&\frac{2}{\pi}\theta(\Delta t)\int\limits_0^{\infty}\Re\lP\sGwl\rP\cos\lP\omega\Delta t\rP\df\omega\\
        =&\frac{2}{\pi}\theta(\Delta t)\int\limits_0^\infty\Im\lP\sGwl\rP\sin\lP\omega\Delta t\rP\df\omega.
     \label{eqn:sGwlFourierIntegral-Im}
\end{align}

An asymptotic analysis for large $\omega\in\mathbb{R}$ in App.~\ref{app:FourierModesAsymptotics} (specifically, see around Eq.~\eqref{eqn:sGwlAsymtotic}) shows that the integrand in Eq.~\eqref{eqn:sGwlFourierIntegral}  decays  exponentially when $r=r'$, resulting in a rapidly convergent integral. By contrast, the integrand in Eq.~\eqref{eqn:sGwlFourierIntegral-Im} decays only as $\omega^{-1}$ (whether $r$ and $r'$ are equal or not). Therefore, from a numerical point of view, it is much more convenient to use Eq.~\eqref{eqn:sGwlFourierIntegral} instead of Eq.~\eqref{eqn:sGwlFourierIntegral-Im} to calculate $\sGwl(r,r'=r;\Delta t)$. 
Thus, we used Eq.~\eqref{eqn:sGwlFourierIntegral}, with a choice of the stepsize in $\omega$ equal to $10^{-4}/M$.

\begin{figure}[tb]
    \centering
        \includegraphics[width=0.9\linewidth]{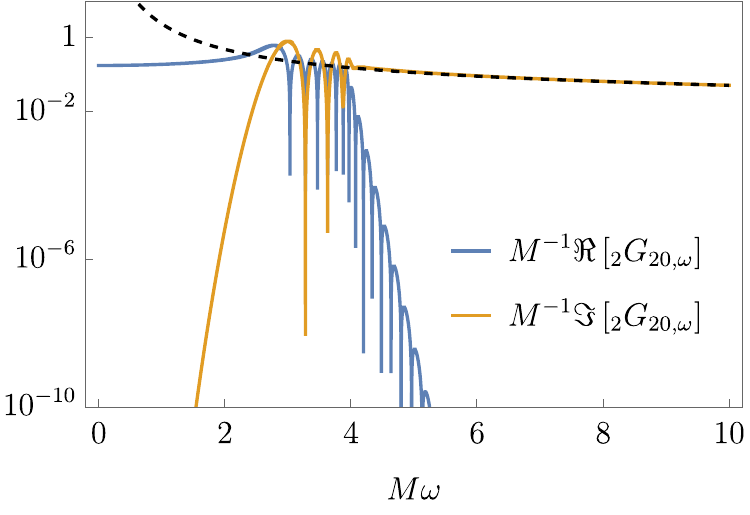}

    \caption{Fourier modes ${}_{2} G_{\ell=20,\omega}$ (continuous curves)  of the RW ${}_{2} G_{\ell=20}$, together with the leading-order in the asymptotic expansion for large $\omega\in\mathbb{R}$ in Eq.~\eqref{eqn:sGwlAsymtotic} of the imaginary part (dashed black line),  as functions of $M\omega$ for $r=r'=6M$.}
    \label{fig:FourierModeAsymptotics}
\end{figure}

The Fourier modes ${}_{2} G_{\ell\omega}$ show a distinct behaviour as functions of $\omega$ which we next describe and which seems to hold for all $\ell$'s. We illustrate this behaviour in the specific case of $\ell=20$:
In Fig.~\ref{fig:FourierModeAsymptotics} we plot 
${}_{2} G_{\ell=20,\omega}$ together   with the leading-order in the  asymptotic expansion for large $\omega\in\mathbb{R}$ in Eq.~\eqref{eqn:sGwlAsymtotic} for its imaginary part,  evaluated at $r=r'=6M$, as a function of $\omega$.
In the behavior of ${}_{2} G_{\ell\omega}$ for arbitrary $\ell(\geq 2)$ as a function of $\omega$, we observe three distinct regions.
In the first region, $\Re\lC\sGwl\rC$ and $\Im\lC\sGwl\rC$ start at $\omega=0$ from, respectively, nonzero and zero values~\cite{2009PhRvD..80h4035D,PhysRevD.103.084021} and  increase monotonically until reaching a new, oscillatory region, where the amplitude does not change by much. This second region starts approximately at the frequency $\omega_\textrm{osc}^{RW}$ such that the zero-order  coefficient  in Eq.~\eqref{eqn:radiialRWEsGwl}  is zero (for $\omega$ smaller/larger than $\omega_\textrm{osc}^{RW}$, the zero order term in \eqref{eqn:RWEFreqDomain} is negative/positive). That is,
the start of the oscillatory region is at the frequency 
\eqn{\label{eq:w-osc}
\omega_\textrm{osc}^{RW}\equiv\sqrt{4\mathcal{Q}_s(r)}.
}
We note that, given the form in Eq.~\eqref{eqn:RW potential} of the RW potential, the frequency $\omega_\textrm{osc}^{RW}$ marking the start of the oscillatory region increases monotonically with $\ell$  (in particular, linearly  for large $\ell$).
The frequency of the oscillations in this second, oscillatory region seems to increase with $M\omega$ until reaching the third regime. 
The third and last region is the  asymptotic regime for large $\omega\in\mathbb{R}$ already commented on: it is given by the asymptotics in Eq.~\eqref{eqn:sGwlAsymtotic}, which, for $r=r'$, shows a non-oscillatory leading-order decay like $\omega^{-1}$ in the imaginary part and exponential decay in the real part. This latter exponential decay is also oscillatory with a frequency that seems to approach $2r_*$ for large-$\omega$.

Observing these three distinct $\omega$-regions is useful for choosing the finite value 
$\omega_\textrm{max}$ at which we truncate the $\omega$-integral in  Eq.~\eqref{eqn:sGwlFourierIntegral}  in its upper limit  in order
to evaluate it  in practise.
In order to obtain meaningful results, we choose $\omega_\textrm{max}$  so that the Fourier integral extends beyond the onset of the exponential decay of $\Re\lC\sGwl\rC$. 
Based on Fig.~\ref{fig:FourierModeAsymptotics}, and observing a similar behaviour for all the $\ell$-modes up to $\ell=90$, we ensure this by making the choice~\footnote{We could have also chosen $\omega_\textrm{max}=2\, \omega_\textrm{osc}^{RW}$, but we ended up using $\omega_\textrm{max}=\max(10/M,2\, \omega_\textrm{osc}^{RW})$ since we initially already calculated the Fourier modes all the way up to a fixed value of $\omega_\textrm{max}=10/M$.} of
  $\omega_\textrm{max}=\max(10/M,2\, \omega_\textrm{osc}^{RW})$.
Since the precisions chosen for $\sgXInUp$ and derivatives were selected so that $\sgGwl$ has 16 digits of precision at $M\omega=10$,
and the precision of $\sgGwl$ worsens with $M\omega$,
it is likely that $\sgGwl$ for $M\omega>10$ has precision somewhat smaller than 16 digits.
However, any potential loss of precision at such large frequencies is expected to be negligible in the GF, particularly when taking into account the large-$\omega$ smoothing that we next describe.

Finally, we note that we  introduce  in the integrand of Eqs.~\eqref{eqn:sGwlFourierIntegral}--\eqref{eqn:sGwlFourierIntegral-Im} a smoothing factor 
\eqn{\label{eq:smooth}
    \frac{1}{2}\lC1-\textrm{Erf}\lP2\lP\omega-\frac{8}{10}\,\omega_\textrm{max}\rP\rP\rC,
}where $\textrm{Erf}(x)$ is the error function~\cite{NIST:DLMF}, in order to suppress spurious oscillations in $ \s G_\ell$ (throughout the time domain)  arising from the finite truncation of the Fourier integral. 
Although we  introduce the smoothing factor in all cases, it is only necessary when the Fourier integral does not converge exponentially. 
Thus, in particular, it would not be necessary to include it when using Eq.~\eqref{eqn:sGwlFourierIntegral} 
in the specific case of $r=r'$, whereas it is necessary when using Eq.~\eqref{eqn:sGwlFourierIntegral}  with $r\neq r'$ or when using 
Eq.~\eqref{eqn:sGwlFourierIntegral-Im} or in the BPT case (as we shall see later in Sec.~\ref{eq:BPT-ell-modes}).
Introducing this smoothing factor does not significantly affect the overall behaviour of $ \s G_\ell$  otherwise -- 
it
was used in~\cite{CDOW13,BUSS2018168} and is further justified in~\cite{CDOW13,Hardy}.


\subsubsection{Comparison between time and frequency domain calculations}\label{sec:comparisonFreqTimeDomain}

\begin{figure}
    \centering
    \includegraphics[scale=0.65]{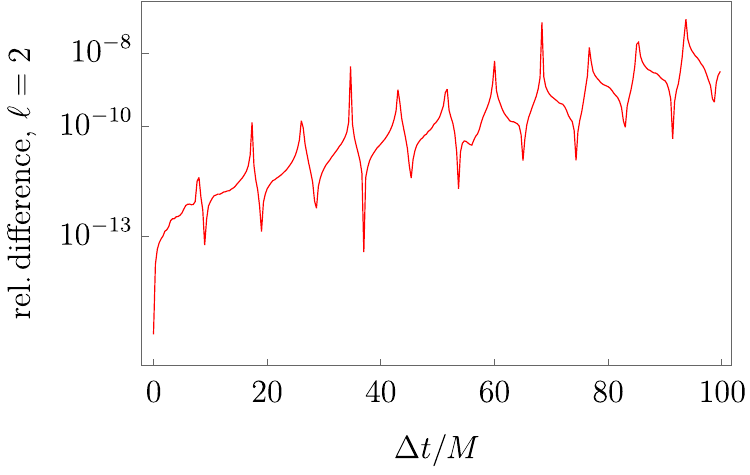}
    \caption{Relative difference  between  $\sg G_\ell$ computed via CID (Eqs.~\eqref{eqn:sGlAnsatz}--\eqref{eq:CID}) and via the Fourier integral in Eq.~\eqref{eqn:sGwlFourierIntegral}, as a function of $\Delta t$ for $r=r'=6M$ and $\ell=2$.}
    \label{fig:RelErrorCIDAndFourierIntegralEll2}
\end{figure}

In Fig.~\ref{fig:RelErrorCIDAndFourierIntegralEll2} we plot the relative difference  between  $\sg G_\ell$ computed via CID (see Eqs.~\eqref{eqn:sGlAnsatz}--\eqref{eq:CID}) and via the Fourier integral in Eq.~\eqref{eqn:sGwlFourierIntegral}.
We plot it as a function of $\Delta t$
for $r=r'=6M$ and $\ell=2$. We checked that the relative difference between the results using the CID and Fourier integral\footnote{The radial derivative of the Fourier modes  $\sGwl(r,r')$, being itself a Green function of the  radial ODE \eqref{eqn:radiialRWEsGwl}, is discontinuous at $r'=r$. For the comparison with CID to work, 
we evaluated
$\left.\partial_r\sGwl\right|_{r'=r}$
as
the average of the limits $\lim_{r'\to r^-}\partial_r\sGwl(r,r')$ and $\lim_{r'\to r^+}\partial_r\sGwl(r,r')$.
See App.~\ref{app:FourierModesAsymptotics} for a justification and further details.} methods for the radial derivative of $\sg G_{\ell=2}$ is similarly small  as for $\sg G_{\ell=2}$ itself in Fig.~\ref{fig:RelErrorCIDAndFourierIntegralEll2}.
We observe agreement between the CID and Fourier integral results to more than ten significant digits at early times ($\Delta t\lesssim 10M$), decreasing to about eight digits at late times (\mbox{$\Delta t\approx100M$}). 
We have seen in other plots which we do not include that, as $\ell$ increases, this minimum agreement drops -- for example, for $\ell=90$ it drops to six digits at early times and five digits at late times. As $\Delta t$ increases, the CID method accumulates larger numerical errors, whereas the Fourier-integral method is expected to remain accurate at late times. The observed increase in the relative error is therefore likely due to the accumulated error in the CID method.
On the other hand, there is a loss of precision in the calculation of $\sgGwl$ which happens for large-$\omega$ but with an onset at smaller values of $\omega$ as $\ell$ increases. This hinders the calculation
of $\sgGl$ via the Fourier integral of $\sgGwl$ for $\ell\gtrsim 90$, which are relevant modes for observing the singular structure of $\sg\Gret$. 
For this reason, we choose to use the CID method to calculate the full RW GF, both for spin-2 and spin-0.
We instead use the Fourier method to: (1) Check on the CID results in the RW case as per Fig.~\ref{fig:RelErrorCIDAndFourierIntegralEll2}; (2) Calculate the BPT GF via a Fourier integral as described in Sec.~\ref{sec:BPT Fourier}.

\subsubsection{Full RW GF in the DP}

Once the $\ell$-modes $\s G_\ell$ are obtained (whether using the time domain or the frequency domain method), we can finally calculate the full GF $\s\Gret$ in the DP by carrying out the $\ell$-sum in Eq.~\eqref{eqn:GretModeSum}.

In practice, the infinite sum in Eq.~\eqref{eqn:GretModeSum} must be truncated at a certain finite value $\ell_\nt{max}$ of $\ell$. 
This produces two  issues. First, similarly to when truncating the Fourier integral at some finite value $\omega_\textrm{max}$, which produced spurious oscillations in  $ \s G_\ell$ that were smoothed out with the factor \eqref{eq:smooth},
sharply cutting off the infinite $\ell$-sum in Eq.~\eqref{eqn:GretModeSum} 
also
produces spurious oscillations (throughout the time domain), on this occasion, on the full GF.
We similarly  smooth out 
these spurious oscillations   by introducing a smoothing factor in the summand, which in this case is 
\eqn{\label{eq:smooth-l}
\exp\lP-\frac{\ell^2}{2\ell_\nt{cut}^2}\rP,}
 for some appropriate choice of the parameter $\ell_\nt{cut}$. The smoothing factor \eqref{eq:smooth-l} was already used in~\cite{CDOW13,CDOWa,PhysRevD.100.104037,2023PhRvD.108d4033C,PhysRevD.89.084021,BUSS2018168,Jonsson:2020npo}  and its use is further justified in~\cite{CDOW13,Hardy,CDOWa,PhysRevD.89.084021}. A second issue with truncating Eq.~\eqref{eqn:GretModeSum} is that the singularities of the GF are effectively smeared out. For example, near  coincidence ($x=x'$), the truncated mode sum approaches the Dirac-$\delta$ singularity in $\s\Gret$ at $x=x'$ (see Eq.~\eqref{eqn:GretHadamarForm}) as a Gaussian distribution which narrows as more $\ell$-modes are included in the sum. 
Here is where the method of matched expansions and the QL calculation comes --partly-- to the rescue:
the DP  is then the region where the truncated sum in Eq.~\eqref{eqn:GretModeSum} (with a smoothing factor in the summand) is accurate enough so that it has a region of overlap with the calculation of the GF (as  $V_s$, and so without including the Dirac-$\delta$ at coincidence, and so with no smearing issues) in the QL region.
However, not only the divergence  of the GF at coincidence is smeared,  but also are its global divergences at light-crossings. 
In particular, the $\delta(\sigma)$ divergences in
Eq.~\eqref{eq:4-fold} become smeared into a Gaussian distribution  (see Eq.~6.3 in~\cite{2023PhRvD.108d4033C}) and the
principal value  $\text{PV}(1/\sigma)$ ones into a  Dawson-like distribution (see Eq.~6.4 in~\cite{2023PhRvD.108d4033C}). Fortunately, such smearings of distributions do not pose a problem in practise -- for example, Refs.~\cite{CDOW13,PhysRevD.89.084021} show in the scalar case that such smearings do not affect the value of the self-force (to within the desired precision).

In order to extend the region where the DP calculation of the GF is accurate enough, to as closely as possible to coincidence, we  implemented a further technique originally introduced in \cite{PhysRevD.100.104037}. This technique consists of subtracting $G^\nt{dir}_\ell$ from $\s G_\ell$ in the mode sum in Eq.~\eqref{eqn:GretModeSum}, thus yielding the new quantity
\eqn{\label{eqn:GretNd}
    \s \Gret^\nt{nd}(x,x')\equiv \frac{1}{r\, r'}
    \sum_{\ell=0}^{\infty}
    (2\ell+1)(\s G_{\ell}-G^\nt{dir}_{\ell})\,P_{\ell}(\cos\gamma),
}
which we refer to as the non-direct part of $\s\Gret$.
The practical benefit of subtracting $G^\nt{dir}_\ell$ from $\s G_\ell$ is that it removes part of the `contamination' near coincidence coming from the Gaussian distribution which results from the smearing of $\delta(\sigma)$ when truncating the infinite $\ell$-sum at some finite value $\ell_\nt{max}$.
Thus, the truncated $\ell$-sum in \eqref{eqn:GretNd} yields a much better representation of the actual GF in the DP but nearer to coincidence than the truncated $\ell$-sum in Eq.~\eqref{eqn:GretModeSum}.

It readily follows from Eqs.~\eqref{eqn:GretNd} and \eqref{eqn:GretHadamarForm} that
the Hadamard form for the non-direct part  of the RW GF is given by  
\eqn{\label{eq:non-dir,Had}
\s \Gret^\nt{nd}(x,x')=-V_{s}(x,x')\theta(-\sigma)\theta(x,x'),\quad x'\in\mathcal{N}(x).} 


\subsection{Full RW Green function}


In this subsection we present our calculation of the spin-0 and spin-2 RW GF, as well as its radial derivative in the spin-2 case, throughout the time domain (i.e., in the DP and QL region), in the settings of a timelike circular geodesic and a static worldline described in Sec.~\ref{sec:Schw}.

\subsubsection{Scalar  GF}\label{sec:scalar}

\begin{widetext}
\begin{figure*}
\captionsetup[subfigure]{labelformat=empty}
    \centering
    \subfloat[]{
        \includegraphics[width=0.49\linewidth]{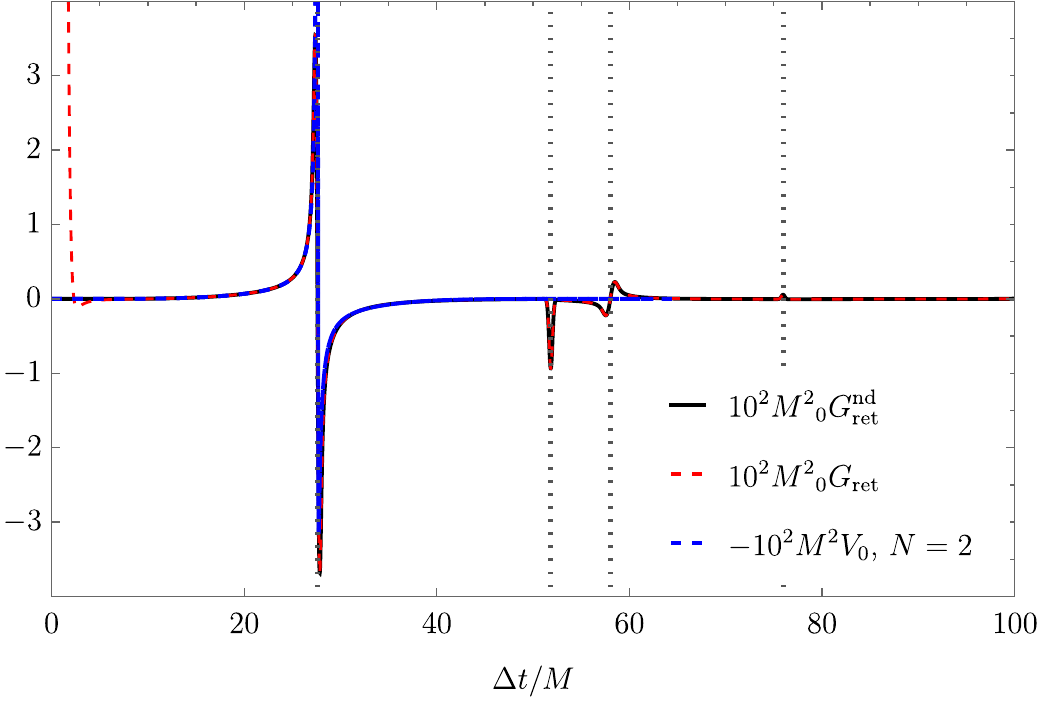}
    }
        \subfloat[]{
        \includegraphics[width=0.49\linewidth]{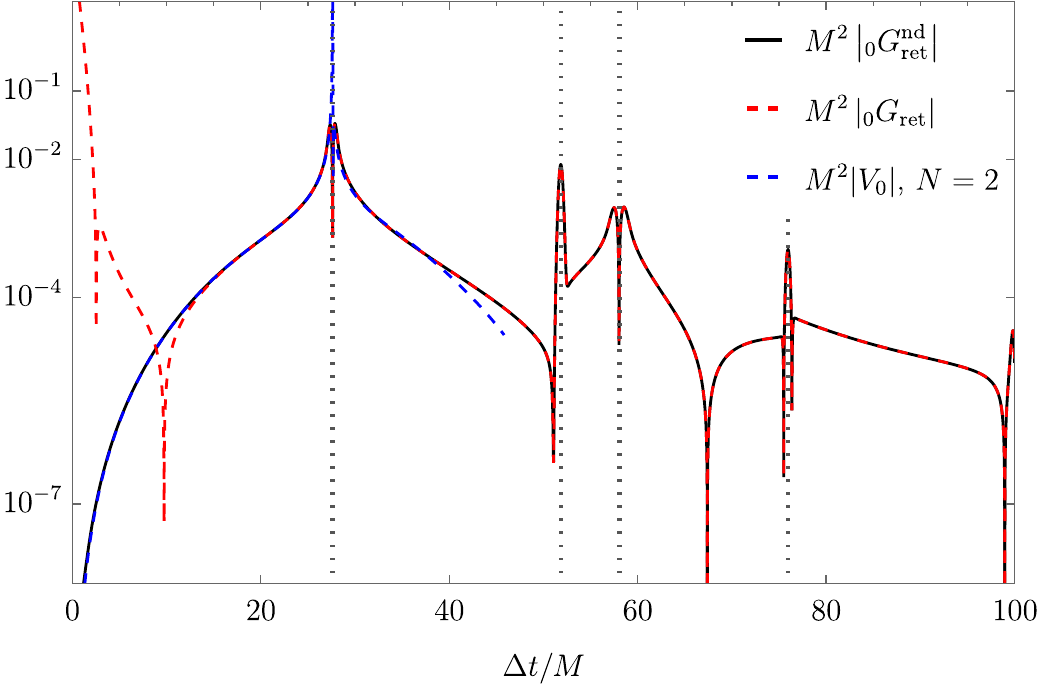}
    }\\
    \subfloat[]{
        \includegraphics[width=0.49\linewidth]{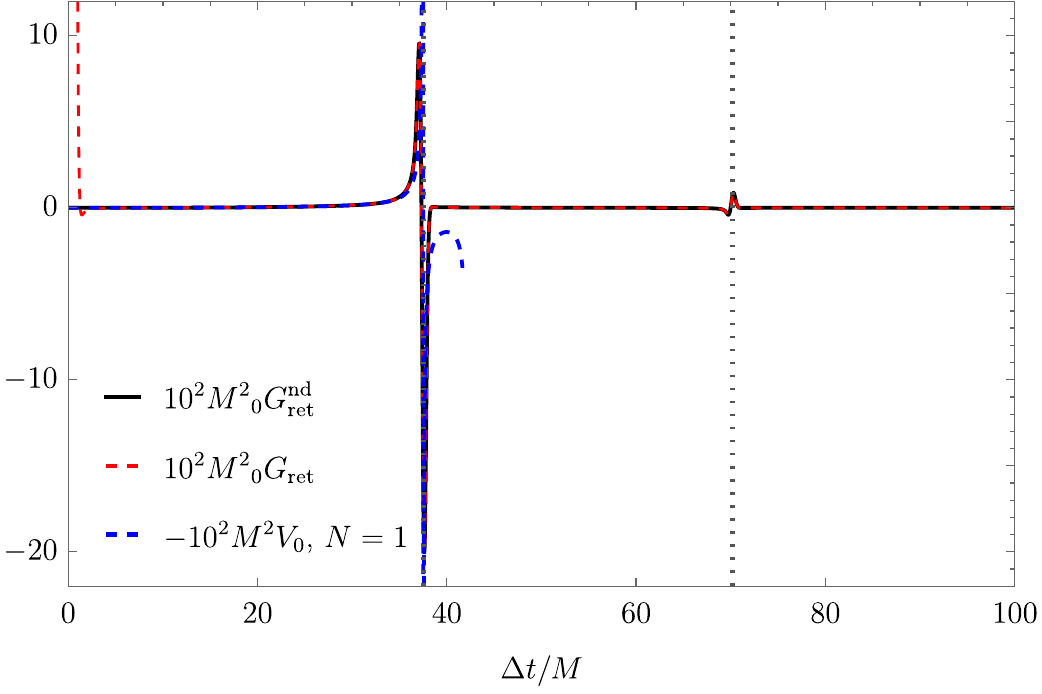}
    }
        \subfloat[]{
        \includegraphics[width=0.49\linewidth]{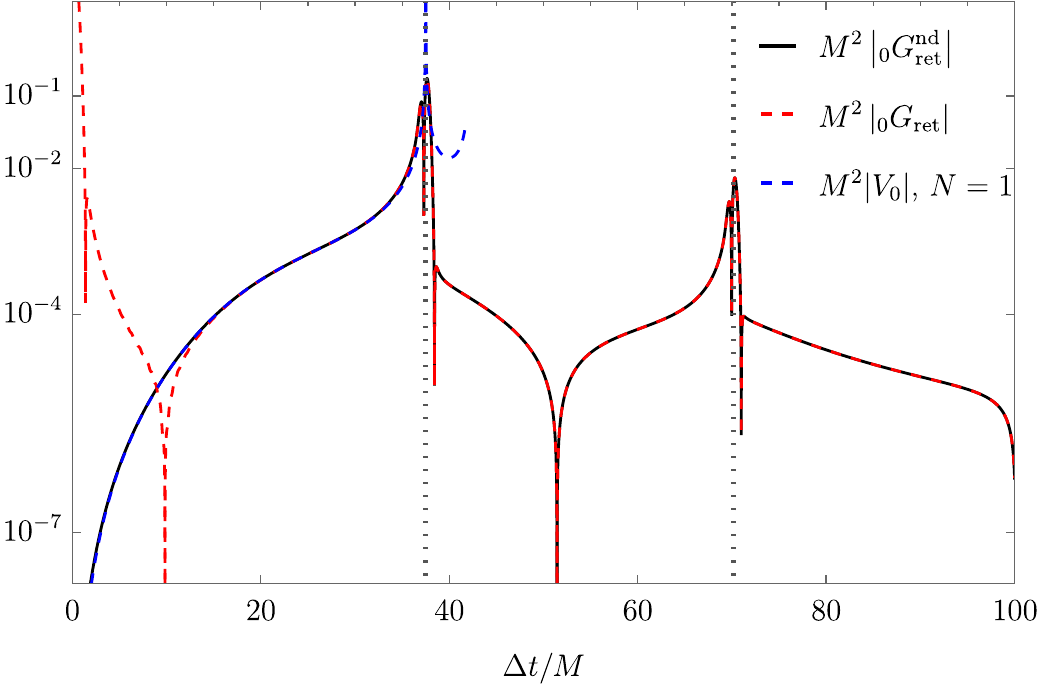}
    }
    \caption{Linear plots  (left) and log-plots  (right)  of the scalar GF ${}_0\Gret$ (dashed red), the non-direct part  ${}_0\Gret^\nt{nd}$ (solid black) and the Hadamard tail  $V_0$ (dashed blue) as functions of $\Delta t$. The points are on $r=r'=6M$ in the cases of: a circular geodesic (top) and static worldline (bottom). The dotted, gray vertical lines correspond to the light-crossings as shown in Fig.~\ref{fig:lightCrossings}.
    }
    \label{fig:PlotScalarGretCircularAndStatic}
\end{figure*}
\end{widetext}

Even though this paper is focused on the gravitational case, and that the full scalar GF was already presented in, e.g.,~\cite{CDOW13,Zenginoglu:2012xe}, here we also present our calculation of the spin-0 GF, as a check against the existing results and for comparison with the spin-2 GF.

In the DP, we calculated the $\ell$-modes ${}_0 G_{\ell}$,  using the CID scheme of Eqs.~\eqref{eqn:sGlAnsatz}--\eqref{eq:CID}. 
For the
$\ell$-modes of the
non-direct part  
in 
Eq.~\eqref{eqn:GretNd}, we calculated $G_\ell^\nt{dir}$ via  Eq.~\eqref{eqn:directModesum} and using small coordinate expansions for $\Delta^{1/2}_\textrm{2d}$ and $\eta$.
We then obtained the full GF ${}_0\Gret$ from Eq.~\eqref{eqn:GretModeSum}  and its non-direct part ${}_0\Gret^\nt{nd}$ from
\eqref{eqn:GretNd}, truncating the mode sums at $\ell_\nt{max}=100$ and including the smoothing factor \eqref{eq:smooth-l} with 
\mbox{$\ell_\nt{cut}=20$}.

In its turn, in the QL region, we calculated $V_0$  using the Hadamard-WKB method introduced in Sec.~\ref{Sec:RW-QL-tail}. 
We then applied the Padé approximation to $\mathcal{V}_0$ in Eq.~\eqref{eq:Pade} for $V_0$. We used  $N=2$  and $N=1$ in Eq.\eqref{eq:Pade} for, respectively, the circular geodesic and static worldline, since these are the values that yielded the best results, as we next see.

We plot the  GF ${}_0\Gret$, its non-direct part ${}_0\Gret^\nt{nd}$ and the Hadamard tail $V_0$ in Fig.~\ref{fig:PlotScalarGretCircularAndStatic}. We plot them for points $x$ and $x'$
at $r=r'=6M$ in two different settings: on a timelike circular geodesic ($\gamma=\sqrt{M/r^3}\Delta t$) in the top plot and in the static ($\gamma=0$) setting in the bottom plot.
In the setting of the circular geodesic,  the considered light-crossings do not occur at caustics and so the singularities of the GF are to follow the 4-fold cycle of Eq.~\eqref{eq:4-fold} (see top plot in Fig.~\ref{fig:lightCrossings}).
On the other hand,
in the static setting, the light-crossings occur at caustics with $\gamma=0$ (see bottom plot in Fig.~\ref{fig:lightCrossings}) and so the singularities of the GF are to follow the 2-fold cycle of Eq.~\eqref{eq:2-fold,g=0}.

\begin{figure}
    \centering
    \includegraphics[width=\linewidth]{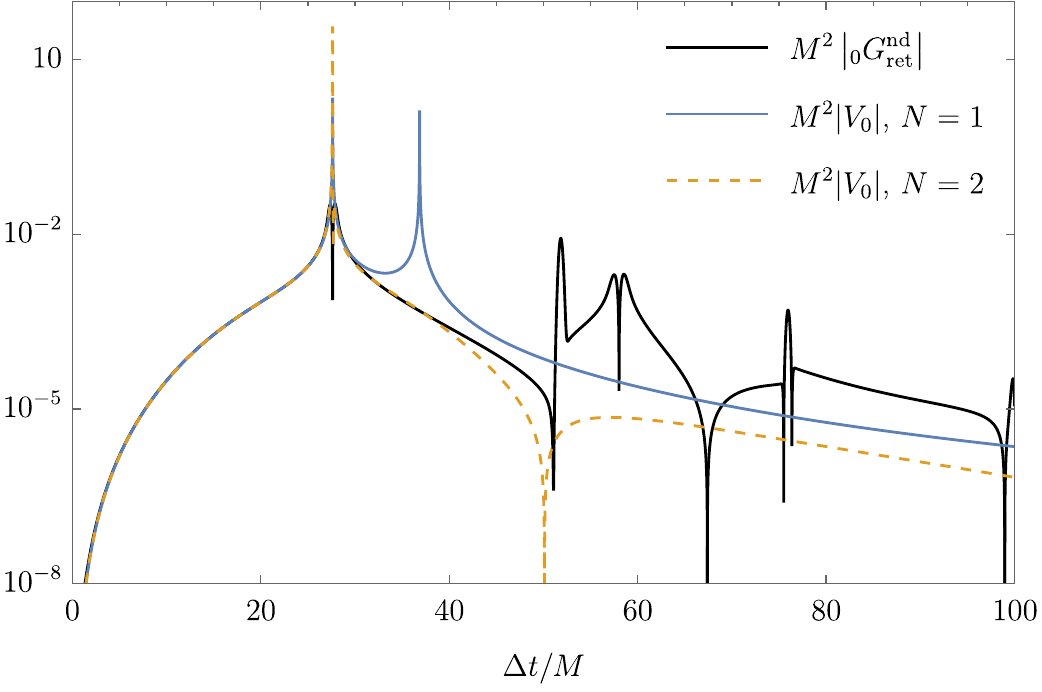}
    \caption{Comparison between $\sz\Gret^\nt{nd}$  (solid black) and $V_0$  obtained using Pad\'e approximants with $N=1$ (solid blue) and $N=2$ (dashed orange) in Eq.~\eqref{eqn:noSingVs}.  The points are along a timelike circular geodesic at $r=r'=6M$. }
    \label{fig:PadeN1N2Comparison}
\end{figure}

Fig.~\ref{fig:PlotScalarGretCircularAndStatic} shows that: (i) there is a region of overlap between the DP and QL calculations of the GF; (ii) the calculation of ${}_0\Gret^\nt{nd}$ is significantly better near coincidence than the calculation of the raw  ${}_0\Gret$; (iii) (smeared) singularities appear at light-crossings in accordance with the cycles described in Sec.~\ref{sec:Schw} (in particular, in the bottom left plot for the static case, an asymmetry can be seen between the `peaks' --smeared divergences-- to the left and to the right of the first light-crossing singularity); (iv) the Pad\'e-extended $V_0$ captures well the singularity of the GF at the first light-crossing and it does well even beyond it (particularly well in the circular setting).
The fact that, in the circular setting,  the Pad\'e-extended $V_0$ with $N=2$ performs  well for so long beyond the end of the normal neighbourhood and that  it performs better than using $N=1$  (in the other situations of static setting and of $s\neq0$, $N=2$ does not do better than $N=1$) may seem surprising. 
We show this in Fig.~\ref{fig:PadeN1N2Comparison}, where we plot  the spin-0 GF together with the Pad\'e-extended $V_0$ using both $N=1$ and $N=2$.

These plots in the scalar case serve not only to show the described features but also as a validation of our methods and codes by comparing with calculations in the literature.
In particular, there is visual agreement between the following:
the top left plot in our Fig.~\ref{fig:PlotScalarGretCircularAndStatic}  with,  the left plot in Fig.~10 in~\cite{CDOW13} and Fig.~1 in~\cite{2023PhRvD.108d4033C};  the top right plot in our Fig.~\ref{fig:PlotScalarGretCircularAndStatic}  with Fig.~2 in~\cite{2023PhRvD.108d4033C};
the bottom right plot in our Fig.~\ref{fig:PlotScalarGretCircularAndStatic}  with the bottom plot in Fig.~13 in~\cite{Zenginoglu:2012xe} (in this latter case, the agreement is only qualitative,  since the results in the two figures are at different radii).


\subsubsection{Gravitational RW GF}\label{sec:RW GF spin-2}

We used the same methods 
to calculate the spin-2 RW $\sg\Gret$ as the ones we used to calculate the scalar ${}_0\Gret$, which are detailed in the previous subsection. In the gravitational case here, we set $N=1$ in the Pad\'e approximant for $V_2$, we truncated the $\ell$-sum at $\ell_\nt{max}=200$ and, in the smoothing factor, we chose $\ell_\nt{cut} = 50$ in the circular setting and $\ell_\nt{cut} = 45$ in the static setting. 

\begin{figure}
    \centering
    \includegraphics[width=\linewidth]{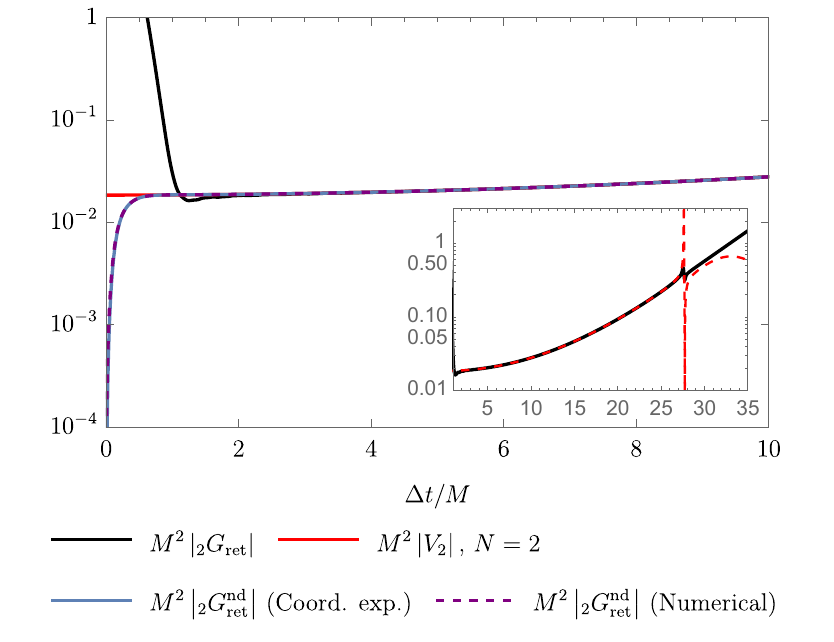}
    \caption{Plot of $\sg\Gret$ (solid black), the non-direct  $\sg\Gret^\nt{nd}$ (in the dashed purple curve, $\eta$ and $\Delta_\nt{2d}^{1/2}$ are calculated using a numerical method  and,  in the solid blue curve, using a small coordinate expansion) and $-V_2$ (solid red) as functions of $\Delta t$. The points are along a timelike circular geodesic at $r=r'=6M$. 
 The subplot is an extension to later times, so as to show the whole matching region between $\sg\Gret$ (solid black) and $-V_2$ (dashed red).
    }
    \label{fig:spin2MatchingRegion}
\end{figure}

In Fig.~\ref{fig:spin2MatchingRegion} we plot $\sg\Gret$, $\sg\Gret^\nt{nd}$  and  $-V_2$ along a timelike circular geodesic at $r=r'=6M$. The black curve corresponds to $\sg\Gret$ and the red curve to $-V_2$ calculated using Eq.~\eqref{eq:Pade}.
 The quantities $\eta$ 
 and
 $\Delta_\nt{2d}$,  
 which feed (via Eq.~\eqref{eqn:directModesum}) into $G_\ell^\nt{dir}$ in the non-direct $\sg\Gret^\nt{nd}$ in Eq.~\eqref{eqn:GretNd}, were calculated in the two different methods described in Sec.~\ref{sec:direct}.
More precisely, the dashed-purple and solid-blue curves correspond to $\sg\Gret^\nt{nd}$ obtained with  $\eta$ and $\Delta_\nt{2d}$  calculated using, respectively, the  numerical method and the small coordinate expansions (described in Ref.~\cite{PhysRevD.100.104037}). When we compare these two results for the non-direct part in the DP with $-V_2$ in the QL region, we find a large region of overlap.
We next make two remarks about the behaviour of the non-direct $\sg\Gret^\nt{nd}$ near coincidence.

First, as pointed out in Sec.~\ref{sec:direct}, a numerical approach for calculating $\eta$ and $\Delta^{1/2}_\nt{2d}$ does not produce precise values near coincidence. We saw (in a plot which we do not  include) that this inaccuracy
leads to values for $\sg\Gret^\nt{nd}$ in the region $0<\Delta t\lessapprox 0.5/M$ in the setting of Fig.~\ref{fig:spin2MatchingRegion}  which are very different from $-V_2$.
For this reason,  we instead extrapolated  to the region $0<\Delta t\leq 0.5/M$  the values   for $\Delta t> 0.5/M$ of the parameters  (namely, the energy and $\eta$) of the  geodesic in $\mathcal{M}_2$. We used these extrapolated values for the geodesic parameters in order  to then solve the  transport equations for $\Delta_\nt{2d}$  in the region $0<\Delta t\leq 0.5/M$. This extrapolation significantly improves the values of $\sg\Gret^\nt{nd}$ near coincidence and is how the dashed-purple  curve in Fig.~\ref{fig:spin2MatchingRegion} was calculated in $0<\Delta t\leq 0.5/M$. Despite such improvement, however, we still expect that, very near coincidence, the small coordinate approach is more precise than performing such extrapolation. 

Second, there is an issue with the non-direct part as coincidence is approached.
As mentioned in Sec.~\ref{subsubsec:time}, analytically one can see  (Eqs.~\eqref{eq:Gldir-LO}, \eqref{eqn:sGlAnsatz} and~\eqref{eq:gsl-LO}) that the modes ${}_{s}G_{\ell}$ and   $G^{\textrm{dir}}_\ell$ agree in the limit to coincidence (from a causal direction) for any RW spin $s$, both going like $1/2+o(1)$. This agreement leads to a cancellation between the modes ${}_{s}G_{\ell}$ and   $G^{\textrm{dir}}_\ell$ in the limit to coincidence when constructing the non-direct part of the GF with Eq.~\eqref{eqn:GretNd}.
However, the Hadamard form for the non-direct part $\sg\Gret^\nt{nd}$ of the GF is given by Eq.~\eqref{eq:non-dir,Had} and the Hadamard tail at coincidence is 
$\left[V_s\right]
=-s^22M/r^3$.  For spin-2, it is 
$\left[V_2\right]
=-4M/r^3\neq 0$, and so  ${}_2\Gret^\nt{nd}$ should not be zero in the limit to coincidence (taking the limit along the timelike  worldline as $\Delta t\to 0^+$), contrasting with the fact that  the modes ${}_{s}G_{\ell}$ and   $G^{\textrm{dir}}_\ell$ have the same limit to coincidence.
The resolution probably lies in the fact 
that 
the $\ell$-mode sum  in Eq.~\eqref{eqn:GretNd} for $\sg\Gret^\nt{nd}$
was truncated at a finite value $\ell_\nt{max}$ and
we are dealing with distributions: 
$\s \Gret^\nt{nd}$ contains a $\theta(-\sigma)$ (see \eqref{eq:non-dir,Had}).
Thus, the small-$\Delta t$ expansion of the $\ell$-modes ${}_{2}G_{\ell}-G^{\textrm{dir}}_\ell$ of $\s \Gret^\nt{nd}$ is probably not uniform in $\ell$: it only holds for {\it fixed} $\ell$.
As a consequence, taking the limits $\Delta t\to 0^+$ and  $\ell_{max}\to\infty$  in Eq.~\eqref{eqn:GretNd} for $\sg\Gret^\nt{nd}$, previously truncated at a finite value $\ell_\nt{max}$, need not commute. 
So first taking $\Delta t\to 0^+$ and then $\ell_{max}\to\infty$ need not yield the right result. 
If, instead, one could first calculate ${}_{2}G_{\ell}-G^{\textrm{dir}}_\ell$ without doing a small-$\Delta t$ expansion and all the way to  $\ell_{max}=\infty$, and afterwards sum them all, then one should recover the correct distributional behaviour for $\s \Gret^\nt{nd}$ and so recover $-V_2$ for $\Delta t>0$ -- but this infinite sum is of course not achievable in practise.
This is why the solid-blue and dashed-purple curves for  $\sg\Gret^\nt{nd}$ in Fig.~\ref{fig:spin2MatchingRegion} do not go  to  $[V_2]=4M/r^3$ as $\Delta t\to 0^+$. 
In its turn, for spin-0, it is $\left[V_0\right]=0$, in accord with the fact that  the modes ${}_{s}G_{\ell}$ and   $G^{\textrm{dir}}_\ell$ have the same limit to coincidence.
Thus, the issue of non-commutativity is a lot less apparent in the spin-0 case and not really visible in Fig.~\ref{fig:PlotScalarGretCircularAndStatic}.
However, even if less visible, this issue is still there in the spin-0 case, as explained below Eq.~(27) in \cite{PhysRevD.100.104037}.

Away from coincidence, Fig.~\ref{fig:spin2MatchingRegion} shows the divergence of the RW GF and of $-V_2$ at the first light-crossing ($\Delta t\approx27.7M$). However, while this divergence is clear in the case of $-V_2$ (where this divergence has been ``massaged in" via \eqref{eq:Pade} and the Pad\'e approximant of $\mathcal{V}_s$), in the case of the GF (where the smoothing factor \eqref{eq:smooth-l} has been included), this divergence appears merely as  a ``bump", as it
is somewhat masked by its $\ell=0,1$ modes as we next see.

Of arguably more physical importance than $\sg\Gret$ is this quantity without including its $\ell=0$ and 1 modes, i.e., without including
\eqn{\label{eqn:G01}
    \sg G_{(01)}(x,x')\equiv \frac{1}{r\, r'}
    \sum_{\ell=0}^{1}
    \sg\mathcal{G}_\ell(x,x').
}
The reason is that the $\ell=0,1$ modes of the RW GF have no physical meaning\footnote{
As detailed in Refs.~\cite{PhysRev.108.1063,PhysRevD.2.2141}, the  $\ell=0,1$ modes of the metric perturbation $\ell$-modes $h_{\mu\nu}^{(\ell)}$
cannot be obtained from the RW equation. The $\ell=0,1$ modes resulting from solving Eq.~\eqref{eqn:1plus1GModes} are not associated to $h_{\mu\nu}^{(\ell)}$.
}. 
We thus add a tilde over a quantity to denote that $ \sg G_{(01)}$ has been subtracted from it:
\begin{align}\label{eq:G-l01}
\sg\tilde{G}_\nt{ret}&\equiv \sg\Gret-\sg G_{(01)},
\\
\sg\tilde{G}^\nt{nd}_\nt{ret}&\equiv \sg\Gret^\nt{nd}-\sg G_{(01)},
\nonumber
\\
-\tilde{V}_2&\equiv -V_2-\sg G_{(01)}.\nonumber
\end{align} 

\begin{figure}
\captionsetup[subfigure]{labelformat=empty}
    \centering
    \subfloat[]{
        \includegraphics[width=\linewidth]{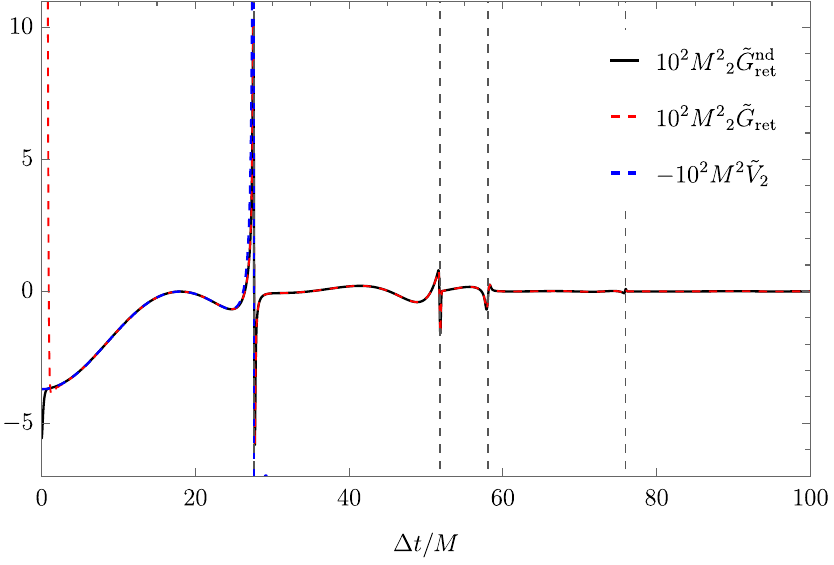}
    }
    \\
    \subfloat[]{
        \includegraphics[width=\linewidth]{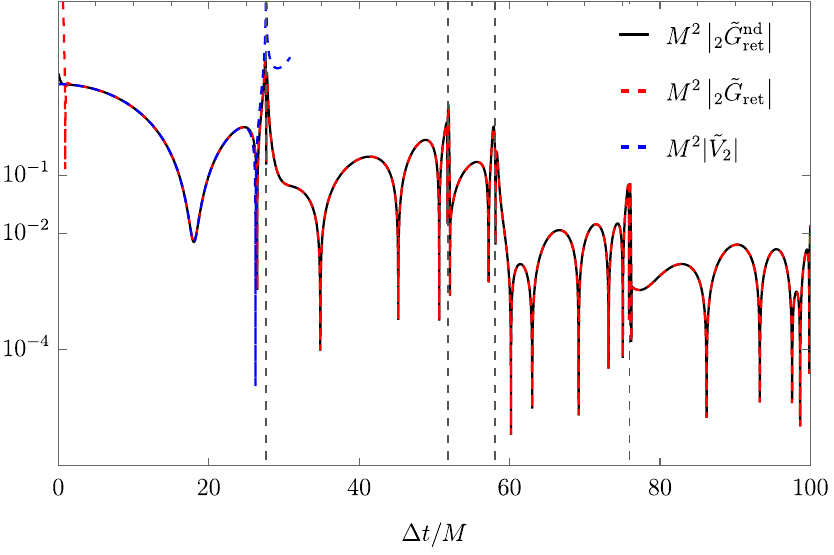}
    }
    \\
    \subfloat[]{
        \includegraphics[width=\linewidth]{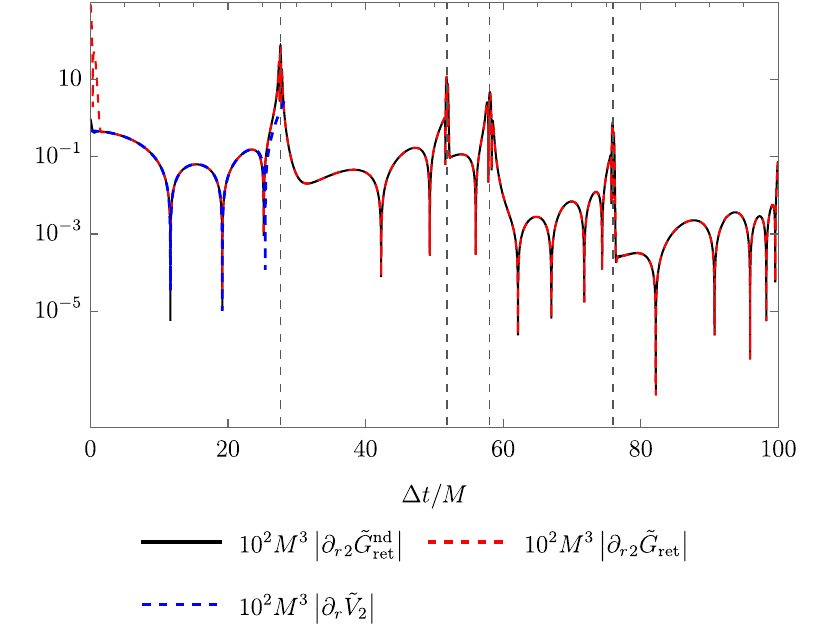}
    }
    \caption{Top: linear plot of 
   $\sg\tilde{G}_\nt{ret}$ (dashed red), $\sg\tilde{G}^\nt{nd}_\nt{ret}$ (solid black) and $-\tilde{V}_2$  (dashed blue), defined in Eq.~\eqref{eq:G-l01},
   as functions of $\Delta t$ for points along a timelike circular geodesic of radius $r=6M$. The  dotted, gray vertical lines correspond to the  light-crossings as per the top plot in Fig.~\ref{fig:lightCrossings}.
   Middle: log-plot version of the top plot.
   Bottom: log-plot of the radial derivatives of the quantities in the top plot.
    }
    \label{fig:LogPlotRWGretCircular}
\end{figure}

Fig.~\ref{fig:LogPlotRWGretCircular} shows
$\sg\tilde{G}_\nt{ret}$, $\sg\tilde{G}^\nt{nd}_\nt{ret}$ and $-\tilde{V}_2$,
as well as their radial derivatives,
along a timelike circular geodesic at $r=r'=6M$. In this setting, we used  $\ell_\textrm{cut}=50$ for $\sg\tilde{G}_\textrm{ret}$ and $\sg\tilde{G}_\textrm{ret}^\textrm{nd}$, $\ell_\textrm{cut}=45$ for $\partial_r\sg\tilde{G}_\textrm{ret}$ and $\partial_r\sg\tilde{G}_\textrm{ret}^\textrm{nd}$ and the same $\ell_\textrm{max}=200$ for all of them.
Similarly to the spin-0 case, (i) there is a large matching region between $-\tilde{V}_2$ in a QL region and both $\sg\tilde{G}_\nt{ret}$ and $\sg\tilde{G}^\nt{nd}_\nt{ret}$ in a DP; (ii) the non-direct $\sg\tilde{G}^\nt{nd}_\nt{ret}$ supposes an improvement nearer to coincidence with respect to the `bare' GF $\sg\tilde{G}_\nt{ret}$.
In the spin-2 case here, the Pad\'e approximant for $-\tilde{V}_2$  reaches the end of the normal neighbourhood, although it does not extend beyond it, as it happened in the spin-0 case.

Let us now turn to the singularity structure at light-crossings.
When comparing  
the spin-2 RW 
$\sg\tilde{G}_\nt{ret}$
in the top plot of Fig.~\ref{fig:LogPlotRWGretCircular}
with the spin-0 GF $\sz\Gret$ in our Fig.~\ref{fig:PlotScalarGretCircularAndStatic} (and in Fig.~2 in~\cite{2023PhRvD.108d4033C}), we observe that, in the  circular setting, the singularity structure of 
$\sg\tilde{G}_\nt{ret}$, and so also of $\sg\Gret$, is the same as that of 
$\sz\Gret$, which is given by the $4$-fold cycle~\eqref{eq:4-fold}.
We note, however, that, in the spin-2  RW case, there seems more of an oscillatory behaviour in-between the light-crossing singularities. For example, before the first light-crossing singularity (a Dawson-like distribution approximating the $\text{PV}(1/\sigma)$ at  $\Delta t /M \approx 27.62$),  there  is a clear  maximum followed by a minimum. Similarly, just before the second light-crossing singularity (a negative-amplitude Gaussian distribution  approximating the $-\delta(\sigma)$ at  $\Delta t /M \approx 51.84$), there is a clear minimum followed by a rather sharp maximum.
These oscillations are genuine, physical features, unaffected by tweaking the values of $\ell_\textrm{cut}$ and $\ell_\textrm{max}$.
In App.~\ref{app:LightCrossingsAnalysis}, we show how the singularities of the 4-fold structure of the spin-2 RW GF $\sg\Gret$ along the circular geodesic arise as resonances between the various $\ell$-mode contributions $\sg\mathcal{G}_\ell$ (see Fig.~\ref{fig:RWEllModesNearLightCrossings}).

The middle and bottom plots in Fig.~\ref{fig:LogPlotRWGretCircular} are log-plots of, respectively, the same quantities as in the top plot and their radial derivatives.
The radial-derivative of the GF shares similar features as the GF, although $-\partial_r\tilde{V}_2$ does not quite reach the end of the normal neighbourhood.

\begin{figure}
\captionsetup[subfigure]{labelformat=empty}
    \centering
        \subfloat[]{
        \includegraphics[width=\linewidth]{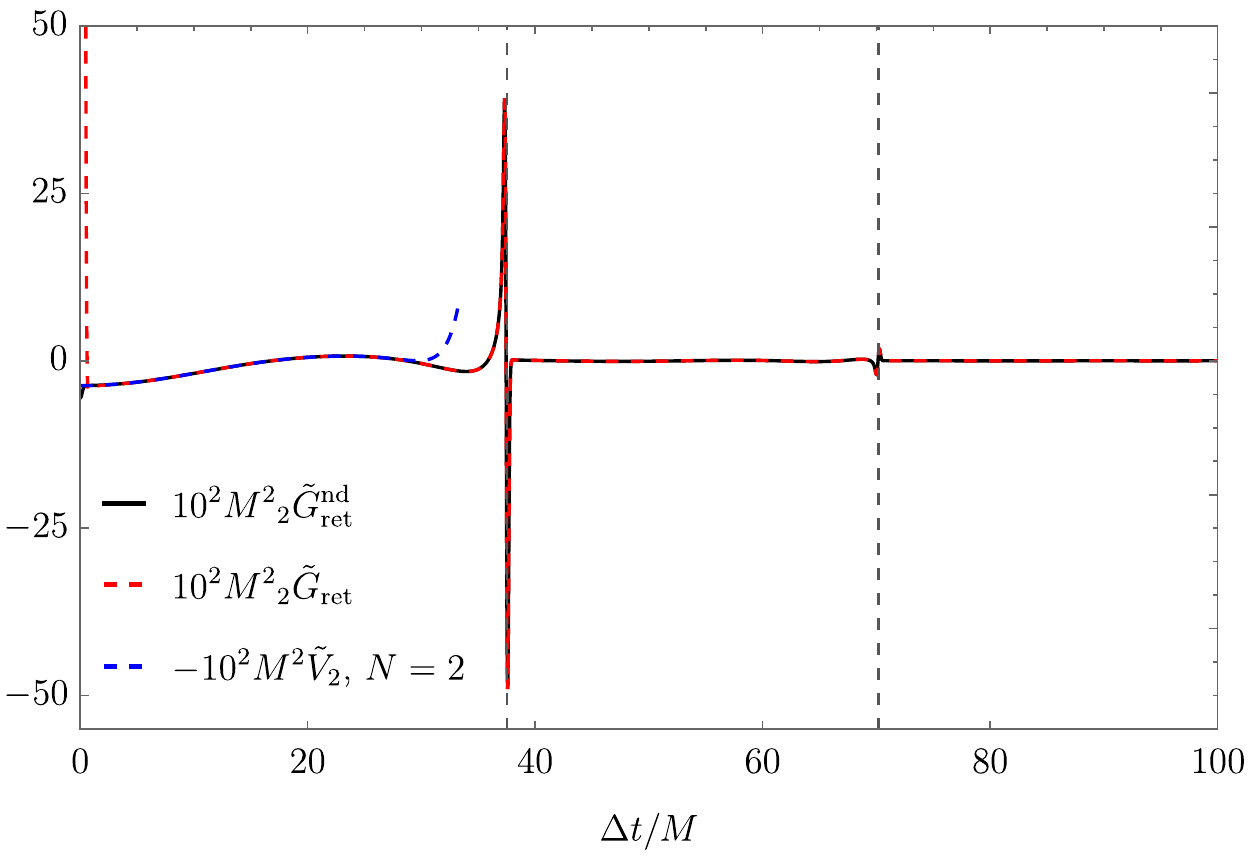}
    }
    \\
    \subfloat[]{
        \includegraphics[width=\linewidth]{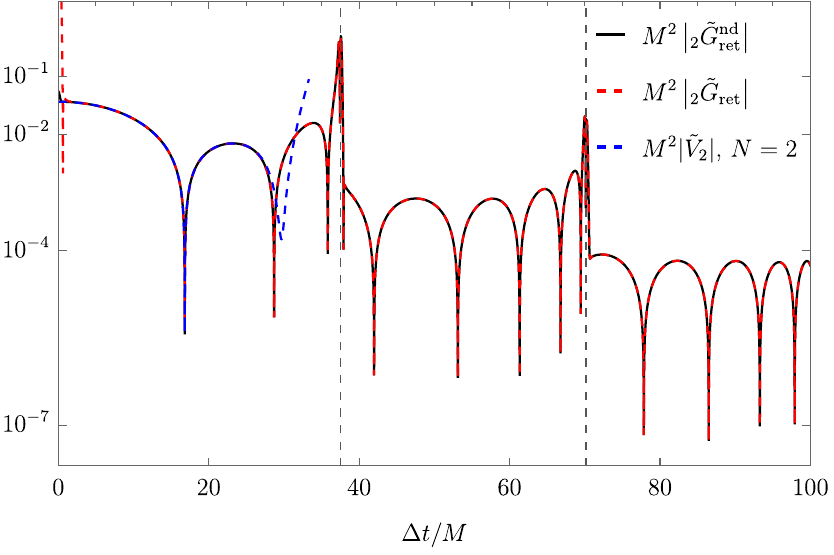}
    }
    \\
    \subfloat[]{
        \includegraphics[width=\linewidth]{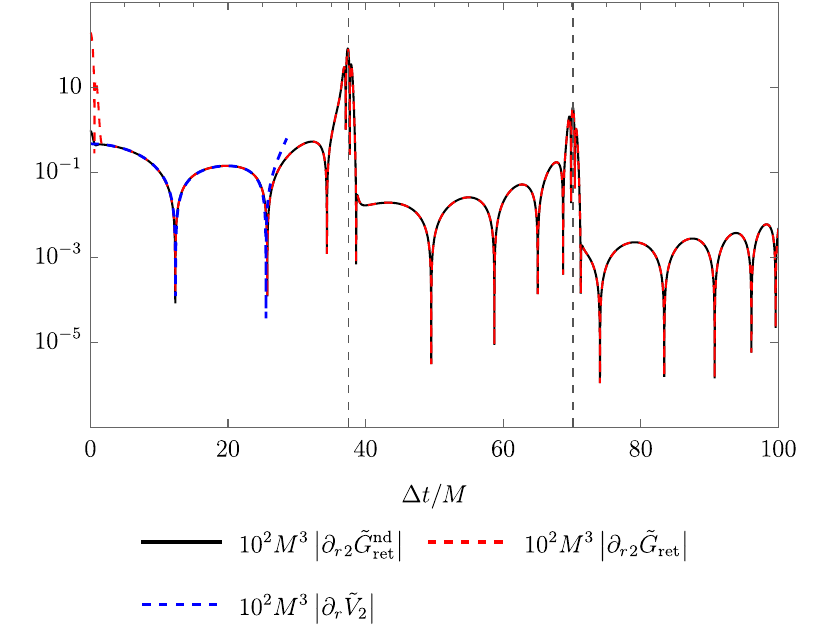}
    }
    
    \caption{
   Top: linear plot of 
   $\sg\tilde{G}_\nt{ret}$ (dashed red), $\sg\tilde{G}^\nt{nd}_\nt{ret}$ (solid black) and $-\tilde{V}_2$  (dashed blue), defined in Eq.~\eqref{eq:G-l01},
   as functions of $\Delta t$ for points  on a static worldline  at $r=6M$.
    The  dotted, gray vertical lines correspond to the  light-crossings as per the bottom plot in Fig.~\ref{fig:lightCrossings}.
      Middle: log-plot version of the top plot.
   Bottom: log-plot of the radial derivatives of the quantities in the top plot.
    }
    \label{fig:PlotsRWGretStatic}
\end{figure}

Fig.~\ref{fig:PlotsRWGretStatic} shows $\sg\tilde{G}_\nt{ret}$, $\sg\tilde{G}^\nt{nd}_\nt{ret}$ and $-\tilde{V}_2$,
as well as their derivatives, similarly to 
Fig.~\ref{fig:LogPlotRWGretCircular}, but on a static setting (instead of a circular geodesic). 
In this setting, we used
$\ell_\textrm{cut}=45$ for $\sg\tilde{G}_\textrm{ret}$ and $\sg\tilde{G}_\textrm{ret}^\textrm{nd}$, $\ell_\textrm{cut}=20$ for $\partial_r\sg\tilde{G}_\textrm{ret}$ and $\partial_r\sg\tilde{G}_\textrm{ret}^\textrm{nd}$ and the same $\ell_\textrm{max}=200$ for all of them.
The singularity structure of the spin-2 RW GF in the top plot of 
Fig.~\ref{fig:PlotsRWGretStatic} 
follows the same 2-fold pattern as the spin-0 GF in Fig.~\ref{fig:PlotScalarGretCircularAndStatic}. As opposed to the circular setting, in the static setting, the performance of the small coordinate expansion plus Pad\'e approximant for $\tilde{V}_2$ falls short of reaching the first light-crossing singularity. This might be related to the singularity structure in the static setting (and so at caustics) being ``stronger" than in the circular setting and containing fractional powers of $\sigma$ (see Eq.~\eqref{eq:2-fold,g=0}).


\section{Bardeen-Press-Teukolsky Green function}\label{sec:BPT}

The so-called Bardeen-Press-Teukolsky (BPT) equation~\cite{teukolsky1972rotating,Teukolsky:1973ha,barden-doi:10.1063/1.1666175},  containing a BPT spin\footnote{In BPT, ``$s$" is really a field  helicity but, following the common abuse of language in the literature, we  will refer to it as a spin.} $s$ parameter, is obeyed by  scalars $\Psi_s$ which represent scalar ($s=0$), electromagnetic ($s=\pm 1$) and gravitational ($s=\pm 2$) field perturbations of Schwarzschild spacetime.
In the case $s=0$, the BPT and RW equations coincide and are just the scalar wave equation.
In the case $s\neq 0$, the various BPT  scalars 
 $\Psi_s$ 
are constructed within the Newman-Penrose formalism \cite{newmanPenroseFormalismRef} by projecting the various field tensors onto a Newman-Penrose tetrad.
In the gravitational case of interest here, it is 
 $\Psi_2=\psi_0$ and  $\Psi_{-2}=r^4\psi_0$, where $\psi_0$ and $\psi_4$ are Weyl scalars --which are projections of the Weyl tensor-- and  $\Psi_{\pm 2}$ are gauge-invariant.
The BPT GF $\sGTret$ obeys the BPT GF equation:
\eqn{\label{eqn:fullTeukolskyEqn}
    \TOp
    \sGTret(x,x')=-4\pi\delta_4(x,x'),
}
where
\begin{align}
    \TOp\equiv \Box+\frac{2s}{r^2}\left(\frac{3M-r}{f}\pd{}{t}+(r-M)\pd{}{r}\right.+\quad\quad\quad \nonumber\\
    \left. i\frac{\cos\theta}{\sin^2\theta}\pd{}{\varphi}+\frac{1-s\cot^2\theta}{2}\right).\label{eq:BPT op}
\end{align}
Similarly to the RW GF, the causality condition on the BPT GF requires that $\sGTret(x,x')$ is the solution of Eq.~\eqref{eqn:fullTeukolskyEqn} such that it is equal to zero if the field point $x$ is not in the causal future of the base point $x'$. 

An important difference between the BPT operator $\TOp$ in \eqref{eq:BPT op} and the RW operator $\RWOp$ in~\eqref{eqn:RW op} is that, while the principal parts of both operators coincide with that of the wave operator  $\Box$ (thus establishing their hyperbolic character), the first-order derivatives in the RW operator coincide with those in $\Box$ but not with those in the BPT operator.
In fact, the coefficient of a first-order derivative (namely, that of $\partial_{\varphi}$)
in $\TOp$ is purely imaginary. The fact that a coefficient  is complex comes from the fact that elements of the Newman-Penrose tetrad on which the Weyl tensor is projected  to construct the BPT scalars $\Psi_s$ are complex-valued.
This of course means that the BPT GF $\sGTret$ will generally be complex-valued, as opposed to the RW GF, which was real-valued.

It will prove useful later to re-express the BPT 
operator in
\eqref{eq:BPT op} in the suggestive form
\eqn{\label{eqn:shortTeukolskyEqn}
    \TOp
    =
    \Box+A^\mu\partial_\mu+B,
}
where
\eqnalgn{
    A^\mu\equiv\,&\frac{2s}{r^2}\lP\lP-r+\frac{M}{f}\rP\delta^\mu_t+(r-M)\delta^\mu_r+i\frac{\cos\theta}{\sin^2\theta}\delta^\mu_\varphi\rP,\nonumber\\
    B\equiv\,&\frac{s}{r^2}(1-s\cot^2\theta).
}

In the next subsection we describe the Chandrasekhar transformation between homogeneous solutions of the RW and BPT equations. In the subsequent two subsections we 
develop the Hadamard form for the BPT GF in the QL region and present a method (in the frequency domain via the Chandrasekhar transformation) for calculating the BPT GF in the DP.
In the fourth, last subsection, we present our calculation of the BPT GF and its radial derivative in the same settings as for the RW GF: a timelike circular geodesic and a static worldline. We note that, in this BPT case, we do not carry out a calculation of the GF in the QL (nor a calculation of its direct part) -- we leave that for future work.


\subsection{Chandrasekhar transformation}\label{sec:Chandr transf}

Remarkably, Chandrasekhar~\cite{chandrasekhar1998mathematical} found that, for $s\leq0$, solutions $\Psi_{\s}$ of the homogeneous  version of the BPT Eq.~\eqref{eqn:fullTeukolskyEqn} (i.e., $\TOp \Psi_{\s}=0$) are related to  solutions $\chi\s$ of the homogeneous version of the RW Eq.~\eqref{eqn:RWEGret} (i.e., $\RWOp\, \chi_{\s}=0$) via 
\eqn{\label{eqn:4DChandrasekhar}
    \Psi\s=\eth^{|s|}\ChOp\,\chi\s,
}
where 
\eqn{\label{eqn:radialChandrasekharOperator}
   \ChOp\equiv \lP rf\rP^{|s|}\lP\pd{}{r}-f^{-1}\pd{}{t}\rP^{|s|}r^{|s|-1},
}
is referred to as the Chandrasekhar operator
and
\eqn{
    \eth=-\lP\Hsd{\theta}+\frac{\ii}{\sin\theta}\Hsd{\varphi}-s\cot\theta\rP
}
is the spin-raising (angular) operator. Similarly, the spin-lowering (angular) operator is
\eqn{
    \bar\eth=-\lP\Hsd{\theta}-\frac{\ii}{\sin\theta}\Hsd{\varphi}+s\cot\theta\rP.
}
When applied to spin-weighted spherical harmonics $\sYlm(\theta,\varphi)$~\cite{goldberg1967spin}, the spin-raising and lowering operators yield
\algn{
    \eth\lP\sYlm\rP=\,&+\sqrt{\ell(\ell+1)-s(s+1)}\,{}_{s+1}\Ylm,\\
    \bar\eth\lP\sYlm\rP=\,&-\sqrt{\ell(\ell+1)-s(s-1)}\,{}_{s-1}\Ylm,
}
and
\eqn{\label{eth^2Y}
\bar\eth\,\eth\,\sYlm-\lP \ell(\ell+1)-s(s+1)\rP\sYlm.
}

We choose the normalization of the spin-weighted spherical harmonics so that the orthonormality relation is
\eqn{
\int_{\mathbb{S}^2}\sYlm(\theta,\varphi)\,\s Y^*_{\ell',m'}(\theta,\varphi)\df\Omega=\delta_{\ell\ell'}\delta_{mm'}
}
and the completeness relation is
\eqn{\label{eq:Compl.Rln}
\sum_{\ell=|s|}^{\infty}\sum_{m=-\ell}^{\ell}\sYlm(\theta,\varphi)\,\s Y^*_{\ell m}(\theta',\varphi')=\delta(\cos\theta-\cos\theta')\delta(\varphi-\varphi').
}

Constructing BPT solutions via the Chandrasekhar operator is advantageous because it maps the BPT equation to the RW  equation, which has a short-range potential. This  simplifies the imposition of boundary conditions at the horizon and at infinity, improves numerical stability, and facilitates both analytical and numerical treatments of the solutions.
Let us next investigate  various ways for calculating the BPT GF $\sGTret$.


\subsection{BPT GF in a Quasi-Local region: Hadamard form}\label{sec:BPT-Had}

In this subsection we carry out an investigation of the Hadamard form of the BPT GF.

Within a normal neighbourhood of $x'$, the Hadamard form for $\s\Gret^T(x,x')$ can be written as (see Th.4.5.1 in \cite{friedlander})
\eqn{\label{eqn:hadamardTGret}
    \s\Gret^T=U^T(x,x')\delta_+(\sigma)-V^T_s(x,x')\theta_+(-\sigma),
}
where $U^T(x,x')$ and $V^T_s(x,x')$ are two smooth biscalars.
In a similar manner as in Sec.~\ref{subsec:quasilocalRegion}, 
Eq.~\eqref{eqn:hadamardTGret}  can be obtained as the $\epsilon\to 0^+$ limit of the functions  
\eqn{\label{eqn:epsilonTeukolskyHadamardForm}
    {}_{s}\Gret^{T,\epsilon}\equiv U^T\delta_+(\sigma+\epsilon)-V^T_s\theta_+(-\sigma-\epsilon).
}
Inserting Eq.~\eqref{eqn:epsilonTeukolskyHadamardForm} in 
Eq.~\eqref{eqn:fullTeukolskyEqn}
(with Eq.~\eqref{eqn:shortTeukolskyEqn})
and taking the limit $\epsilon\to0^+$ using the identities
\eqref{eq:ids}, \eqref{eqn:diracDeltaDerivativesIdentity} and \eqref{eq:ids-eps}, yields
\begin{widetext}
\begin{align}
\notag
    \TOp
    {}_{s}\Gret^{T}(x,x')=\,&-4\pi U^T\delta_4(x,x')+
    \delta_+(\sigma)\left\lbrace
    \TOp\,
    U^T
    +\lP2\sigma^\mu\partial_\mu+{\sigma^\mu}_\mu+\sigma_\mu A^\mu-2\rP V^T_s\right\rbrace+
    \notag\\    
    &
    \delta'_+(\sigma)
   \lP2\sigma^\mu\partial_\mu+{\sigma^\mu}_\mu +\sigma_\mu A^\mu -4\rP U^T
    +\theta_+(-\sigma)
     \TOp\,V^T_s.
        \label{eqn:TeukolskyGretEqnDecomp}
\end{align}
\end{widetext}
By comparing Eqs.~\eqref{eqn:TeukolskyGretEqnDecomp} and~\eqref{eqn:fullTeukolskyEqn},
we readily find that 
\eqn{\label{eqn:UTInitialCondition}
    \lC U^T
    \rC=
    1.
}
From  the term involving $\delta_+(\sigma)$ in~\eqref{eqn:TeukolskyGretEqnDecomp} and its absence in~\eqref{eqn:shortTeukolskyEqn}, we obtain the constraint for $V^T_s$ as a transport equation along null geodesics:
\begin{align}\label{eqn:VTBoundaryConditions}
    2\sigma^\mu\partial_\mu\check{V}^T_s+({\sigma^\mu}_\mu+\sigma_\mu A^\mu-2)\check{V}^T_s=
    -\left.
    \lP
 \TOp\,
    U^T\rP\right|_{\sigma=0},
\end{align}
where $\check{V}^T_s\equiv \left.V^T_s\right|_{\sigma=0}$.
The initial condition for the above transport equation is obtained by evaluating it at coincidence and imposing regularity. This yields, using \eqref{eqn:shortTeukolskyEqn} and~\eqref{eqn:UTInitialCondition},
\eqn{\label{eqn:hatVTsInitialCondition}
    \lC\check{V}_s^T\rC=\lC V_s^T\rC=-\frac{1}{2}\lC\Box U^T\rC-\frac{1}{2} \lC A^\mu\Hsd{\mu}U^T\rC-\frac{1}{2} B.
}

From the terms involving $\delta'_+(\sigma)$ and $\theta_+(-\sigma)$  in~\eqref{eqn:TeukolskyGretEqnDecomp} and their absence in~\eqref{eqn:shortTeukolskyEqn}, we obtain two independent differential equations for $U^T$ and $V^T_s$, 
\eqnalgn{\label{eqn:transportUT}
    \lP2\sigma^\mu\partial_\mu+{\sigma^\mu}_\mu+\sigma_\mu A^\mu -4\rP U^T&=0,\\
    \label{eqn:VTs}
    \TOp\,
    V^T_s&=0.
}
 Eq.~\eqref{eqn:transportUT} for $U^T$ was already written (in a different form) in Eq.88 in~\cite{2023PhRvD.108l5017I}.

In the case $s=0$, Eqs.~\eqref{eqn:VTBoundaryConditions}--\eqref{eqn:VTs}
 (so with $A^\mu=0$ and $B=0$), together with \eqref{eqn:UTInitialCondition}, agree (taking cognizance of $[\Box U]=0$) with Eqs.~\eqref{eqn:UTranspEq}--\eqref{eq:Vs x=x'}, as expected.

On the other hand, by comparing the transport equations \eqref{eqn:transportUT} and \eqref{eqn:UTranspEq}  for, respectively, $U^T$ and $U$, it follows that,
for $s\neq 0$, 
$U^T$ and $U=\Delta^{1/2}$ will generically  be different. This is due to the fact that the first-order derivatives in the BPT and RW operators are different.
Since $A^\mu$ is complex-valued, the BPT Hadamard biscalars $U^T$ and $V^T_s$ will generally be complex-valued, as opposed to the RW Hadamard biscalars $U$ and $V_s$, which were real-valued.

Let us briefly comment on potential methods for calculating $U^T$ and $V^T_s$.
To the best of our knowledge, at the moment there do not exist practical methods for calculating these BPT Hadamard biscalars. 
The analysis and calculation of $U^T$ will be addressed in a forthcoming paper.
Regarding $V^T_s$, by comparing Eqs.~\eqref{eqn:VTBoundaryConditions}--\eqref{eqn:hatVTsInitialCondition} and \eqref{eqn:VTs} for it, with Eqs.~\eqref{eq:PDE-V-RW}--\eqref{eqn:TranspEqVHat} and \eqref{eq:Vs x=x'} for the RW Hadamard tail $V_s$,  it becomes apparent that $V^T_s$ satisfies a characteristic initial value problem similar --but different-- to that satisfied by $V_s$. Thus, the methods we detailed in Sec.~\ref{Sec:RW-QL-tail} for calculating $V_s$ could be adapted for $V_s^T$. Unfortunately, this adaptation is not straightforward. For instance, the Hadamard-WKB method introduced in \cite{CDOWb} (and generalised to arbitrary RW spin $s$ in Sec.~\ref{Sec:RW-QL-tail}) for obtaining a small-coordinate expansion for $V_s$ does not readily apply to $V^T_s$ since Eq.~\eqref{eqn:VTs} lacks spherical symmetry. 
Lastly, it is worth noting that Ref.~\cite{2023PhRvD.108l5017I} provides transport equations for coefficients in a covariant expansion for  $V^T_s$  
as well as expressions for the first few coefficients in covariant expansions for $U^T$ and $V^T_s$.
Unfortunately, the transport equations are notoriously difficult to solve to high order even in the scalar case~\cite{Ottewill:2009uj} and  expressions for the coefficients are also difficult to obtain to an order high enough that could be useful in practise for our purposes. This is apart from the fact that $\sigma^\mu$ would need to be separately calculated.

Thus, we shall henceforth focus on calculating $\sGTret$ in the DP and leave  for future work the calculation of: (i) $V_s^T$, and so $\sGTret$, in the QL region; (ii) the $\ell$-modes of $U^T$, and so the non-direct part of $\sGTret$, in order to extend the DP closer to coincidence.


\subsection{Calculation of BPT GF in a Distant Past: Frequency domain}

In order to calculate the BPT GF $\sGTret$ we perform again a multipolar mode decomposition~\cite{PhysRevD.92.124055}, i.e.,
\begin{align}
\label{eqn:TGretEllModeDecomp}
    \s\Gret^T(x,x')&=
    \left(\Delta_S(r')\right)^s
    \sum\limits_{\ell=|s|}^{\infty}
\s\mathcal{G}^T_\ell(x,x'),
\\
\s\mathcal{G}^T_\ell(x,x')&\equiv
4\pi \s{}{G}^{T}_{\ell}\lP r,r',\Dt\rP
\sum\limits_{m=-\ell}^{\ell}\sYlm(\theta,\varphi)\sYlm^*(\theta',\varphi'),\nonumber 
\end{align}
where 
 $\s{}{G}^{T}_{\ell}$ are the $\ell$-modes of $\s\Gret^T$. From Eq.~\eqref{eqn:fullTeukolskyEqn}, and using Eqs.~\eqref{eth^2Y} and~\eqref{eq:Compl.Rln}, it may be shown that these $\ell$-modes are Green functions of the equation 
\eqn{\label{eq:GF eq-l}
    {}_s\hat{\mathcal{O}}^T_\ell\sGTl(r,r',\Dt)=-\lP\Delta_S(r')\rP^{-s}\delta(\Delta r)\delta(\Dt ),
}
where
\eqnalgn{\notag
    {}_s\hat{\mathcal{O}}^T_\ell\equiv\,&-\frac{r^2}{f(r)}\pd{^2}{t^2}+\pd{}{r}r^2f(r)\pd{}{r}+2s\lP-r+\frac{M}{f(r)}\rP\pd{}{t}\\
    &+2s(r-M)\pd{}{r}+s(s+1)-\ell(\ell+1).
    \label{eq:radial BPT op}
}

The $m$-sum in Eq.~\eqref{eqn:TGretEllModeDecomp} can be carried out in closed form by using the addition law for the spin-weighted spherical harmonics~\cite{Michel2020},
\begin{align}
&\sum\limits_{m=-\ell}^{\ell}\sYlm(\theta,\varphi)\sYlm^*(\theta',\varphi')=
\\&
(-1)^{s}\sqrt{\frac{2\ell+1}{4\pi}}e^{-i s\alpha}\s Y_{\ell,-s}\lP\gamma,\beta\rP,
\nonumber
\end{align}
where
\eqnalgn{\notag
\cot{\alpha}\equiv \,&\cos{\theta}\cot{(\varphi-\varphi')}-\cot{\theta'}\sin{\theta}\csc{(\varphi-\varphi')},\\\notag
\cot{\beta}\equiv\,&\cos{\theta'}\cot{(\varphi-\varphi')}-\cot{\theta}\sin{\theta'}\csc{(\varphi-\varphi')}.
}
We can thus rewrite $\s\mathcal{G}^T_\ell$ as
\eqn{\label{eqn:reducedTGretEllModeDecomp}
\s\mathcal{G}^T_\ell=
\sqrt{
4\pi(2\ell+1)
}\,
\sGl^T\lP r,r',\Dt\rP\,(-1)^{s}e^{-i s\alpha}\s Y_{\ell,-s}\lP\gamma,\beta\rP,
}

For calculating the $\ell$-modes  of the RW GF in Sec.~\ref{subsec:distantPastRegion}, we had two methods: one (CID) in the time-domain and one in the frequency-domain. For calculating the $\ell$-modes $\sGl^T$ of the BPT GF we will follow two methods in the frequency domain in the following subsection.
While progress towards a  time-domain method for calculating $\sGl^T$ was made in~\cite{o2022green}, there were some difficulties, which, to the best of our knowledge, remain unresolved.

\subsubsection
{BPT Fourier modes $\sGTwl$}
\label{sec:BPT Fourier}

The $\ell$-modes $\sGl^T$  of the BPT GF may be (inverse-)Fourier decomposed as
\eqn{
    \label{eqn:BPTFourierIntegral}
    \sGTl(r,r',\Dt)=
    \frac{1}{2\pi}
    \int_{-\infty}^{\infty}\sGTwl(r,r')e^{-\ii\omega\Dt}\df\omega.
}
The BPT Fourier modes $\sGTwl$ are solutions to the radial BPT GF equation,
\begin{align}
\OpOTw{s}{\ell\omega}\,
    \sGTwl(r,r')=
    -
    \lP\Delta_S(r')\rP^{-s}
    \delta(\Delta r),
    \label{eqn:radialTeukolskyEqn}
\end{align}
where
\begin{align}\label{eq:BPT op}
 \OpOTw{s}{\ell\omega} & \equiv\,
 r^2f(r)\dd{^2}{r^2}+2(r-M)(s+1)\dd{}{r}+4i\omega s r+
 \nonumber\\& 
 s(s+1)-
\ell(\ell+1)
    +\frac{r^2\omega^2-2i\omega s(r-M)}{f(r)}.
\end{align}
The boundary conditions obeyed by $\sGTwl$ are  inherited from the causality boundary conditions of the full GF $\sGTret$ as follows.

We construct the BPT Fourier modes $\sGTwl$ from suitable homogenous solutions to the radial BPT equation, similarly to how we constructed the RW Fourier modes $\sGwl$ in Sec.~\ref{sec:RWEFreq}.
 Let $\s{R}^{\nt{in}}_{\ell\omega}$ and $\s{R}^{\nt{up}}_{\ell\omega}$ be homogeneous solutions to Eq.~\eqref{eqn:radialTeukolskyEqn},   i.e.,
$\OpOTw{\s}{\ell\omega}\,\gRInUps=0$, with the following asymptotic conditions:
\algn{\label{eqn:sRInConditions}
    \s{R}^\nt{in}_{\ell\omega}\sim\left\lbrace
        \begin{aligned}
             &\s{R}_{\ell\omega}^{\nt{in,tra}}
              \Delta_S^{-s}
             e^{-i\omega r_*},\, r_*\to-\infty,\\
             &\frac{\s{R}_{\ell\omega}^{\nt{in,inc}}}{r}e^{-i\omega r_*}+\frac{\s{R}_{\ell\omega}^{\nt{in,ref}}}{r^{2s+1}}e^{i\omega r_*},\, r_*\to\infty
        \end{aligned}
    \right.
}
and
\algn{\label{eqn:sRUpConditions}
    \s{R}^\nt{up}_{\ell\omega}\sim\left\lbrace
        \begin{aligned}
            &\s{R}_{\ell\omega}^{\nt{up,inc}}e^{i\omega r_*}+
            \frac{\s{R}_{\ell\omega}^{\nt{up,ref}}}{\Delta_S^{s}}
            e^{-i\omega r_*},\, r_*\to-\infty,\\
            &\frac{\s{R}_{\ell\omega}^{\nt{up,tra}}}{r^{2s+1}}e^{i\omega r_*},\, r_*\to\infty,
        \end{aligned}
    \right.
}
where $\s{R}^{\nt{in,inc/ref/tra}}_{\ell\omega}$ and $\s{R}^{\nt{up,inc/ref/tra}}_{\ell\omega}$ are the incidence/reflection/transmission coefficients of, respectively, the `in' and `up' solutions. 

We have chosen the boundary conditions for the solutions $\gRInUps$ so that the causality condition for the full GF $\s\Gret^T$ means that its Fourier modes are given in terms of these homogeneous solutions as
\eqn{
    \label{eqn:TeukolskyFourierMode}
    \s{G}^T_{\ell\omega}(r,r')=-\frac{\s{R}^{\nt{in}}_{\ell\omega}(r_<)\s{R}^{\nt{up}}_{\ell\omega}(r_>)}{\mathbb{W}^T(\s{R}^{\nt{in}}_{\ell\omega},\s{R}^{\nt{up}}_{\ell\omega})},
}
where the BPT ``Wronskian"
\begin{align}\label{eqn:TEWronskian}
    &\mathbb{W}^T(\s{R}^{\nt{in}}_{\ell\omega},\s{R}^{\nt{up}}_{\ell\omega})\equiv 
    \Delta_S^{s+1}
    \lP\s{R}^{\nt{in}}_{\ell\omega}\dd{}{r}\s{R}^{\nt{up}}_{\ell\omega}-\s{R}^{\nt{up}}_{\ell\omega}\dd{}{r}\s{R}^{\nt{in}}_{\ell\omega}\rP
    \nonumber\\& =
        \s{R}_{\ell\omega}^{\nt{in,tra}}\s{R}_{\ell\omega}^{\nt{up,tra}}W^T,
   \\&
  W^T\equiv 2\ii \omega\,
   \frac{\s{R}_{\ell\omega}^{\nt{in,inc}}}
    {\s{R}_{\ell\omega}^{\nt{in,tra}}},
    \nonumber
\end{align}
is constant in $r$.

It is manifest from the symmetries of 
Eq.~\eqref{eqn:radialTeukolskyEqn}
as well as Eqs.~\eqref{eqn:sRInConditions}--\eqref{eqn:TeukolskyFourierMode},   that 
\eqn{\s{G_{\ell,-\omega}^T}(r,r')=\left(\s{G_{\ell\omega}^T}(r,r')\right)^*.}
Together with Eq.~\eqref{eqn:BPTFourierIntegral}, this implies that $\sGl^T$ is real-valued. In this way, the angles $\alpha$ and $\beta$ in Eq.~\eqref{eqn:reducedTGretEllModeDecomp} determine whether ${}_{-2}\Gret^T$ is real or complex. 
Henceforth we consider $\theta=\theta'=\pi/2$,
corresponding to the worldline settings we are interested in in this paper. This 
yields $\alpha=\beta=\pi/2$, and so a real-valued BPT GF ${}_{-2}\Gret^T$.

Let us next describe how we calculate $\s{G}^T_{\ell\omega}$ in practise from Eq.~\eqref{eqn:TeukolskyFourierMode}.
Its value at $\omega=0$ we know analytically, from~\cite{2024PhRvD.110d4007B}
\eqnalgn{
    {}_sR_{\ell\omega}^\nt{in}(r)=\,&\lP\Delta_s(r)\rP^{-s/2}P_\ell^s\lP \frac{r}{M}-1\rP,\\
    {}_sR_{\ell\omega}^\nt{up}(r)=\,&\lP\Delta_s(r)\rP^{-s/2}Q_\ell^s\lP \frac{r}{M}-1\rP,
}
leading to
\eqnalgn{\notag
    {}_{s}G^T_{\ell,0}(r,r')=\,&-\frac{\Gamma(\ell+1-s)}{\Gamma(\ell+1+s)}\lP\Delta_S(r)\Delta_S(r')\rP^{-s/2}\cdot \\\label{eq:GlT w=0}
    &P_\ell^{s}\lP\frac{\rL}{M}-1\rP Q_\ell^{s}\lP\frac{\rG}{M}-1\rP,
}
where $P_\ell^{s}$ and $Q_\ell^{s}$ are associated Legendre functions\footnote{If evaluating $Q_\ell^{s}$ in Eq.~\eqref{eq:GlT w=0} using Mathematica, one should use ``type 3" for its branch cut.}.

In order to calculate the various 
quantities appearing in the expression  \eqref{eqn:TeukolskyFourierMode} for the GF Fourier modes $\s{G}^T_{\ell\omega}$ for $\omega>0$  we follow two alternative ways. The first way uses the MST method in the BHPT for calculating the homogeneous solutions $\gRInUps$ and $\mathbb{W}^T$ via the last expression in \eqref{eqn:TEWronskian}.

The second way relies on the Chandrasekhar transformation detailed in Sec.~\ref{sec:Chandr transf}. 
The Chandrasekhar transformation in Eq.~\eqref{eqn:4DChandrasekhar} at the level of spacetime, yields a corresponding radial Chandrasekhar  transformation at the level of the Fourier modes. 
Let us write it explicitly in the case of BPT spin  $s=-2$, on which we shall henceforth focus~\footnote{The so-called Teukolsky-Starobinsky identities~\cite{chandrasekhar1998mathematical} relate homogeneous solutions of the BPT equation for spin $+s$  to those for spin $-s$.
}.
If 
${}_sX_{\ell\omega}(r)$ 
is a homogeneous solution of the spin-2 radial RW equation \eqref{eqn:RWEFreqDomain}, i.e.,
$\OpORW{2}{\ell\omega}{}_sX_{\ell\omega}=0$, then
\eqn{\label{eq:RW-BPT slns}
    {}_{-2}{R}_{\ell\omega}(r)
    \equiv \OpCh{-2}{r}{\omega}\,
    {}_{2}X_{\ell\omega},
}
where~\cite{chandrasekhar1998mathematical}
\eqn{\label{eqn:radialChandrasekharOp}
    \OpCh{-2}{r}{\omega}\equiv \lP rf\rP^{2}\lP\dd{}{r}+i\frac{\omega}{f}\rP^{2}r,
}
is a homogeneous solution of the $s=-2$ radial BPT equation, i.e.,
$\OpOTw{-2}{\ell\omega}\,{}_{-2}{R}_{\ell\omega}=0$.
Since $\gXInUp$, defined via Eqs.~\eqref{eqn:RWEFreqDomain}--\eqref{eqn:sXUpConditions}, are homogeneous solutions of the radial RW equation, it follows  that the functions
\eqnalgn{\label{eq:TRin/up Chandr}
 {}_{-2}{\mathcal{R}}^{\nt{in/up}}_{\ell\omega}(r)\equiv
    \OpCh{-2}{r}{\omega}\gXInUp(r)=\OpChR{-2}{r}{\omega}\gXInUp(r),
}
where
\eqnalgn{\label{eq:ChandrOp-R}
&
   \OpChR{-2}{r}{\omega}\equiv 2 r f\lC r(1+i\omega r)-3 M\rC\dd{}{r}+
    \ell(\ell+1)r f+
     \nonumber  \\&
    \frac{12 M^2}{r}-6 M (1+i\omega r)+2 r^2\ii \omega  (1+i\omega r),
}
are homogeneous solutions of the $s=-2$ radial BPT equation, i.e., $\OpOTw{\s}{\ell\omega}\,{}_{-2}{\mathcal{R}}^{\nt{in/up}}_{\ell\omega}=0$.
In the second equality in Eq.~\eqref{eq:TRin/up Chandr} we have used the fact that 
$\gXInUp$, are  solutions of the homogeneous radial RW equation (Eq.~\eqref{eqn:RWEFreqDomain}) in order to express their second-order  derivatives as a linear combination of $\gXInUp$ and their first-order  derivatives. Thus, the second-order operator $\OpCh{-2}{r}{\omega}$ has been reduced to the first-order operator $\OpChR{-2}{r}{\omega}$ when acting on homogeneous RW solutions.
Furthermore, ${}_{-2}{\mathcal{R}}^{\nt{in}}_{\ell\omega}$ and ${}_{-2}{\mathcal{R}}^{\nt{up}}_{\ell\omega}$ obey the boundary conditions, respectively, \eqref{eqn:sRInConditions} and \eqref{eqn:sRUpConditions} with the following specific values for the tranmission coefficients $\s{R}_{\ell\omega}^{\nt{in,tra}}$ and $\s{R}_{\ell\omega}^{\nt{up,tra}}$ which are given in Eq.~(B3) in~\cite{PhysRevD.92.124055}:
\algn{\label{eqn:tRInTra}
    \begin{aligned}
        {}_{-2}{\mathcal{R}}^\nt{in,tra}_{\ell\omega}=\,&\frac{2(\alpha_++2M^2\beta_+)}{M},\\
        {}_{-2}{\mathcal{R}}^\nt{up,tra}_{\ell\omega}=\,&-4\omega^2,
    \end{aligned}
    }
    where
\begin{widetext}
\algn{
    \alpha_+ \equiv\,&\frac{\lP\ell(\ell+1)-3\rP}{4M^2(1-4M\ii\omega)},\\
    \beta_+ \equiv\,&\frac{12+\ell^2(\ell+1)^2-\ell(\ell+1)(2-16M\ii\omega)-4(2\ell(\ell+1)+20M\ii\omega-3)+20M\ii\omega}{64M^4(1-2M\ii\omega)(1-4M\ii\omega)}. \nonumber
}
\end{widetext}
    The specific values in  Eq.~\eqref{eqn:tRInTra} are determined from Eqs.~\eqref{eqn:sXInConditions}--\eqref{eqn:sXUpConditions} having unit RW transmission coefficients.

We can now rewrite the Fourier mode of the BPT GF in Eq.~\eqref{eqn:TeukolskyFourierMode} as (following~\cite{PhysRevD.92.124055})
\begin{widetext}
\eqnalgn{\nonumber
    {}_{-2}{G}^T_{\ell\omega}(r,r')=\,&-\frac{{}_{-2}{\mathcal{R}}^\nt{in}_{\ell\omega}(r_<){}_{-2}{\mathcal{R}}^\nt{up}_{\ell\omega}(r_>)}{{}_{-2}{\mathcal{R}}^\nt{in,tra}_{\ell\omega}{}_{-2}{\mathcal{R}}^\nt{up,tra}_{\ell\omega}W^T}
       =-\frac{W}{{}_{-2}{\mathcal{R}}^\nt{in,tra}_{\ell\omega}{}_{-2}{\mathcal{R}}^\nt{up,tra}_{\ell\omega}W^T}\,\frac{\lP\OpChR{-2}{r}{\omega}\gXIn(r)\rP_{r=r_<}\lP\OpChR{-2}{r}{\omega}\gXUp(r)\rP_{r=r_>}}{W}
    \\\notag
    =\,&-\frac{W}{{}_{-2}{\mathcal{R}}^\nt{in,tra}_{\ell\omega}{}_{-2}{\mathcal{R}}^\nt{up,tra}_{\ell\omega}W^T}\,\OpChR{-2}{r_<}{\omega}\OpChR{-2}{r_>}{\omega}\frac{\gXIn(r_<)\gXUp(r_>)}{W}\\\label{eqn:teukolskyFourierModes}
    =\,&\frac{W}{{}_{-2}{\mathcal{R}}^\nt{in,tra}_{\ell\omega}{}_{-2}{\mathcal{R}}^\nt{up,tra}_{\ell\omega}W^T}\,\OpChR{-2}{r_<}{\omega}\OpChR{-2}{r_>}{\omega}\sg{G}_{\ell\omega}(r,r'),
}
\end{widetext}
where $W$ is the RW Wronskian  in Eq.~\eqref{eqn:RWEWronskian} and we have used Eq.~\eqref{eqn:RWGellModes} in the last equality.
 In \eqref{eqn:teukolskyFourierModes}, the Wronskian ratio is given in Eq.~(B4) in Ref.~\cite{PhysRevD.92.124055}:
\algn{\label{eqn:WdWT}
    \frac{W}{W^T}=\,&\frac{\alpha_++2M^2\beta_+}{M\beta_\infty}
}
where
\eqn{
    \beta_\infty\equiv \,\frac{\ell(\ell+1)(2-\ell(\ell+1))+12M\ii\omega}{8\omega^2}.
}

\begin{figure}
    \centering
        \includegraphics[width=0.48\textwidth]{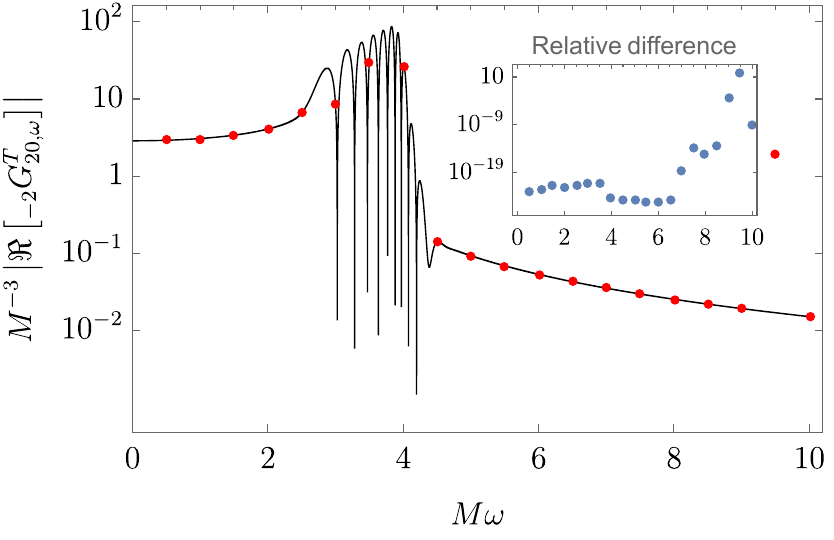}
        \includegraphics[width=0.48\textwidth]{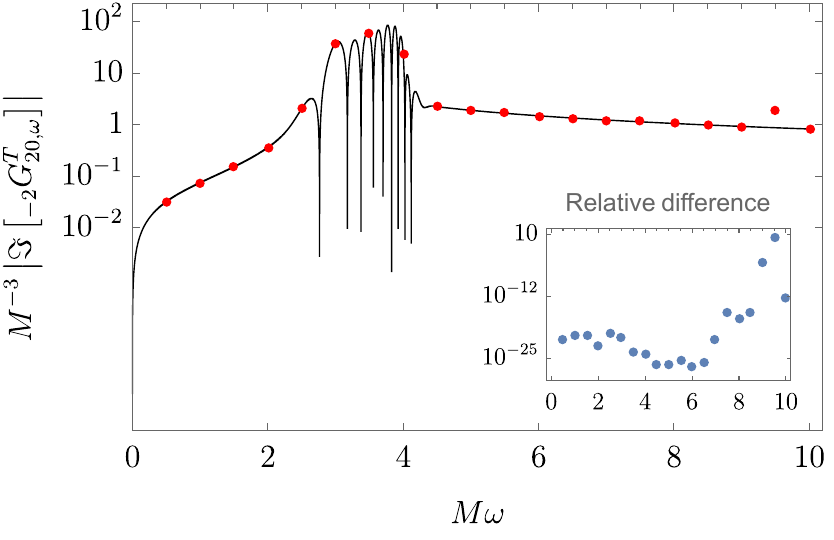}
    \caption{Absolute values of the real (top) and imaginary (bottom) parts of ${}_{-2}{G}^T_{\omega,20}$ as functions of $M\omega$ for $r=r'=6M$. The black solid line is obtained from  Eq.~\eqref{eqn:teukolskyFourierModes}    
    and the red dots 
    from Eq.~\eqref{eqn:TeukolskyFourierMode} with $\s{R}^{\nt{in/up}}_{\ell\omega}$ and $\mathbb{W}^T$ using the MST method in the BHPT.}
    \label{fig:teukolskyEll2FourierMode}
\end{figure}

In Fig.~\ref{fig:teukolskyEll2FourierMode} we plot ${}_{-2}{G}^T_{\ell\omega}$ for $\ell=20$  obtained in the two different ways just mentioned.
The black curve is ${}_{-2}{G}^T_{\omega,20}$ obtained via the second expression in Eq.~\eqref{eqn:teukolskyFourierModes}. 
In Eq.~\eqref{eqn:teukolskyFourierModes},
we use \eqref{eqn:tRInTra} for ${}_{-2}{\mathcal{R}}^\nt{in/up,tra}_{\ell\omega}$,  \eqref{eqn:WdWT} for $W/W^T$, and we calculate the various RW quantities 
as detailed in Sec.~\ref{sec:RWEFreq}, namely: Jaff\'e series for $\gXIn$ and its radial derivative, the numerical integration method in the BHPT for $\gXUp$ and its radial derivative, and the top line in \eqref{eqn:RWEWronskian} for  $W$. 
The red dots in Fig.~\ref{fig:teukolskyEll2FourierMode} correspond to values of ${}_{-2}{G}^T_{20,\omega}$ obtained directly with Eq.~\eqref{eqn:TeukolskyFourierMode} using the BHPT: its MST method  for ${}_{-2}{R}^{\nt{in/up}}_{\ell\omega}$ and $\mathbb{W}^T$ via \eqref{eqn:TEWronskian}.
The figure shows generically very good agreement between the two methods for calculating ${}_{-2}{G}^T_{20,\omega}$, thus validating both methods.
The relative difference in the results from the two methods, which is plotted in insets, 
shows a steady agreement in the precision up until $M\omega\approx7$. Thereafter, however, the relative difference starts increasing for the following reason. 
Internally, the BPHT for BPT relies on the MST method for calculating $\mathbb{W}^T$ as well as the boundary conditions for $\gRIn$.
Unfortunately,  as  $M\omega$ becomes larger, this implementation of the MST method 
requires a higher 
working precision
and becomes slower --
e.g., we start finding some performance difficulties for $M\omega\gtrsim7$ when using MST.
At even higher frequencies, around the region $M\omega\in(9,10)$ we just failed to obtain accurate enough values for $\sGTwl$ using the BHPT in this particular case -- see the couple of red dots off the curve at high frequencies in Fig.~\ref{fig:teukolskyEll2FourierMode} (we leave these reds dots in precisely to illustrate this point).
We thus decided to proceed with the calculation of  ${}_{-2}{G}^\nt{T}_{\ell\omega}$ 
by following the second method (namely, via  Eq.~\eqref{eqn:teukolskyFourierModes}
and $\gXInUp$ using Jaff\'e series/numerically using the BHPT)
rather than the first method (namely, via Eq.~\eqref{eqn:TeukolskyFourierMode} and $\s{R}^{\nt{in/up}}_{\ell\omega}$ numerically using the BHPT).

In addition to ${}_{-2}{G}^T_{\ell\omega}$, we also calculate its radial derivative, $\partial_r\lP{}_{-2}{G}^T_{\ell\omega}\rP$, by differentiating Eq.~\eqref{eqn:teukolskyFourierModes} and reducing arising second-order derivatives with $\OpORW{2}{\ell\omega}\lP\sgXInUp\rP=0$.
We note that the radial-coincidence limit of $\partial_r\lP{}_{-2}{G}^T_{\ell\omega}\rP$ depends on the direction of the limit, as expected since ${}_{-2}{G}^T_{\ell\omega}$ is a Green function of the ODE \eqref{eqn:TGretEllModeDecomp}, and as shown explicitly in Eq.~\eqref{eq:r-deriv GlwT,r=r'}.


\subsubsection{
BPT $\ell$-modes $\sgGTl$}\label{eq:BPT-ell-modes}

\begin{figure}
    \centering
        \includegraphics[width=0.9\linewidth]{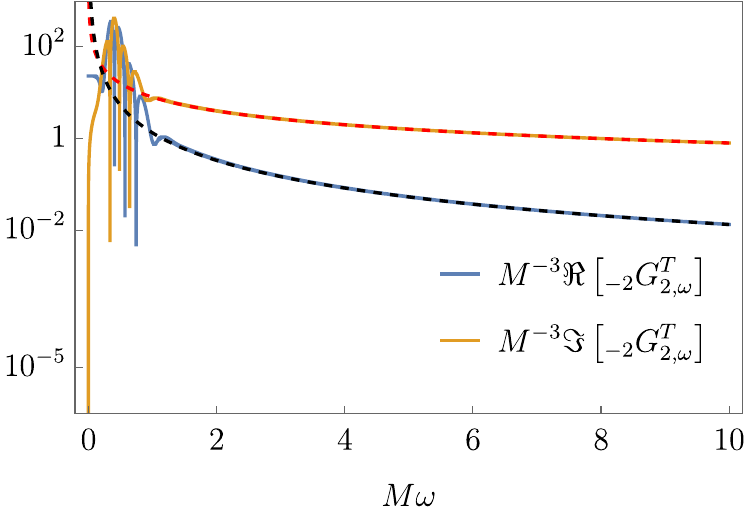}
    \\
        \includegraphics[width=0.9\linewidth]{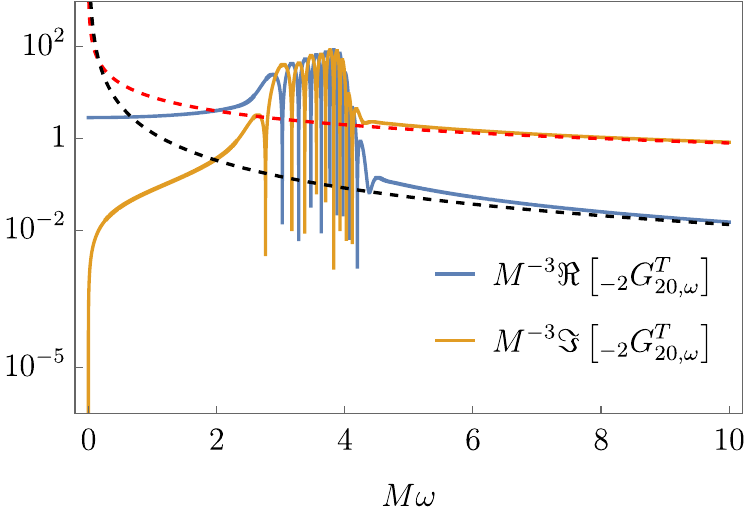}
    \caption{Fourier modes  (continuous curves) ${}_{-2}{G}_{\ell\omega}^T$, for $\ell=2$ (top) and $\ell=20$ (bottom), as well as their  asymptotic expansions (dashed curves) for large $\omega\in\mathbb{R}$  in Eq.~\eqref{eq:large-w BPT Fourier} (so leading orders for the real and imaginary parts),   as functions of $M\omega$ for $r=r'=6M$.}
    \label{fig:teukolskyFourierModesAndAsymptotics}
\end{figure}

Let us here analyse the convergence of the BPT Fourier integral in Eq.~\eqref{eqn:BPTFourierIntegral}.
Eq.~\eqref{eq:large-w BPT Fourier} shows that the real and imaginary parts of the $s=-2$ BPT Fourier modes ${}_{-2}{G}^T_{\ell\omega}(r,r)$ for $r'=r$ are of order $1/\omega^2$ and $1/\omega$, respectively,  for large $\omega$.
The behaviour of the real part is in sharp contrast to the RW case in Sec.~\ref{sec:RWEFreq}, where there was exponential decay. 
The exponential decay in the real part (but not imaginary part) of the RW modes with $r=r'$  is lost in the BPT modes since the (reduced) radial Chandrasekhar  operator \eqref{eq:ChandrOp-R} is complex-valued and so it mixes the real and imaginary parts of the RW modes.
This difference between the asymptotics of the RW and BPT Fourier modes at $r=r'$ can probably be traced to the radial RW operator \eqref{eq:radial RW op} being real-valued while the radial BPT operator \eqref{eq:radial BPT op} being complex-valued.
The large-$\omega$ behaviour of the BPT ${}_{-2}{G}^T_{\ell\omega}(r,r)$  as $1/\omega$ is, in a way, of no surprise, given that ${}_{-2}{G}^T_{\ell\omega}(r,r')$  has the distributional behaviour in \eqref{eqn:TeukolskyFourierMode}, which may be rewritten in terms of $\theta(\Delta r)$ and  $\theta(-\Delta r)$.
Any convergence  of the Fourier modes of the GF being faster than $1/\omega$ comes as an extra benefit.

In Fig.~\ref{fig:teukolskyFourierModesAndAsymptotics} we plot the first two leading orders in the asymptotics for large $\omega\in\mathbb{R}$ given in Eq.~\eqref{eq:large-w BPT Fourier} together with ${}_{-2}{G}^T_{\ell\omega}$ (obtained using Eq.~\eqref{eqn:teukolskyFourierModes}) for $\ell=2$ and $20$. 
The plots show agreement between the numerical $\Re(\sgGTwl)$   and $\Im(\sgGTwl)$ and their leading orders
in the large-$\omega$ regime.
We are able to differentiate three frequency regimes (more distinguishable as $\ell$ increases), similarly to the RW Fourier modes in Sec.~\ref{sec:RWEFreq}: first, a  monotonically-increasing region whose size grows as $\ell$ increases; second, a region where the functions rapidly oscillate 
with the frequency;  third, the asymptotic, power-law ($1/\omega$ and and $1/\omega^2$ for the imaginary and real parts respectively) decaying region. 
This third region is similar to that for the imaginary part of the RW Fourier modes; the exponential decay in the real part of the RW Fourier modes does not happen in the BPT case, as already mentioned.

For calculational purposes, we shall see that it is more convenient  to express the Fourier integral in Eq.~\eqref{eqn:teukolskyFourierModes} as
\eqnalgn{\notag
    {}_{-2}G^T_\ell(r,r',\Dt)=\,&\frac{2}{\pi}\theta(\Dt)\int\limits_{0}^{\infty}\Re\lP{}_{-2}{G}^T_{\ell\omega}(r,r')\rP\cos(\omega\Dt)\df\omega\\\label{eqn:realTGellFourierModeIntegral}
    =\,&\frac{2}{\pi}\theta(\Dt)\int\limits_{0}^{\infty}\Im\lP{}_{-2}{G}^T_{\ell\omega}(r,r')\rP\sin(\omega\Dt)\df\omega,
}
where, similarly  to the RW $\sGl$, we have used the properties
\algn{
    \sgGTl(r,r';\Dt)=\,&0,\quad &\forall\Dt<0,\\
    \lP{}_{-2}{G}^{T}_{\ell,-\omega}(r,r')\rP^*=\,&\sgGTwl(r,r'),\quad &\forall\omega\in\mathbb{R}.
    \nonumber
}

\begin{figure}
    \centering
    \includegraphics[scale=0.6]{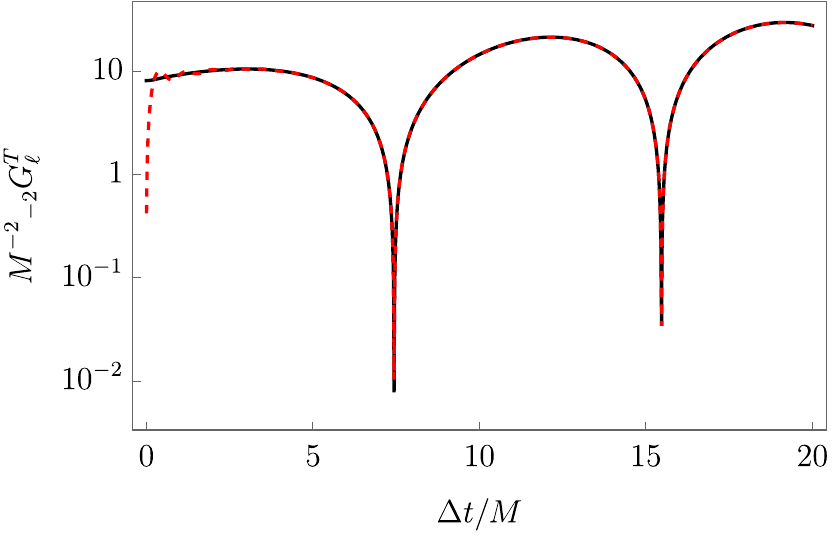}
    \caption{BPT $\ell=2$ mode ${}_{-2}{G}^{T}_{2}(r,r,\Dt)$ with $r=6M$ calculated using 
    the top (black solid line) and the bottom (red-dashed line) integrals in Eq.~\eqref{eqn:realTGellFourierModeIntegral}.
    }
    \label{fig:reducedTeukolskyEllMode}
\end{figure}

In practice, the upper limit of the integrals in Eq.~\eqref{eqn:realTGellFourierModeIntegral} must be truncated at a finite value.
Since we calculate ${}_{-2}{G}^{T}_{\ell,\omega}$ in Eq.~\eqref{eqn:realTGellFourierModeIntegral} from the RW solutions $\gXInUp$, we choose the same  upper limit $\omega_{\nt{max}}$ as in Sec.~\ref{sec:RWEFreq} for the RW case.
This cap will result in a less accurate value of $\sgGTl$ for $\Dt$ small, as   the large-$\omega$ region is responsible for the behaviour of $\sgGTl$ at early times.
We also include 
the large-$\omega$ smoothing factor in Eq.~\eqref{eq:smooth}, just as we did in the RW case.
In Fig.~\ref{fig:reducedTeukolskyEllMode} we plot ${}_{-2}{G}^{T}_{\ell=2}$ as resulting from the two integrals in Eq.~\eqref{eqn:realTGellFourierModeIntegral} after capping the upper limit and including the smoothing factor, for not too large values of $\Dt$. 
We observe that, after introducing the cap and the smoothing factor, there are still  spurious oscillations in the bottom integral but not in the top one. This is probably  because the top integral contains the real part of the Fourier modes, which decay like $1/\omega^2$, whereas the bottom integral has their imaginary part, which decays like $1/\omega$, and so slower.
We thus choose to use the top integral in Eq.~\eqref{eqn:realTGellFourierModeIntegral} for further calculations.
\begin{figure}
    \centering
    \includegraphics[width=\linewidth]{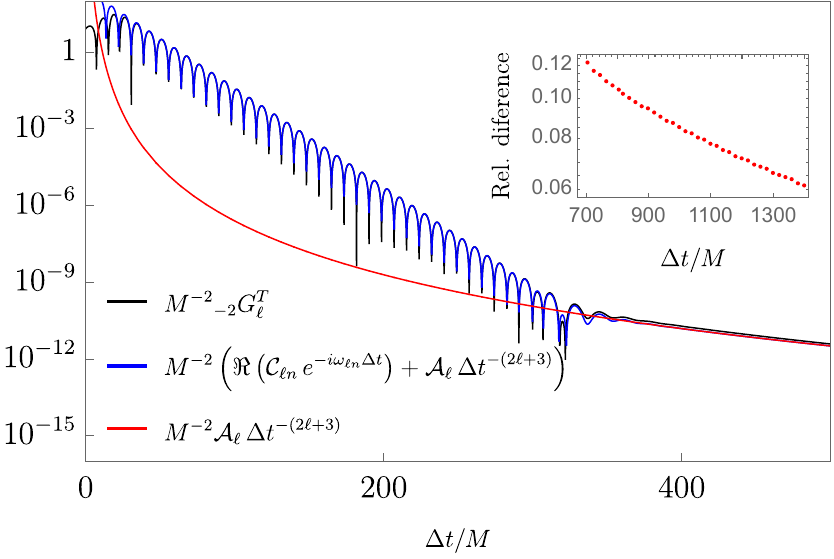}
    \caption{The black curve is ${}_{-2}{G}^{T}_{\ell}(r,r,\Dt)$ with $r=6M$ calculated with (the top integral in) Eq.~\eqref{eqn:realTGellFourierModeIntegral};
    the red curve is the leading late-time decay $\mathcal{A}_{\ell}\lP\Delta t\rP^{-(2\ell+3)}$ (see App.~\ref{sec:late-time});
    the blue curve
    is $\mathcal{A}_{\ell}\lP\Delta t\rP^{-(2\ell+3)}$ plus  the fundamental quasinormal mode $\Re\lP\mathcal{C}_{\ell,0}e^{-\ii\omega_{\ell,0}}\rP$ (see App.~\ref{sec:QNM}); 
    they are all for $\ell=2$.}
    \label{fig:latetimeTeukolskyEllMode2}
\end{figure}

We next investigate the behaviour of ${}_{-2}G^T_\ell$ at later times. 
It is well-known (e.g.,~\cite{Leaver:1986}) that, at later times, the $\ell$-modes of the GF should have, first, an oscillatory exponentially-decaying regime (the so-called {\it ringdown}), followed by a late-time asymptotic power-law decay regime.
Regarding the ringdown,
the time behaviour of ${}_{-2}G^T_\ell$
 can be modelled by 
$e^{-\ii\omega_{\ell n}\Dt}$, where the quasinormal mode (QNM) frequencies $\omega_{\ell n}\in\mathbb{C}$  are poles of the radial Green function $\sGTwl$ and --for a given value of $\ell$-- are labelled by the overtone number $n$;
typically, the dominant quasinormal mode is the fundamental overtone $n=0$. 
In its turn, the late-time power-law decay towards future timelike infinity $i^+$ is  $\Dt^{-(2\ell+3)}$~\cite{PhysRevD.5.2439,Price:1971fb}, and may be seen to arise from the small-frequency asymptotics of a branch cut of $\sGTwl$.

In App.~\ref{sec:spec} we derive the dominant contribution to ${}_{-2}G^T_\ell$ from the QNMs as well as its leading late-time tail, for $s=-2=-\ell$, $n=0$ and $r=r'=6M$. In App.~\ref{sec:QNM} it is seen that the oscillatory and exponentially-damped time behaviour due to the dominant QNM is 
$\Re\lP\mathcal{C}_{20}(6M,6M) e^{-\ii \omega_{2,0}\Dt}\rP$,
where $\omega_{2,0}$ is given in Eq.~\eqref{eq:QNN f} and $\mathcal{C}_{20}(6M,6M)$  in Eq.~\eqref{eq:C,QNM}.
As for the late-time asymptotics, App.~\ref{sec:late-time} yields the behaviour $\mathcal{A}_2\,
\lP\Delta t\rP^{-7}$, with
and ${}_{-2}\hat{R}^{in,0}_{2}(r)$ given in Eq.~\eqref{eq:Rin0,s=-2=-l}.

We have checked that our  ${}_{-2}G^T_{\ell}$ numerically calculated with the top integral in Eq.~\eqref{eqn:realTGellFourierModeIntegral} has the expected analytic behaviours in these two regimes.
In Fig.~\ref{fig:latetimeTeukolskyEllMode2}, the black curve corresponds to the numerically-calculated $\sgGTl(r,r',\Dt)$ for $\ell=2$ and $r=r'=6M$ over $\Dt:0 \to 600M$. 
In its turn, The red curve corresponds to the late-time asymptotic  $\mathcal{A}_2\,
\lP\Delta t\rP^{-7}$.
The blue curve corresponds to
$\mathcal{A}_2\,
\lP\Delta t\rP^{-7}$ plus
$\Re\lP\mathcal{C}_{20}(6M,6M) e^{-\ii \omega_{2,0}\Dt}\rP$.
Fig.~\ref{fig:latetimeTeukolskyEllMode2} shows 
good agreement between the numerical $\sgGTl$
and the analytic behaviours
in their corresponding regimes, thus offering a validation of our numerical calculation of $\sgGTl$.

Our Fig.~\ref{fig:latetimeTeukolskyEllMode2}
is for the BPT GF mode ${}_{-2}G^T_{\ell=2}$, which, to the best of our knowledge, is the first time that has been calculated. For comparison, the corresponding (but note that different radial factors are pulled out of the $\ell$-sums in Eqs.~\eqref{eqn:GretModeSum} and \eqref{eqn:TGretEllModeDecomp}) RW mode ${}_2 G_{\ell=2}$ has been plotted in Fig.~8 in~\cite{otoole2020characteristic} (see also Figs.~4--6 in~~\cite{2026arXiv260122015S}, although those are at $r=30M$, $r'=10M$).

Finally, we also calculate the radial derivative $\partial_r\lP{}_{-2}{G}^T_{\ell}\rP$ from the Fourier modes $\partial_r\lP{}_{-2}{G}^T_{\ell\omega}\rP$ by integrating them similarly as in Eq.~\eqref{eqn:realTGellFourierModeIntegral}.
Although the radial-coincidence limit of $\partial_r\lP{}_{-2}{G}^T_{\ell\omega}\rP$ depends on the direction of the limit (as mentioned in Sec.~\ref{sec:BPT Fourier}),
that is not the case for 
$\partial_r\lP{}_{-2}{G}^T_{\ell}\rP$  for $\Delta t>0$.
Still, the distributional nature of $\partial_r\lP{}_{-2}{G}^T_{\ell}\rP$   poses some challenges with the convergence of its Fourier integral and
in App.~\ref{sec:int dGlwdr-BPT} we explain how we deal with them.


\subsection{BPT GF  ${}_{-2}\Gret^T$ and radial derivative}

After describing how to calculate the $\ell$-modes ${}_{-2}G^T_{\ell}$, we move on to calculate the full BPT GF ${}_{-2}\Gret^T$ via the $\ell$-mode sum in Eq.~\eqref{eqn:TGretEllModeDecomp}. 
We truncate this sum at the finite value $\ell_\nt{max}=90$. 
Similarly to the RW case, 
this leads to 
spurious
oscillations in $\sgGTret$ which we  smooth out by introducing the smoothing factor in Eq.~\eqref{eq:smooth-l}.
We proceed similarly with the radial derivative of $\sgGTret$.

\begin{figure}
    \centering
    \subfloat[]{
        \includegraphics[width=\linewidth]{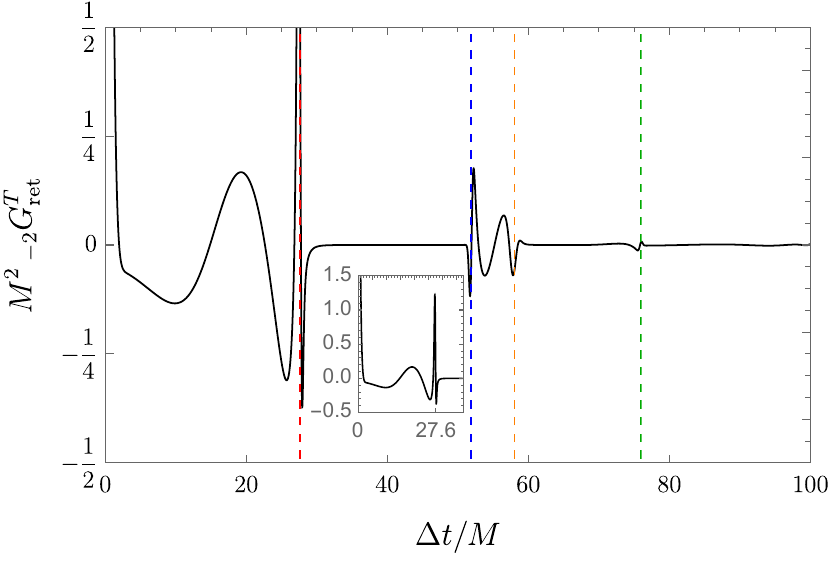}
    }\\
    \subfloat[]{
        \includegraphics[width=\linewidth]{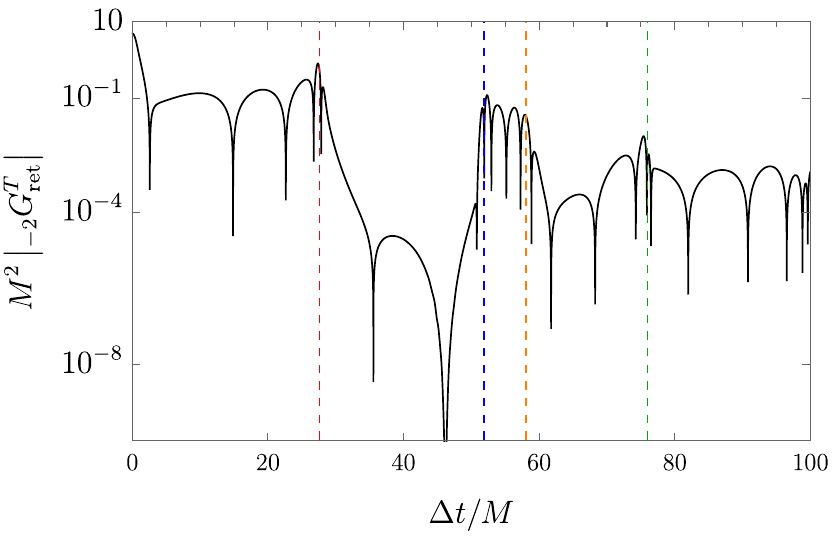}
    }\\
    \subfloat[]{
        \includegraphics[width=\linewidth]{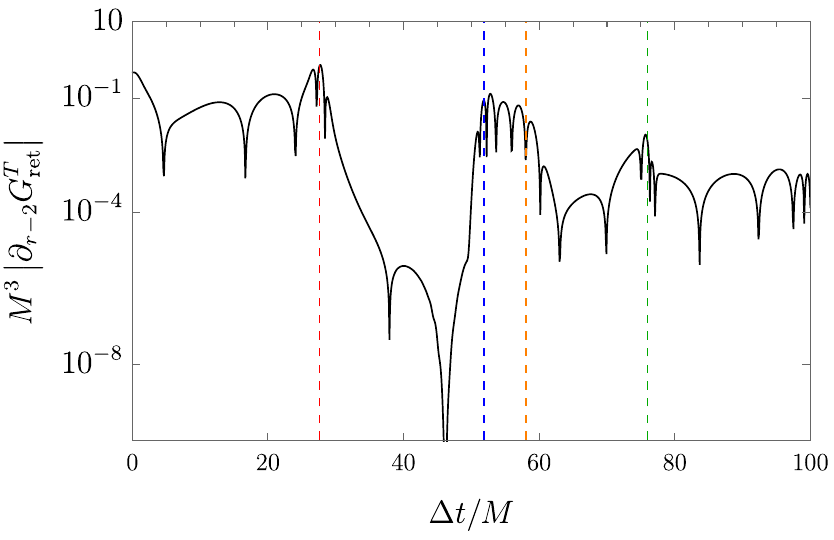}
    }
    \caption{
    Linear- (top) and log- (middle) plots of the BPT GF $\sgGTret$, together with  its radial derivative (bottom), for points on a timelike circular geodesic  at $r = 6M$.  
    The inset plot in the top shows the full scale of the (smeared) divergence at the first light-crossing.
    The  dashed vertical lines  correspond to the  light-crossings as per the top plot in Fig.~\ref{fig:lightCrossings}.
 }
    \label{fig:BPTGretCircular}
\end{figure}
\begin{figure}
    \centering
    \subfloat[]{
        \includegraphics[width=\linewidth]{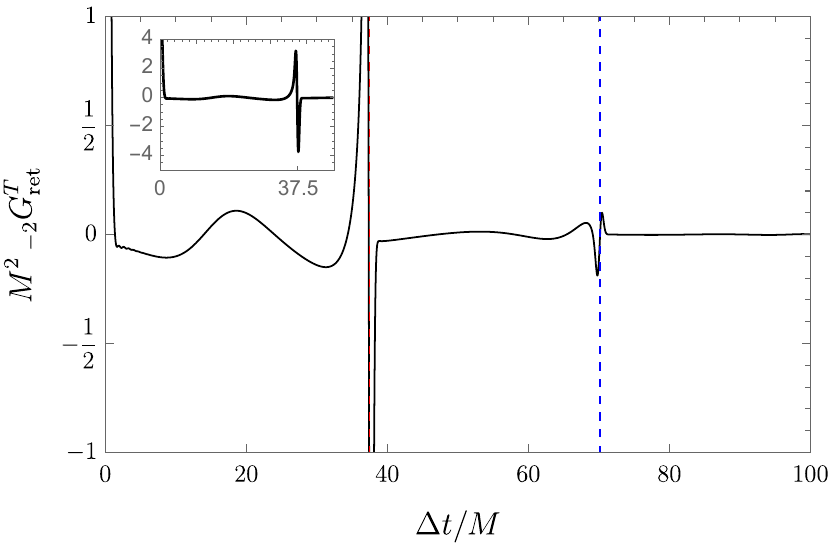}
    }\\
    \subfloat[]{
        \includegraphics[width=\linewidth]{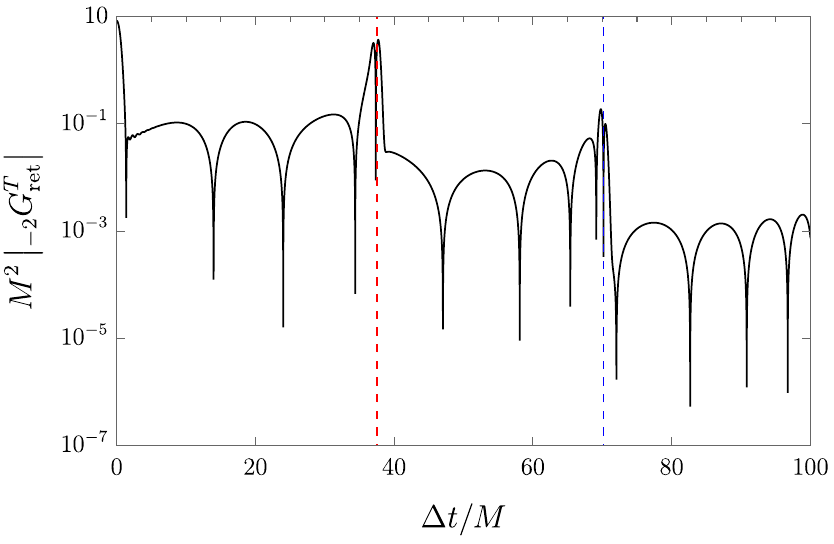}
    }\\
    \subfloat[]{
        \includegraphics[width=\linewidth]{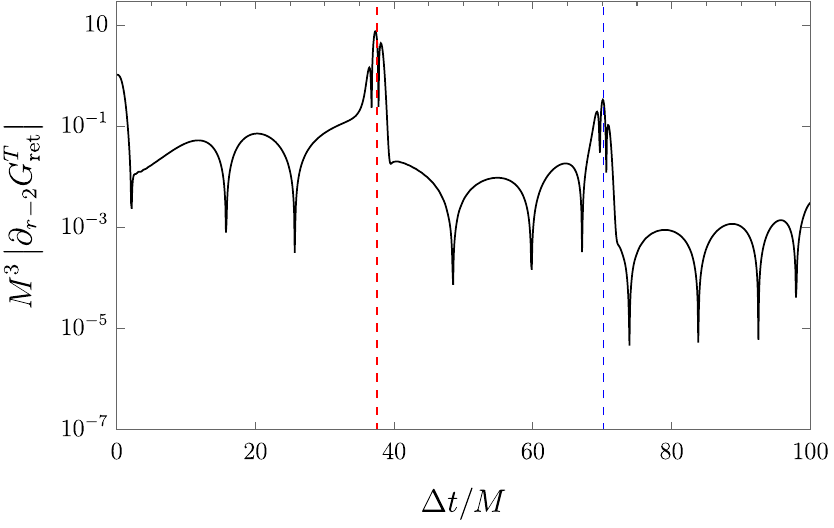}
    }
    \caption{Linear- (top) and log- (middle) plots of the  BPT GF $\sgGTret$, together with  its radial derivative (bottom), for points on a static worldline  at $r = 6M$.
    The inset plot in the top shows the full scale of the (smeared) divergence at the first light-crossing.
     The  dashed vertical lines  correspond to the  light-crossings as per the bottom plot in Fig.~\ref{fig:lightCrossings}.
    }
    \label{fig:BPTGretStatic}
\end{figure}

\begin{figure}
    \centering
    \subfloat{
        \includegraphics[width=0.98\linewidth]{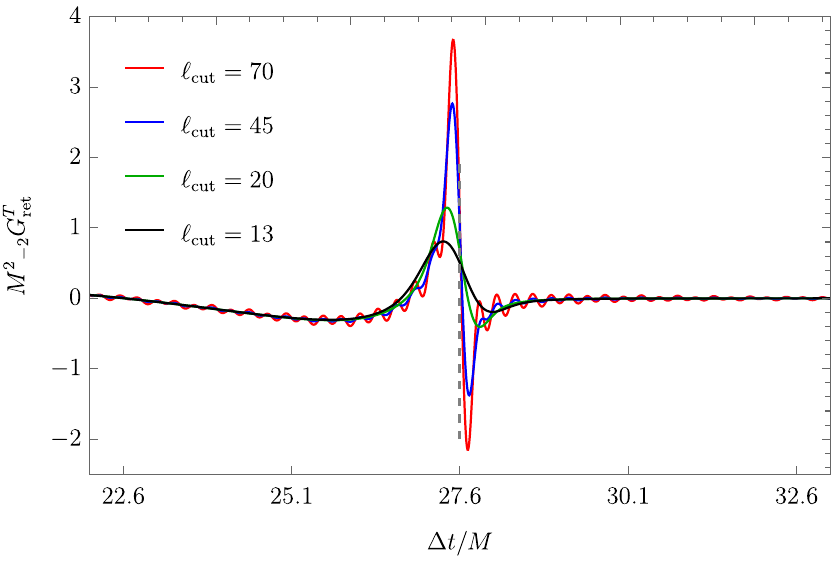}
    }\\
    \subfloat{
        \includegraphics[width=0.98\linewidth]{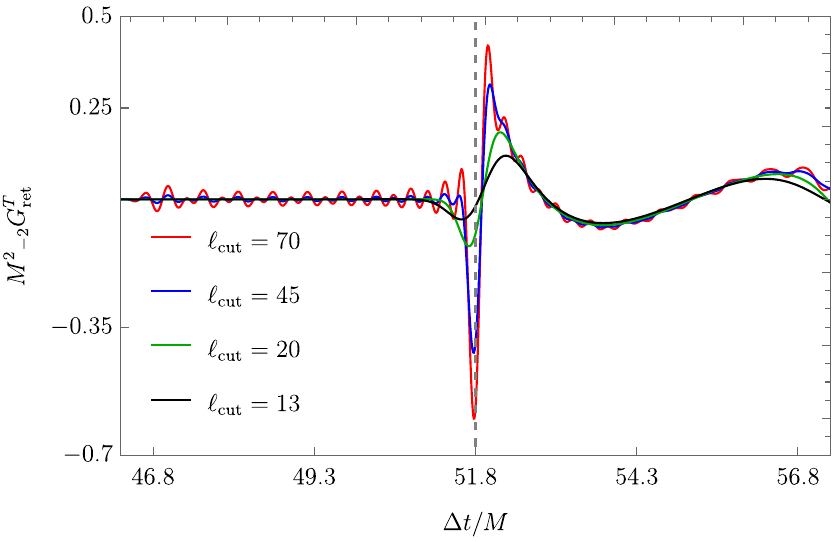}
    }
    \\
    \subfloat{
        \includegraphics[width=0.98\linewidth]{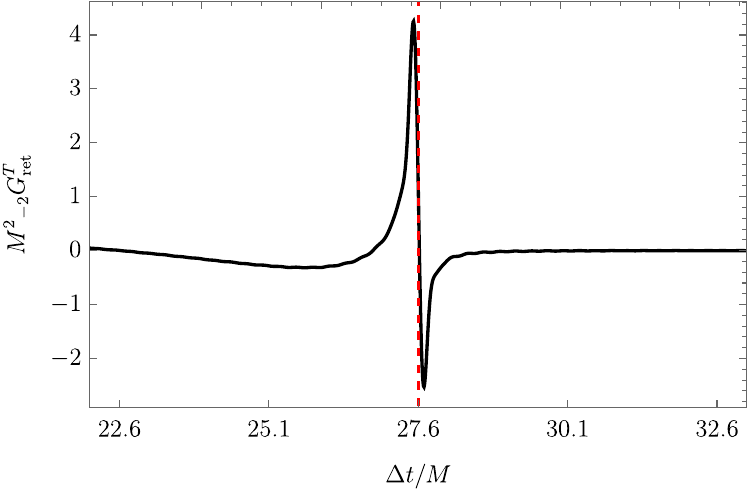}
    }
    \caption{BPT GF in the same setting as in Fig.~\ref{fig:BPTGretCircular} with different smoothing factors.
    Top and middle plots are 
    near the, respectively, first and second light-crossings  for varying values of $\ell_\nt{cut}$ in \eqref{eq:smooth-l}; bottom plot is near the first light-crossing using the smoothing factor in \eqref{eqn:secondSmoothingFactor}.}
    \label{fig:PlotBPTCircularGretNearSings}
\end{figure}

\begin{figure}
    \centering
    \subfloat{
        \includegraphics[width=0.98\linewidth]{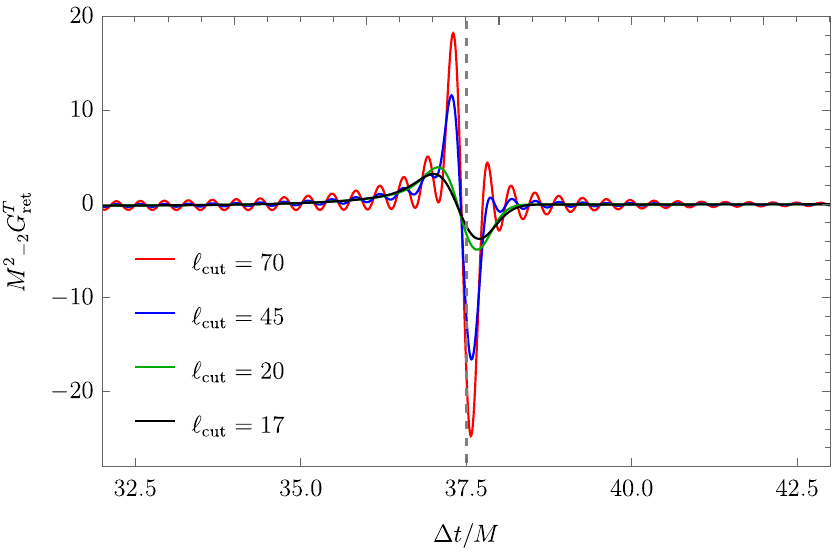}
    }\\
    \subfloat{
        \includegraphics[width=0.98\linewidth]{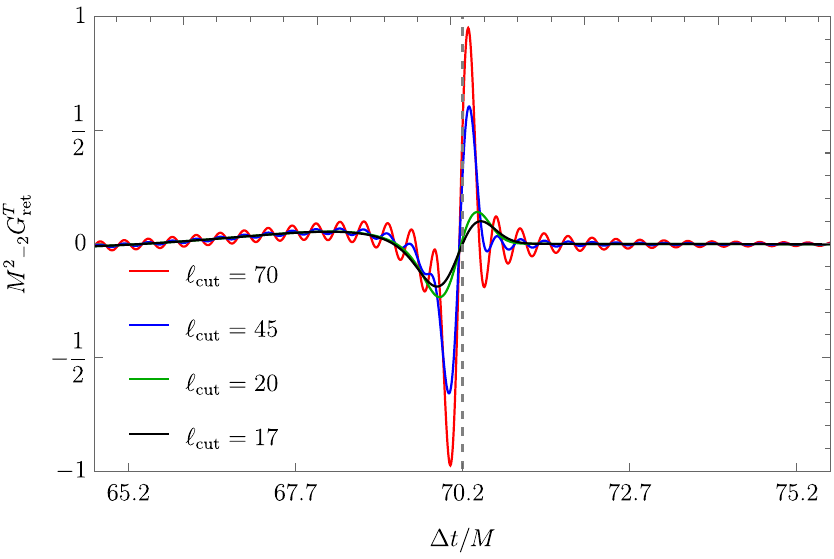}
    }
    \\
    \subfloat{
        \includegraphics[width=0.98\linewidth]{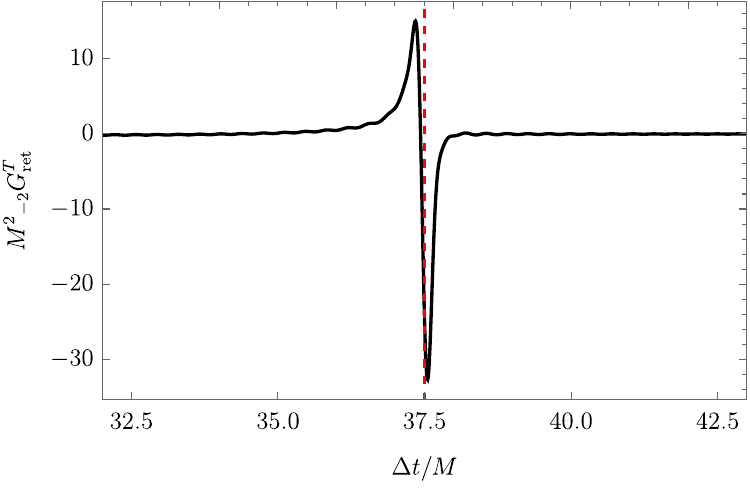}
    }
    \caption{
    BPT GF in the same setting as in Fig.~\ref{fig:BPTGretStatic} with different smoothing factors.
    Top and middle plots are 
    near the, respectively, first and second light-crossings  for varying values of $\ell_\nt{cut}$ in \eqref{eq:smooth-l}; bottom plot is near the first light-crossing using the smoothing factor in \eqref{eqn:secondSmoothingFactor}.
    }
\label{fig:PlotBPTStaticGretNearSings}
\end{figure}

In Figs.~\ref{fig:BPTGretCircular} and~\ref{fig:BPTGretStatic} we plot $\sgGTret$ and its radial derivative for the previous settings of points on, respectively, a timelike circular geodesic and a static worldline, with $r=r'=6M$ in both cases. 
We chose $\ell_\nt{cut}=13$ and $\ell_\nt{cut}=17$  in the smoothing factor  Eq.~\eqref{eq:smooth-l} for the circular  and static cases, respectively. We chose a  lower value for $\ell_\nt{cut}$ for the  BPT $\sgGTret$ than for the RW $\s\Gret$   because the spurious oscillations  (when not including the smoothing factor) are more marked  in the BPT GF than in the RW GF.
We note that the values 
near coincidence (more specifically for $\Dt$ approximately less than $3.6M$ and $6.2M$ for, respectively $\sgGTret$ and $\partial_r\lP\sgGTret\rP$ in Fig.~\ref{fig:BPTGretCircular} and than $2$ and $2.9$ for, respectively $\sgGTret$ and $\partial_r\lP\sgGTret\rP$ in Fig.~\ref{fig:BPTGretStatic}) should not be trusted, as they are ``contaminated" by the smeared divergence at coincidence (as mentioned, we leave the calculation near coincidence for future work).

It is useful to directly compare the results for the BPT, RW and spin-0 GFs.
More specifically, the BPT GF  $\sgGTret$  in the circular geodesic case plotted in Fig.~\ref{fig:BPTGretCircular} can be compared with the corresponding spin-0 GF ${}_0\Gret$ on the top row of Fig.~\ref{fig:PlotScalarGretCircularAndStatic} and the spin-2 RW GF $\sg\Gret$ on the top and middle plots of Fig.~\ref{fig:LogPlotRWGretCircular}.
Similarly, in the static case, the BPT GF in Fig.~\ref{fig:BPTGretStatic} can be compared with the spin-0 GF on the bottom row of Fig.~\ref{fig:PlotScalarGretCircularAndStatic} and the spin-2 RW GF on the top and middle plots of Fig.~\ref{fig:PlotsRWGretStatic}.

 Figs.~\ref{fig:BPTGretCircular} and~\ref{fig:BPTGretStatic} show that the BPT GF presents some true, physical oscillations near the divergences, which already started appearing in the spin-2 RW GF, while not really present in the spin-0 GF, and which are more enhanced  (in their frequency and amplitude) in the BPT case with respect to the RW case.

 Let us now turn to the divergences at the light-crossings themselves.
As mentioned earlier, the inclusion of Eq.~\eqref{eq:smooth-l}  as a factor in the $\ell$-summand, not only smoothes out the spurious oscillations but it also smears out the divergences of the GF at light-crossings.
Unfortunately, the small values of $\ell_\nt{cut}=17$ and $13$ make the magnitudes of some of the smeared divergences small enough that they are not properly distinguishable from the local extrema of the physical oscillations near the divergences. In order to be able to distinguish the smeared divergences from the local extrema, in Figs.~\ref{fig:PlotBPTCircularGretNearSings} and~\ref{fig:PlotBPTStaticGretNearSings}, we plot the BPT GF for increasing values of $\ell_\nt{cut}$ near the first and second light-crossing divergences. These plots show that, as $\ell_\nt{cut}$ is increased from $13$ (circular setting) or $17$ (static setting) to $70$, the magnitude of the peaks nearest to the light-crossing times are indeed enhanced, thus revealing that they are indeed smeared singularities, whereas the magnitudes of the peaks further from the light-crossing times barely change at all, thus revealing that they are local maxima of the physical oscillations.
 Of course, as $\ell_\nt{cut}$ increases, spurious oscillations appear on the top of the 
physical, lower-frequency oscillations (which do not become smoothed out even with the small values of $\ell_\nt{cut}=17$ and $13$).
Additionally, the bottom plots in Figs.~\ref{fig:PlotBPTCircularGretNearSings} and \ref{fig:PlotBPTStaticGretNearSings} show the singularity near the first light-crossings using the following  smoothing factor different from that in \eqref{eq:smooth-l}:
\begin{align}\label{eqn:secondSmoothingFactor}
&
\exp\lP-\frac{\ell^2}{2\ell_\textrm{cut}^2}\rP\lC1-\exp\lP-\frac{(\Dt-t_\textrm{NN})^2}{\delta}\rP\rC+
\\&
\exp\lP-\frac{(\Dt-t_\textrm{NN})^2}{\delta}\rP,
\nonumber
\end{align}
where we have introduced a new parameter $\delta>0$ and $t_\textrm{NN}$ was already introduced in \eqref{eq:Pade} as the first light-crossing time (namely, $27.62M$ and $37.5M$ in the circular and static settings, respectively).
We used \eqref{eqn:secondSmoothingFactor} with the parameter values $\ell_\textrm{cut}=35$ and $\delta=30$. This smoothing factor allows us to smear the spurious oscillations without dampening the singularity at the first light-crossing as much as using \eqref{eq:smooth-l}.

After having identified the (smeared) divergences in the BPT GF, we can see, by comparing Figs.~\ref{fig:PlotBPTCircularGretNearSings} and \ref{fig:PlotBPTStaticGretNearSings} with Figs.~\ref{fig:PlotScalarGretCircularAndStatic}, \ref{fig:LogPlotRWGretCircular}  and \ref{fig:PlotsRWGretStatic}, that its singularity structure is qualitatively the same as that for the RW GF and the spin-0 GF. That is, the singularity structures of all three GFs are
in agreement with the $4$-fold cycle in Eq.~\eqref{eq:4-fold} in the setting of the timelike circular geodesic (and so with light-crossings not at caustics) and with the corresponding $2$-fold cycle  in the setting of the  static worldline (and so with light-crossings at caustics).


\section{Discussion}\label{sec:discussion}

In this paper, we have achieved the calculation of the {\it full} spin-2 RW and BPT GFs.
We have seen that the singularity structures of these GFs follow the same $4$- and $2$-fold patterns as the spin-0 GF, given in, respectively, Eqs.~\eqref{eq:4-fold} and \eqref{eq:2-fold,g=0}, for when the divergences occur away from caustics and at caustics.
This result for the global divergences is probably not surprising, given that all these GFs admit the Hadamard form in normal neighbourhoods  (even if the direct and tail Hadamard bitensors may differ between these GFs), so that they all display the same form  for the local divergence.
On the other hand, new physical oscillations near the light-crossing divergences appear for the spin-2 RW GF, which were not present in the spin-0 case, and which become enhanced for the spin-2 BPT GF.

While for RW we achieved the calculation in both the QL region and DP,
for BPT
we achieved it
in a DP but only gave the equations obeyed by the Hadamard biscalars in a QL region.
Thus, in the BPT case, more
work still needs to be done with the Hadamard form for the calculation in a QL region to be matched  with a DP calculation.
We next give some ideas of how that could be done.

One idea is to make progress on the Hadamard form for BPT.
In particular, and following the calculations in RW, it would be good to do the following two calculations in BPT.
First, given the improvement in extending the DP  in the RW case by subtracting the $\ell$-modes of the direct part from those of the GF, it would be important to be able to calculate the  $\ell$-modes of the direct part in the BPT case and subtract them from the full BPT GF.
Second, in order to calculate the BPT GF in the QL region, one could try to calculate $V_s^T$ by carrying out an expansion in the coordinate distance, similarly to the way it is done in the RW case in~\cite{PhysRevD.77.104002,CDOWb} for spin-0 and in Sec.~\ref{Sec:RW-QL-tail} here for spin-2.

A valuable, alternative method for calculating GFs and which offers some level of analytical control was originally suggested in Ref.~\cite{Leaver:1986} and was already indicated in Sec.~\ref{eq:BPT-ell-modes}. It consists of first decomposing the GF into Fourier modes, then continuing the integration contour on to the complex frequency plane and finally using the residue theorem to pick up the contributions from the poles of the GF Fourier modes.
This method is long-known  to be very useful in the DP region,  where the contour is closed on the lower plane and contributions to the GF come from poles (quasinormal modes) and a branch cut integral along the negative imaginary axis.
Indeed, this method was applied to calculate the full scalar GF in the DP  in~\cite{CDOW13} and in App.~\ref{sec:spec} here we have applied it to the one BPT $\ell=2$ mode using only the dominant contributions from the branch cut an QNM.
Interestingly, a variation of this method has been recently suggested in~\cite{2025arXiv251018956A} 
and applied to one $\ell$(=2) mode of the spin-2 RW GF in~\cite{2026arXiv260122015S} to calculate the GF in a QL region (rather than DP). In this variation, the contour is closed on the {\it upper} (instead of lower) plane and the contribution to the GF comes from a branch cut along the positive imaginary axis (which arises in the Schwarzschild limit of the so-called Matsubara poles in the case of Schwarzschild-de Sitter spacetime). We leave the calculation of the {\it full} GF in the QL region using a `Matsubara branch cut' for future work.

Going beyond the calculation of the full GFs and looking at applications, an important one is in the calculation of the self-force. In the scalar case, the self-force may be obtained via the worldline integral in \eqref{eqn:scalarChargeTailIntegral}, where the GF has been `regularized' by subtracting from it a singular (Detweiler-Whiting) GF~\cite{Detweiler-Whiting-2003,lrr-2011-7}. While a similar worldline integral is known for calculating the {\it gravitational} self-force from a regularized GF of the {\it metric perturbation} equations (in the Lorenz gauge), it is not yet known how the RW or BPT GFs should be regularized in order to obtain (via some metric reconstruction procedure~\cite{PhysRevD.11.2042}) a regularized metric perturbation (in some radiation-like gauge). We leave the important task of appropriate regularization of the RW and BPT GFs for gravitational self-force purposes for future work.

\section*{Acknowledgments}

MC thanks Brien Nolan for helpful discussions.


\appendix
\begin{widetext}
\section{$\s v_{ijk}$ coefficients for $V_s(x,x')$}\label{app:sVijkCoefficients}

We  truncated the expansion in Eq.~\eqref{eqn:VCoordExpAnzats} for the RW Hadamard tail $V_s$ in the following way (see \cite{CDOWb}):
$$\sum\limits_{i,j,k=0}^{\infty}\to\sum\limits_{j=0}^{j_\nt{max}}\, \sum_{i=0}^{j_\nt{max}-j}\, \sum_{k=0}^{j_\nt{max}-j-i},$$
where we took $j_\nt{max}=21$. This value for $j_\nt{max}$ was determined by the computational power we had for calculating the general-RW-spin coefficients $\s v_{i,j,k}$. Following the Hadamard-WKB method provided in \cite{CDOWb} for the spin-0 case, and which we generalized to arbitrary RW spin $s$ and provide in \cite{SchwHadamardV}, we calculated the  required coefficients $\s v_{ijk}$. Here we give the form of the first seven  coefficients:
\begin{align}
    \s v_{0,0,0}&=-\frac{M s^2}{r^3},\nn
    \s v_{1,0,0}&=-\frac{M s^2 f}{4r^6}\left[M \left(s^2-4\right)+r\right],\nn
    \s v_{1,1,0}&=\frac{M^2f}{1680r^7}\left[2 M \left(70 s^6-665 s^4+784s^2+81\right)+9 r \left(35 s^4-56 s^2-9\right)\right],\nn
    \s v_{1,1,1}&=\frac{M^2}{1680 r^9}\left[-16 M^2 \left(70 s^6-665 s^4+784 s^2+81\right)\right.\nn
    &\quad\quad\quad\left.+14M r \left(7 \left(5 s^4-70 s^2+92\right) s^2+81\right)+27 r^2\left(35 s^4-56 s^2-9\right)\right],\nn
    \s v_{2,0,0}&=-\frac{M f}{1344 r^{10}}\left[4 M^3 \left(9-14 s^2 \left(s^4-20s^2+70\right)\right)+4 M^2 r \left(7 s^2 \left(s^4-30s^2+131\right)-9\right)\right.\nn
    &\quad\quad\quad\left.+M r^2 \left(147 s^4-1036s^2+9\right)+84 r^3 s^2\right],\nn
    \s v_{2,1,0}&=-\frac{M^2f}{4032 r^{11}}\left[6 M^3 \left(-2438 s^2+7 \left(s^4-34s^2+272\right) s^4-405\right)\right.\nn
    &\quad\quad\quad+M^2 r \left(-21 s^8+1078s^6-10808 s^4+15088 s^2+2853\right)\nn
    &\quad\quad\quad\left.-9 M r^2 \left(21 s^6-343s^4+543 s^2+119\right)+18 r^3 \left(-14 s^4+27s^2+7\right)\right],\nn
    \s v_{2,2,0}&=\frac{M^2f}{887040 r^{12}}\left[66 M^4 \left(14 s^{10}-735 s^8+10752s^6-41990 s^4+16974 s^2-13365\right)\right.\nn
    &\quad\quad\quad-22 M^3 r \left(21s^{10}-1610 s^8+28728 s^6-117690 s^4+21161 s^2-61560\right)\nn
    &\quad\quad\quad-15M^2 r^2 \left(23276 s^2+77 \left(5 \left(s^2-28\right)s^2+587\right) s^4+51050\right)\nn
    &\quad\quad\quad\left.+3 M r^3 \left(-3465 s^6+8250s^4+67551 s^2+62794\right)+150 r^4 \left(33 s^2\left(s^2-5\right)-113\right)\right]\nonumber.
\end{align}

\end{widetext}

\section{Jaff\'e series for the ingoing RW radial solution}\label{app:Jaffe}

In this appendix we give the details of the Jaffé series method for calculating the ingoing RW radial solution $\sXIn$, defined by Eqs.~\eqref{eqn:RWEFreqDomain} and \eqref{eqn:sXInConditions}.
We express the  solution as 
\eqn{\label{eqn:sXInAnsatz}
    \sXIn=e^{i\omega (r-4M)}\lP\frac{r}{2M}\rP^{1+s-\frac{1}{2}B_2-i\eta}\lP\frac{r}{2M}-1\rP^{-i2M\omega}
    u(f),
}
with $u(f)$ satisfying
\eqn{
f\,(1-f)^2u''(f)+(c_1+c_2f+c_3f^2)u'(f)+(c_4+c_5f)u(f)=0,
}
where 
\begin{align*}
    c_1\equiv \,&B_2+\frac{B_1}{2M},\\
    c_2\equiv \,&-2\lP c_1+1+i(\eta-2M\omega)\rP,\\
    c_3\equiv \,&c_1+2(1+i\eta),\\
    c_5\equiv\,&\lP\frac{1}{2}B_2+i\eta\rP\lP\frac{1}{2}B_2+i\eta+1+\frac{B_1}{2M}\rP,\\
    c_4\equiv \,&-c_5-\frac{1}{2}B_2\lP\frac{1}{2}B_2-1\rP+\eta(i-\eta)+i2M\omega c_1+B_3,
\end{align*}
and
\begin{align*}
    B_1\equiv&-2M(2s+1),\\
    B_2\equiv&\,2(1+s-i2M\omega),\\
    B_3\equiv&\,s(s+1)-\ell(\ell+1)-i2M(2s+1)\omega+8M^2\omega^2,\\
    \eta=&-2M\omega.
\end{align*}
Finally, we calculate $u(f)$  as a Taylor series expansion about $f=0$:
\eqn{\label{eqn:uJaffeAnzats}
    u(f)=\sum\limits_{n=0}^{\infty}a_n f^n,
}
where $a_0=1$ and the coefficients $a_{n>0}$  satisfy the following three-term recurrence relations:
\begin{align}
    &\alpha_0 a_1+\beta_0 a_0=0,\\
    &\alpha_n a_{n+1}+\beta_n a_n+\gamma_n a_{n-1}=0,\,n>0,
    \nonumber
\end{align}
with
\begin{align}
    \alpha_n\equiv\,&(n+1)\lP n+B_2+\frac{B_1}{2M}\rP,\\\notag
    \beta_n\equiv\,&-2n^2-2\lP B_2+i(\eta-2M\omega)+\frac{B_1}{2M}\rP n\\\notag
    \,&-\lP\frac{1}{2}B_1+i\eta\rP\lP B_2+\frac{B_2}{2M}\rP\\
    \,&+i\omega\lP B_1+2MB_2\rP+B_3,\\
    \gamma_n\equiv\,&\lP n-1+\frac{1}{2}B_2+i\eta\rP\lP n+\frac{1}{2}B_2+i\eta+\frac{B_1}{2M}\rP.
\end{align}

The radial derivative of $\sXIn$ is readily obtained by differentiating Eq.~\eqref{eqn:sXInAnsatz}. 


\section{Analysis of light-crossings as $\ell$-mode resonances}
\label{app:LightCrossingsAnalysis}

The full GFs diverge at light-crossings  as per the $4$-fold cycle in Eq.~\eqref{eq:4-fold} in the setting of the circular geodesic (so with light-crossings not at caustics)
and the $2$-fold cycle 
in Eq.~\eqref{eq:2-fold,g=0}
in the setting of the static worldline (so with light-crossings  at caustics) - see also Fig.~\ref{fig:lightCrossings}. In order to analyze these singularity structures, we turn to 
the summands
$\s\mathcal{G}_\ell$ and $\s\mathcal{G}^T_\ell$
in, respectively, Eqs.~\eqref{eqn:GretModeSum} and~\eqref{eqn:TGretEllModeDecomp}.
In Fig.~\ref{fig:RWEllModesNearLightCrossings} we plot $\sg \mathcal{G}_\ell$    for multiple values of $\ell$ near the first four light-crossings in the setting of the timelike circular geodesic at $r=6M$. In Fig.~\ref{fig:BPTEllModesNearLightCrossings}
we do similarly but for ${}_{-2}\mathcal{G}^T_\ell$ (instead of $\sg \mathcal{G}_\ell$) and near the first  light-crossing;
in Fig.~\ref{fig:BPTEllModesNearLightCrossingsStatic} we do similarly  for ${}_{-2}\mathcal{G}^T_\ell$ but in the static setting.
Figs.~\ref{fig:RWEllModesNearLightCrossings}--\ref{fig:BPTEllModesNearLightCrossingsStatic} 
show how the singularities  at light-crossings  of the full GFs arise as resonances between the summands 
in Eq.~\eqref{eqn:GretModeSum}
for different $\ell$'s. 

For definiteness, let us look deeper at Fig.~\ref{fig:BPTEllModesNearLightCrossings} for   ${}_{-2}\mathcal{G}^T_\ell$  in the setting of the  circular geodesic.
The smeared singularity of the full GF $\sgGTret$ at the first light-crossing at $\Delta t= t_1\approx 27.62M$ is of $\textrm{PV}(1/\sigma)$-type, and so with a  positive peak just before $\Delta t= t_\textrm{NN}$ followed by a second, negative peak just after $\Delta t= t_\textrm{NN}$. This can be seen in Fig.~\ref{fig:BPTGretCircular}, which also shows that the first peak is larger in magnitude than the second one.
Fig.~\ref{fig:BPTEllModesNearLightCrossings} shows that the $\ell$-summands ${}_{-2}\mathcal{G}^T_\ell$ oscillate in such a way that their last extremum before $\Delta t= t_\textrm{NN}$ is a local maximum and the first extremum after $\Delta t= t_\textrm{NN}$ is a local minimum.
As $\ell$ increases, the locations of these two extrema approach $\Delta t= t_\textrm{NN}$, with the amplitude of the maximum before $\Delta t= t_\textrm{NN}$ decreasing and 
the amplitude of the minimum after $\Delta t= t_\textrm{NN}$ increasing.
This shows how the $\text{PV}(1/\sigma)$ singularity of the GF $\sgGTret$ at $\Delta t = t_\textrm{NN}$  arises as resonances between  ${}_{-2}\mathcal{G}^T_\ell$  for increasing values of $\ell$.
Fig.~\ref{fig:BPTEllModesNearLightCrossings} also shows an asymmetry about how the peaks to the left and to the right of $\Delta t= t_\textrm{NN}$ arise.
For the peak to the left, the magnitude of the $\ell$-mode contribution diminishes with $\ell$ whereas, for the peak to the right, the magnitude increases.
This implies that using a smoothing factor like \eqref{eq:smooth-l} will affect the peak to the right more strongly than that to the left.
Also, the distances between the extrema of ${}_{-2}\mathcal{G}^T_\ell$  for consecutive $\ell$'s is smaller before $\Delta t= t_\textrm{NN}$ than after $\Delta t= t_\textrm{NN}$, with these distances decreasing with $\ell$.
This might account for the fact that 
the peak of the GF $\sgGTret$ just before  $\Delta t= t_\textrm{NN}$ is larger in magnitude than that after  $\Delta t= t_\textrm{NN}$.

\begin{widetext}
\begin{figure*}
    \centering
    \includegraphics[width=\linewidth]{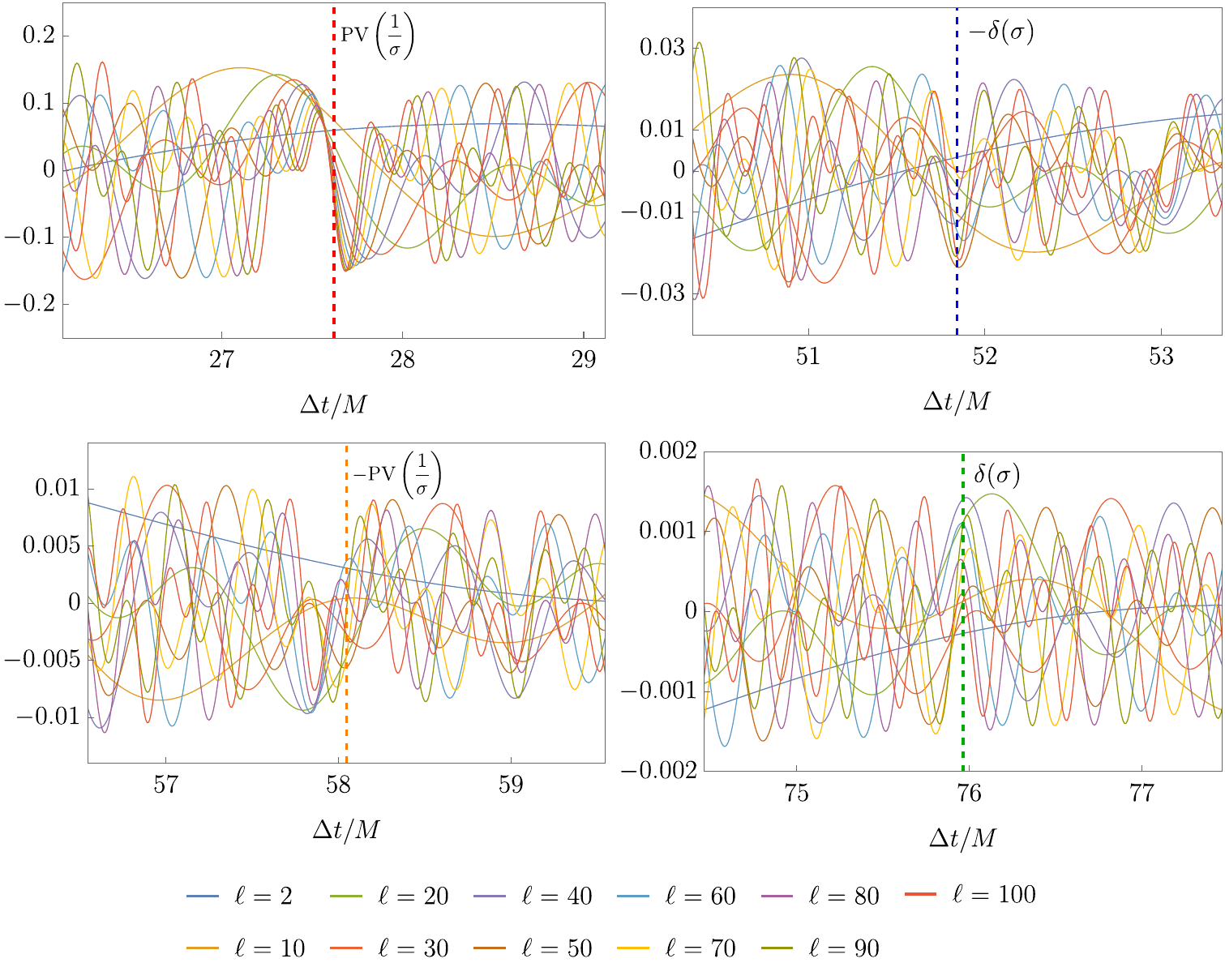}
    \caption{Plot of the RW summand  $\sg \mathcal{G}_\ell$ in Eq.~\eqref{eqn:GretModeSum} as a function of $\Delta t$ around the light-crossings  for various values of $\ell$. 
     The points are along a timelike circular geodesic of radius $r=6M$.
    We also include the expected singularity type corresponding to each light-crossing as given by Eq.~\eqref{eq:4-fold}.}
    \label{fig:RWEllModesNearLightCrossings}
\end{figure*}
\begin{figure*}
    \centering
    \includegraphics[width=\linewidth]{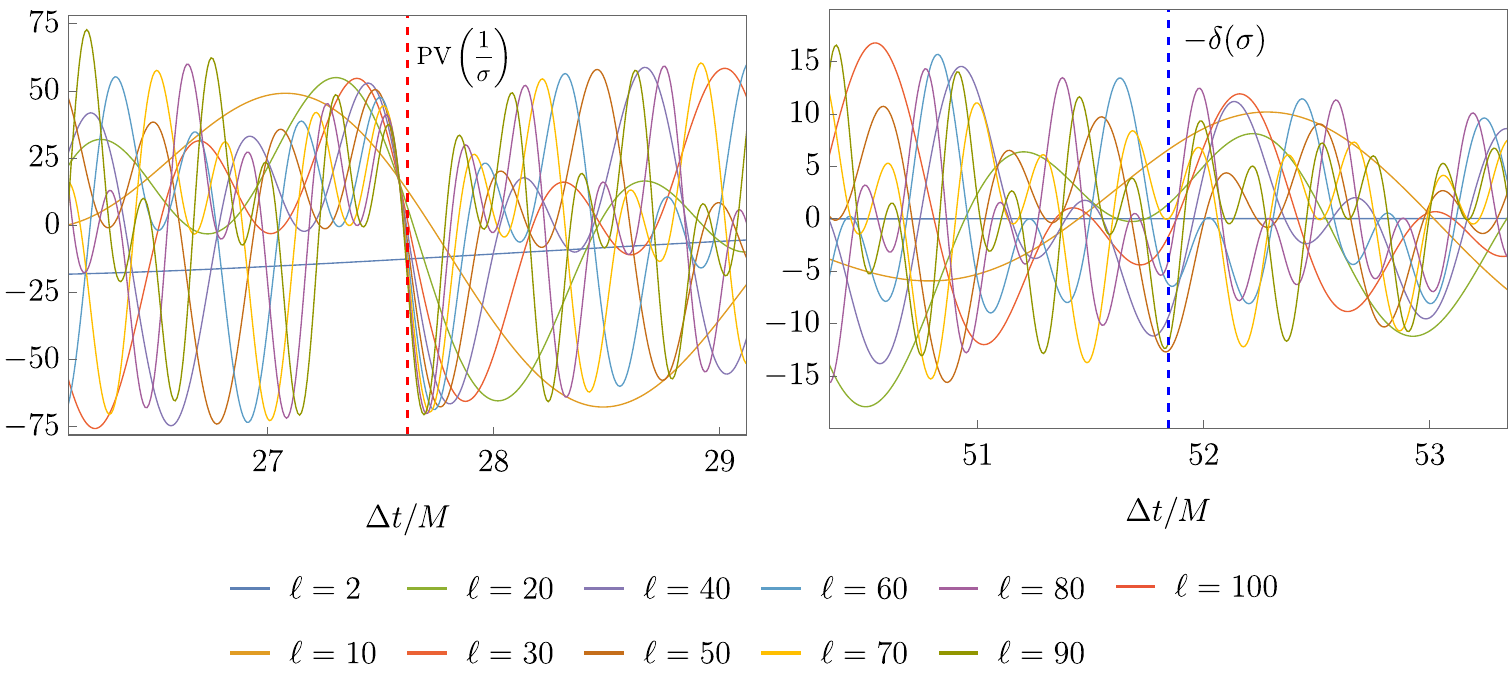}
    \caption{Plot of the BPT summand  ${}_{-2}\mathcal{G}_\ell^T$ (times $M^{-2}$) in Eq.~\eqref{eqn:TGretEllModeDecomp} as a function of $\Delta t$ near the  first (left) and second (right) light-crossings.  The points are along a timelike circular geodesic of radius $r=6M$. 
   }
    \label{fig:BPTEllModesNearLightCrossings}
\end{figure*}

\begin{figure*}
    \centering
    \subfloat{
        \includegraphics[width=\linewidth]{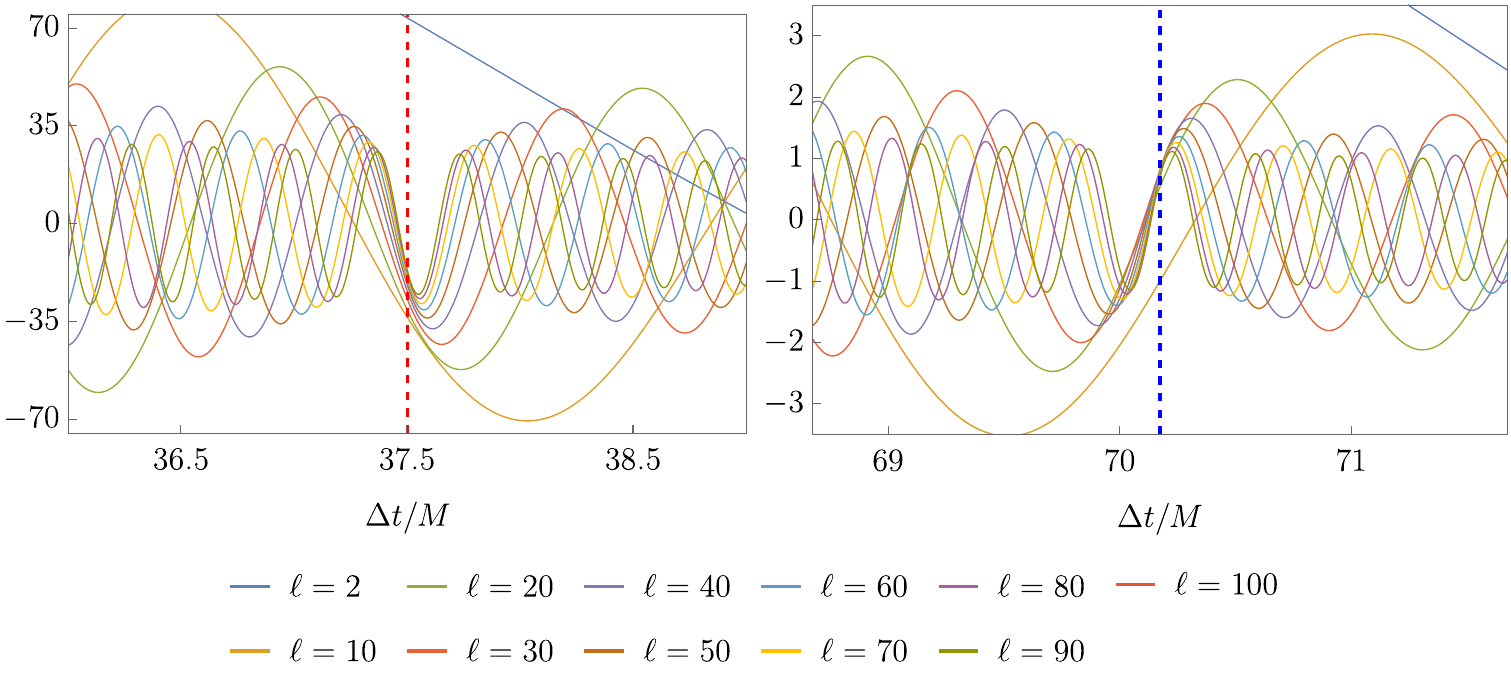}
    }
    \caption{Plot of the BPT summand ${}_{-2}\mathcal{G}^T_\ell$ (times $M^{-2}$)  in Eq.~\eqref{eqn:TGretEllModeDecomp} as a function of $\Delta t$ near the first (left) and second (right)  light-crossings. 
    The points are along a static worldline at the radius $r=6M$.}
    \label{fig:BPTEllModesNearLightCrossingsStatic}
\end{figure*}
\end{widetext}

\section{Asymptotics for large $\omega$ of the RW and BPT Fourier modes and integral convergence}\label{app:FourierModesAsymptotics}

In this appendix we provide asymptotics for large $\omega\in\mathbb{R}$ of the Fourier modes of the RW GF and its radial derivative as well as of the  BPT GF.
 These asymptotics are useful for assessing the rate of convergence of the Fourier integrals
and we also use them to provide a numerically-amenable expression for the Fourier integral for the radial derivative of the RW GF.

\subsection{RW Fourier modes}

A simple way of investigating the convergence of the Fourier integrals 
\eqref{eqn:sGlFourierDecomposition}, \eqref{eqn:sGwlFourierIntegral-Im} and \eqref{eqn:sGwlFourierIntegral}
of the RW Fourier modes $\sGwl$ can be achieved by providing  asymptotic expansions for large $\omega\in\mathbb{R}$ for $\sXIn$ and $\sXUp$. These expansions were given in \cite{Levi:2016esr} for the case of $s=0$.
Here we extend their results to the case of arbitrary RW spin $s$.
The forms of the expansions are:

\begin{align}
    \label{eqn:XInAsymptExpansion}
    \sXIn(r)=&\,e^{-i\omega r_*}\lP1+\sum\limits_{n=1}^{\infty}\frac{a^{\nt{in}}_n(r)}{\omega^n}\rP+[...],\\\label{eqn:XUpAsymptExpansion}
    \sXUp(r)=&\,e^{i\omega r_*}\lP1+\sum\limits_{n=1}^{\infty}\frac{a^{\nt{up}}_n(r)}{\omega^n}\rP+[...],
\end{align}
where $[...]$ denotes terms with exponential decay and $a_n^{\nt{in/up}}\in\mathbb{C}$. The coefficients $a_n^{\nt{in/up}}$ can be computed recursively by inserting Eqs.~\eqref{eqn:XInAsymptExpansion}--\eqref{eqn:XUpAsymptExpansion}  into Eq.~\eqref{eqn:RWEFreqDomain}, yielding:
\begin{align}
    \label{eqn:aInN}
    a_n^\nt{in}(r)=\,&-\frac{i}{2}f(r)\dd{}{r}a_{n-1}^\nt{in}(r)+2i\int\limits_{\infty}^r\frac{\mathcal{Q}_s(\rho)}{f(\rho)}a_{n-1}^\nt{in}(\rho)\df\rho,\\
    \label{eqn:aUpN}
    a_n^\nt{up}(r)=\,&\frac{i}{2}f(r)\dd{}{r}a_{n-1}^\nt{up}(r)-2i\int\limits_{2M}^r\frac{\mathcal{Q}_s(\rho)}{f(\rho)}a_{n-1}^\nt{up}(\rho)\df\rho.
\end{align}
In this way, the first four coefficients are given by
\begin{widetext}
    \begin{align}\label{eq:a1in}
    a^\nt{in}_1=\,&-\frac{i \left(M \hat{s}+\Lambda  r\right)}{2 r^2},\\
    a^\nt{in}_2=\,&-\frac{\Lambda  r (4 M+(\Lambda -2) r)+M \hat{s} \left(M \hat{s}+8 M+2 (\Lambda -2) r\right)}{8 r^4},\\
    a^\nt{in}_3=\,&\frac{i}{240 r^6}\left[5 \Lambda  r \left(72 M^2+15 (\Lambda -4) M r+(\Lambda -6) (\Lambda -2) r^2\right)\right.\nn
    \,&+M \hat{s} \left(960 M^2+12 (19 \Lambda -70) M r+M \hat{s} \left(5 M \hat{s}+160M+3 (5 \Lambda -28) r\right)\right.\nn
    \,&\left.\left.+15 (\Lambda -6) (\Lambda -2) r^2\right)\right],\\
    a^\nt{in}_4=\,&\frac{1}{1920 r^8}\left\lbrace
    5 \Lambda  r \left[2880 M^3+48 (11 \Lambda -70) M^2 r+12 (\Lambda  (3 \Lambda -40)+104) M r^2\right.\right.\nn
    \,&\left.+(\Lambda -12) (\Lambda -6) (\Lambda -2) r^3\right]\nn
    \,&+M \hat{s} \left[46080 M^3+M\hat{s} \left(8640 M^2+136 (7 \Lambda -60) M r\right.\right.\nn
    \,&\left.+M \hat{s} \left(5 M \hat{s}+400 M+4 (5 \Lambda -54) r\right)+6 (\Lambda  (5 \Lambda -86)+320) r^2\right)\nn
    \,&+480 (21 \Lambda -118)M^2 r+12 (\Lambda  (61 \Lambda -780)+1880) M r^2\nn
    \,&\left.\left.+20 (\Lambda -12) (\Lambda -6) (\Lambda -2) r^3\right]
    \right\rbrace,
\end{align}
and
\begin{align}
    a^\nt{up}_1=\,&-\frac{i f \left(\hat{s} (2 M+r)+2 \Lambda  r\right)}{8 M r},\\
    a^\nt{up}_2=\,&\frac{f}{128 M^2 r^3}\left[\hat{s} \left(64 M^3+16 \Lambda  M^2 r-r \hat{s} (2 M+r)^2 f-4 \Lambda  r^3\right)
    -4 \Lambda  r \left(\Lambda  r^2 f-8M^2\right)\right],\\
    a^\nt{up}_3=\,&\frac{i f}{15360 M^3r^5}\left\lbrace
    3840 \Lambda  M^3 r (3 M-r)+30 \hat{s} \left(128 M^4 (8 M-3 r)+\Lambda  r \hat{s} \left(r^2-4 M^2\right)^2\right)\right.\nn
    \,&-4 r f \left[5 \Lambda ^2 r^2 \left(60 M^2+4 Mr+r^2\right)+6 \Lambda  r \hat{s} \left(152 M^3+32 M^2 r+4 M r^2+r^3\right)\right.\nn
    \,&\left.+2 \hat{s}^2 \left(320 M^4+152 M^3 r+12 M^2 r^2+4 M r^3+r^4\right)\right]\nn
    \,&\left.+5 r^2 f^2\left[12 \Lambda ^2 r^2 \hat{s} (2 M+r)+\hat{s}^3 (2 M+r)^3+8 \Lambda ^3 r^3\right]
    \right\rbrace,\\
    a^\nt{up}_4=\,&\frac{f}{491520 M^4 r^7}\left\lbrace
    \hat{s} \left[16 \left(960 M^3 \left(-\Lambda  r^2 \left(-42 M^2+5 M r+r^2\right)f\right.\right.\right.\right.\nn
    \,&\left.-8 M^2 \left(48 M^2-35 M r+6 r^2\right)\right)\nn
    \,&\left.+\Lambda ^2 r^4 f^2\left(-1464 M^3-484 M^2 r+10 \Lambda  r^2 (2M+r) f-78 M r^2-17 r^3\right)\right)\nn
    \,&+r \hat{s} f \left\lbrace120 \left(\Lambda ^2 r^2 \left(r^2-4M^2\right)^2+128 M^4 \left(36 M^2+2 M r-3 r^2\right)\right)\right.\nn
    \,&+r f \left(-32 \Lambda  r \left(952 M^4+592 M^3 r+114 M^2 r^2+26 M r^3+5 r^4\right)\right.\nn
    \,&-32 \hat{s}\left(200 M^4+92 M^3 r+12 M^2 r^2+4 M r^3+r^4\right) (2 M+r)\nn
    \,&\left.\left.\left.+5 r \hat{s} (2 M+r)^3\left(2 M \hat{s}+r\hat{s}+8 \Lambda  r\right)f\right)\right\rbrace
    \right]\nn
    \,&+80 \Lambda  r\left[-23040 M^6+384 (40-11 \Lambda ) M^5 r-288 ((\Lambda -10) \Lambda +8) M^4 r^2\right.\nn
    \,&\left.\left.-8 \Lambda  ((\Lambda -32) \Lambda +48) M^3 r^3+12 (\Lambda -4) \Lambda ^2 M^2 r^4-6 \Lambda^3 M r^5+(\Lambda -2) \Lambda ^2 r^6
    \right]
    \right\rbrace,\label{eq:a4up}
\end{align}
where $\Lambda\equiv \ell(\ell+1)$ and $\hat{s}\equiv 1-s^2$.

Inserting Eqs.~\eqref{eqn:XInAsymptExpansion}--\eqref{eqn:XUpAsymptExpansion}, together with Eqs.~\eqref{eq:a1in}--\eqref{eq:a4up}, for $\gXInUp$  into
Eq.~\eqref{eqn:RWGellModes} for $\sGwl$ (using 
the top expression in Eq.~\eqref{eqn:RWEWronskian} for $W$), we obtain 
\eqn{
\begin{aligned}\label{eqn:sGwlAsymtotic}
    \sGwl(r,r')=\,& \frac{e^{i\omega({r_*}_{\scriptscriptstyle >}-{r_*}_{\scriptscriptstyle <})}}{2\omega}\left\lbrace\ii+\frac{\lP\rG-\rL\rP(\Lambda\, r\, r'+M(r+r')\hat{s})}{2{(r \cdot r')}^2\omega}\right.\\
    &-\frac{\ii}{8(r \cdot r')^4\omega^2}\left[
    2 \Lambda \, r\, r'\left(2 M \Sigma_3+M
   \hat{s} (r+r') \lP\Delta r\rP^2- r\, r'
   \Sigma_2\right)\right.\\
   &\left.\left.+M \hat{s} \left(8 M
   \Sigma_4+M \hat{s}
   \left(r^2-(r')^2\right)^2-4\, r\, r'
   \Sigma_3\right)+\Lambda ^2 (r \cdot r')^2
   \lP\Delta r\rP^2
    \right]+\order{\frac{1}{M^3\omega^3}}\right\rbrace,
\end{aligned}
}
where $\Sigma_{k}\equiv r^k+(r')^k$.

By looking at up to 5 terms in the expansion, we identify the following pattern in the expressions within the square brackets. 
Terms proportional to $\omega^{-(2n-1)}$, with $n\in\mathbb{Z}_{>0}$, are purely real and vanish when $r=r'$, whereas terms proportional to $\omega^{-2n+2}$ are purely imaginary. As a consequence,  $\Re\lC\sGwl(r=r',r)\rC$ decays exponentially for $\omega\to \pm \infty$, whereas $\Im\lC\sGwl(r=r',r')\rC$ exhibits a power-law decay, $\mathcal{O}(1/\omega)$.

Let us now look at the asymptotics for large $\omega\in\mathbb{R}$ of the radial derivative of the RW Fourier modes as radial coincidence ($r'=r$) is approached.
It follows from Eq.~\eqref{eqn:sGwlAsymtotic} that the radial derivative of the GF modes is not continuous at radial coincidence:
\eqn{\label{eq:r-deriv Glw}
    \lim_{r\to{r'}^\pm}\pd{}{r}\sGwl(r,r')=\mp\frac{1}{2f(r')}+\ii\frac{M (8 M - 3 r') \hat{s} + (3 M - r') r' \Lambda}{4(r')^5\omega^3}+\order{\frac{1}{M^5\omega^5}}.
}
In the expansion in Eq.~\eqref{eq:r-deriv Glw}, only the leading, $\order{\omega^0}$ term, is discontinuous at radial coincidence, as we next argue. 
Below Eq.~\eqref{eq:dr-CID} we see that $\s G_\ell$ is smooth other than along $\Dt= \pm \Delta r_*$. 
This means that none of the orders which are not explicitly included in Eq.~\eqref{eq:r-deriv Glw} can be discontinuous at radial coincidence (and the term of order $\omega^{-3}$ is clearly not discontinuous either).
Furthermore, we checked the continuity of the higher order terms explicitly up to 5 orders and also numerically by plotting the two limits as functions of $\omega$, which only differ by a constant shift (coming from the first term in Eq.~\eqref{eq:r-deriv Glw}). 
The discontinuity coming from the first term in Eq.~\eqref{eq:r-deriv Glw} is as expected since $\sGwl$ is itself a Green function of the radial ODE \eqref{eqn:radiialRWEsGwl}.
We note that taking the average of the limits in the two directions in Eq.~\eqref{eq:r-deriv Glw} effectively amounts to removing the first term in the expansion.

\subsection{Fourier integral for  the radial derivative of the RW GF}\label{sec:int dGlwdr}

The  leading-order term in \eqref{eq:r-deriv Glw} leads to a Dirac delta distribution in the time domain. Therefore, 
for the purposes of numerical calculations, it is convenient to 
separate out
this leading-order term from the Fourier integral. In this way, we express the radial derivative of the $\ell$-modes $\sGl$ in the radial-coincidence limits as:
\eqnalgn{\notag
    \lim_{r\to{r'}^\pm}\pd{}{r}\sGl(r,r';\Dt)=\,&\frac{1}{2\pi}\int\limits_{-\infty}^{\infty}\lP\pd{}{r}\sGwl(r,r')\pm\frac{1}{2f(r')}\rP_{r=r'} e^{-\ii\omega\Dt}\df\omega\mp\frac{1}{2\pi}\int\limits_{-\infty}^{\infty}\frac{1}{2f(r')}e^{-\ii\omega\Dt}\df\omega\\
    =\,&\frac{1}{2\pi}\int\limits_{-\infty}^{\infty}\lP\pd{}{r}\sGwl(r,r')\pm\frac{1}{2f(r')}\rP_{r=r'} e^{-\ii\omega\Dt}\df\omega\mp\frac{\delta(\Dt)}{2f(r')},
    \label{eq:r-deriv Glw,delta}
}
Now, the properties $\s G_{\ell,-\omega}=\sGwl^*$ and $\sGl=0$ for $\Dt<0$ (see above Eq.~\eqref{eqn:sGwlFourierIntegral}) imply
\begin{align}
    &\pd{}{r}\s G_{\ell,-\omega}=\lP\pd{}{r}\s G_{\ell\omega}\rP^*,\nonumber\\
    &\pd{}{r}\sGl=0,\,\forall\Dt<0.
\end{align}
By using these properties together with \eqref{eq:r-deriv Glw,delta} we obtain, similarly to Eqs.~\eqref{eqn:sGwlFourierIntegral}--\eqref{eqn:sGwlFourierIntegral-Im},

\eqnalgn{\nonumber
    \lim_{r\to{r'}^\pm}\pd{}{r}\sGl(r,r';\Dt)=\,&\frac{2}{\pi}\theta(\Dt)\int\limits_{0}^{\infty}\lP\Re\lP\pd{}{r}\sGwl(r,r')\rP\pm\frac{1}{2f(r')}\rP_{r=r'}\cos(\omega\Dt)\df\omega\mp\frac{\delta(\Dt)}{2f(r')}\\
    =\,&\frac{2}{\pi}\theta(\Dt)\int\limits_{0}^{\infty}\Im\lP\pd{}{r}\sGwl(r,r')\rP_{r=r'}\sin(\omega\Dt)\df\omega\mp\frac{\delta(\Dt)}{2f(r')}.
    \label{eq:r-deriv G}
}
The radial-coincidence  limit of the radial derivative of $\sGl$ is thus independent of the direction of the limit for $\Dt >0$, as already observed within the time-domain (CID) method below Eq.~\eqref{eq:dr-CID}.
The subtraction of the second term inside the parenthesis in the top integral in Eq.~\eqref{eq:r-deriv G} helps improve its convergence but should not change its value for $\Dt >0$.
Similarly to $\sGl$ in Eqs.~\eqref{eqn:sGwlFourierIntegral}--\eqref{eqn:sGwlFourierIntegral-Im}, we choose to operate with the top integral in \eqref{eq:r-deriv G} since its integrand decays exponentially, whereas the integrand in the bottom integral only decays polynomially.
As noted below Eq.~\eqref{eq:r-deriv Glw}, the subtraction from the radial derivative of the Forier modes of the second term inside the parenthesis on the top integral in Eq.~\eqref{eq:r-deriv G} for $\Dt>0$ can be effectively implemented by taking the average of the limits in the two directions.

\subsection{BPT Fourier modes}

We here carry out an  asymptotic analysis for large $\omega\in\mathbb{R}$ of the 
BPT Fourier modes ${}_{-2}{G}^T_{\ell\omega}$ appearing in the Fourier integrals \eqref{eqn:BPTFourierIntegral} and \eqref{eqn:realTGellFourierModeIntegral} and as
given by the expression in Eq.~\eqref{eqn:teukolskyFourierModes}.
From Eqs.~\eqref{eqn:tRInTra} and ~\eqref{eqn:WdWT} we find that the factor
\eqnalgn{\label{eqn:wronskianPrefactors}
    \frac{W}{{}_{-2}{\mathcal{R}}^\nt{in,tra}_{\ell\omega}{}_{-2}{\mathcal{R}}^\nt{up,tra}_{\ell\omega}W^T}&=\frac{1}{(\ell-1)\ell(\ell+1)(\ell+2)-12iM\omega}
}
in Eq.~\eqref{eqn:teukolskyFourierModes}
is order $\omega^{-1}$ for large $\omega\in\mathbb{R}$. 

We still have to deal with the other factor in Eq.~\eqref{eqn:teukolskyFourierModes}, namely,
\eqn{\label{BPT-RW,r-fac}
-\frac{\lP\OpChR{-2}{r}{\omega}\gXIn(r)\rP_{r=r_<}\lP\OpChR{-2}{r}{\omega}\gXUp(r)\rP_{r=r_>}}{W}.
}

By inserting the  asymptotics for large $\omega\in\mathbb{R}$ of  
$\gXInUp$ given in Eqs.~\eqref{eqn:XInAsymptExpansion}--\eqref{eqn:XUpAsymptExpansion} into Eq.~\eqref{BPT-RW,r-fac}, and combining  it together with Eq.~\eqref{eqn:wronskianPrefactors} into Eq.~\eqref{eqn:teukolskyFourierModes}, we obtain that the behaviour for large $\omega \in \mathbb{R}$ of the BPT Fourier modes is given by

\eqnalgn{\label{eq:large-w BPT Fourier,r<>r'}
    {}_{-2}{G}_{\ell\omega}^T(r,r')=\,&\frac{f(\rL)^2\rG e^{i\omega({r_*}_{\scriptscriptstyle >}-{r_*}_{\scriptscriptstyle <})}}{2\rL\omega}\left[\ii (\rG)^2+\frac{3M\left(5\rG^2-\rL^2\right)+r r'\left(\rG(\Lambda-6)-\rL(\Lambda-2)\right)}{2(\rL)^2\omega}+\order{\frac{1}{M^2\omega^2}}\right].
}

In the case of radial coincidence $r'=r$, which is the case in the two settings that we deal in this paper, the above expression yields
 \eqn{\label{eq:large-w BPT Fourier}
     {}_{-2}{G}_{\ell\omega}^T(r,r)=\frac{\ii f(r)^2{r}^2}{2\omega}-\frac{f(r)^2(r-3M)}{\omega^2}+\order{\frac{1}{M^3\omega^3}}. }

Below Eq.~\eqref{eqn:sGwlAsymtotic} we noted that, for $r=r'$, the imaginary part of the RW Fourier modes $\sGwl$ decays polynomially whereas their real part decays exponentially. Applying the (reduced) radial Chandrasekhar operator in Eq.~\eqref{eq:ChandrOp-R}, as well as including the radius-independent factor Eq.~\eqref{eqn:wronskianPrefactors} in Eq.~\eqref{eqn:teukolskyFourierModes}, will clearly mix up the real and imaginary parts, explaining why both real and imaginary parts of the BPT Fourier modes ${}_{-2}{G}_{\ell\omega}^T$ at radial coincidence  decay polynomially  in Eq.~\eqref{eq:large-w BPT Fourier}.

From Eq.~\eqref{eq:large-w BPT Fourier,r<>r'} we can obtain the large $\omega\in \mathbb{R}$ asymptotics of the radial derivative of the BPT Fourier modes as radial coincidence is approached:

\eqn{\label{eq:r-deriv GlwT,r=r'}
    \lim_{r\to {r'}^\pm}\pd{}{r}\lP\sgGTwl(r,r')\rP=\mp\frac{1}{2}\Delta_S(r')+\ii\frac{\Delta_S(r')}{2r'\omega}+\frac{8M^2(2r'-3M)-(r')^3}{2(r')^3\omega^2}+\order{\frac{1}{M^3\omega^3}}.
}
As in the RW case, the radial derivative is discontinuous at radial coincidence, as expected from the fact that $\sgGTwl$ is a Green function of the radial ODE \eqref{eqn:radialTeukolskyEqn},
and this discontinuity is due only to the first term in the expansion in Eq.~\eqref{eq:r-deriv GlwT,r=r'}, as per arguments similar to those  below Eq.~\eqref{eq:r-deriv Glw}.

\subsection{Fourier integral for  the radial derivative of the BPT GF}\label{sec:int dGlwdr-BPT}

It is convenient to extract the leading-order term from the Fourier integral for the radial derivative of the $\ell$-modes $\sGTl$ in the radial-coincidence limits in a manner similar to how we did it in the RW case in App.~\ref{sec:int dGlwdr}.
That is, we re-express
\eqnalgn{\nonumber
    \lim_{r\to {r'}^\pm}\pd{}{r}\sGTl(r,r';\Dt)=\,&\frac{2}{\pi}\theta(\Dt)\int\limits_{0}^{\infty}\lP\Re\lP\pd{}{r}\sGTwl(r,r')\rP\pm\frac{1}{2}\Delta_S(r')\rP_{r=r'}\cos(\omega\Dt)\df\omega\mp\frac{1}{2}\Delta_S(r')\delta(\Dt)\\
    =\,&\frac{2}{\pi}\theta(\Dt)\int\limits_{0}^{\infty}\Im\lP\pd{}{r}\sGTwl(r,r')\rP_{r=r'}\sin(\omega\Dt)\df\omega\mp\frac{1}{2}\Delta_S(r')\delta(\Dt).
    \label{eq:r-deriv GlT,r=r'}
}
The radial-coincidence  limit of the radial derivative of $\sGTl$ is thus independent of the direction of the limit for $\Dt >0$.
The subtraction of the second term inside the parenthesis in the top integral in Eq.~\eqref{eq:r-deriv GlT,r=r'} helps improves its convergence but should not change its value for $\Dt >0$. Again, we choose the top integral in Eq.~\eqref{eq:r-deriv GlT,r=r'} since its integrand decays as $1/\omega^2$ compared to $1/\omega$ for the bottom integrand.
In practise, when calculating the radial derivative of $\sGTl$ at radial coincidence for $\Dt >0$ in this paper, we chose to take the limit $r\to r'^+$ in Eq.~\eqref{eq:r-deriv GlT,r=r'}.

\end{widetext}


\section{Spectroscopy of ${}_{s}G^T_{\ell}$}\label{sec:spec}

An alternative to calculating the $\ell$-modes of a GF directly via its (inverse-)Fourier integral along real frequencies $\omega$, is deforming the integration contour in the complex-$\omega$ plane and applying the residue theorem.
As shown in Ref.~\cite{Leaver:1986},
in Schwarzschild, this altenative method yields the following contributions to the $\ell$-modes: an integral along the branch cut down the negative imaginary axis; a sum over the residues at the poles (QNMs) of the Fourier modes; an integral along a high-frequency arc.
Essentially, each one of these contributions yields the following features in the time domain:
the branch cut integral for frequencies near $\omega=0$ yields the power-law decay at late times; the sum over QNMs yields the ringdown (an oscillatory, exponentially-decaying stage); the high-frequency arc is negligible for ``sufficiently late" times.
Essentially, this spectroscopical splitting is only useful for calculational purposes for $\Dt >|r_*|+|r_*|$, as shown in Ref.~\cite{Casals:2011aa}\footnote{It readily follows from Eqs.~9, 37, 46 and 57 in \cite{Casals:2011aa} that, as $|\omega|\to\infty$ with $\Im(\omega<0)$,
the Fourier modes of the GF go like $e^{|\omega|(|r_*|+|r'_*|)}$, leaving aside  subdominant factors behaving polynomially in $|\omega|$.}. 

In the next two subsections, we provide the  contributions 
to the BPT $\ell$-mode ${}_{s}G^T_{\ell}$ 
from the QNMs and the branch cut. We provide the final values for the specific case of $s=-2=-\ell$ and only for their dominant terms, namely, for the leading-order for small-frequency in the branch cut case and for the 
slowest-decaying mode in the QNM case.
We used these expressions as checks of our  $\sgGTl$ in Sec.~\ref{eq:BPT-ell-modes}.

\subsection{QNM -- Ringdown}\label{sec:QNM}

Here we derive the expression for the main contribution to ${}_{-2}G^T_{\ell}$ in the ringdown. This main contribution is known to arise from the pole  of the Fourier modes ${}_{s}{G}^T_{\ell\omega}$ on the complex-$\omega$ plane  which lies nearest to the real axis.
This dominant QNM is the so-called fundamental overtone, $n=0$, where the overtone index $n$ labels the QNMs for a given $\ell$.

We start with the expression for the contribution to ${}_{s}G^T_{\ell}$ from a QNM frequency $\omega_{\ell n}$ (e.g.,~\cite{Leaver:1986}):
\begin{align}\label{eq:QN mode}
&
{}_sG_{\ell n}^{QN}(r,r',\Dt)=
\text{Re}\lP
\mathcal{C}_{\ell n}(r,r')e^{-i\omega_{\ell n}\Dt}\rP,
\\&
\mathcal{C}_{\ell n}(r,r')\equiv \frac{2\,\mathcal{B}_{\ell n}}{\lP {}_{s} R^\nt{in,ref}_{\ell\omega_{\ell n}}\rP^2}\,
 {}_{s}R^{in}_{\ell \omega_{\ell n}}(r)\,
 {}_{s}R^{in}_{\ell \omega_{\ell n}}(r'),
 \nonumber
\end{align}
where the coefficient $\mathcal{B}_{\ell n}\in\mathbb{C}$ is the so-called excitation factor.

We now give the needed values to evaluate Eq.~\eqref{eq:QN mode} in our case. 
For $s=-2=-\ell$ and $n=0$, 
 Ref.~\cite{PhysRevD.73.064030} gives \eqn{\label{eq:QNN f}
 M\omega_{2,0}\approx 0.3736716844-0.08896231569\,\ii,}
table V in~\cite{Berti:Cardoso:2006} provides the value for 
$\mathcal{B}_{20}$ and the MST method in the BHPT provides the value for ${}_{s}R^{in}_{\ell \omega_{20}}/{}_{s}R^\nt{in,ref}_{\ell\omega_{20}}$
at $r=6M$.
The result is
\eqn{\label{eq:C,QNM}
\mathcal{C}_{20}(6M,6M)\approx 271.9725 + 
136.0837\,\ii.
}

\subsection{Branch cut -- Late  time tail}\label{sec:late-time}

Here we derive the expression for the leading-order late-time tail of the $\ell$-modes ${}_{-2}G^T_{\ell}$ of the BPT GF.
The late-time asymptotics in the $\ell$-modes ${}_{s}G^T_{\ell}$ may be seen to arise from the small-frequency behaviour along the branch cut down the negative imaginary axis of the complex-$\omega$ plane which the Fourier modes ${}_{s}{G}^T_{\ell\omega}$ possess -- see~\cite{PhysRevD.92.124055}, from which we will take the various small-frequency asymptotics.

We start with the expression for the contribution to ${}_{s}G^T_{\ell}$ from the branch cut as given in Eq.~(2.19)  in~\cite{PhysRevD.92.124055} (this equation is for RW but readily applies  to BPT):
\eqn{\label{eq:GTBC}
{}_sG_{\ell}^{T,BC}(r,r',\Dt )=\frac{1}{2\pi i}\int_0^{\infty} d\sigma\, \delta{}_sG_{\ell}^T(r,r';\sigma) e^{-\sigma \Delta t},
}
where the discontinuity across the branch cut of ${}_{s}{G}^T_{\ell\omega}$ is given  by Eq.~(2.37) in~\cite{PhysRevD.92.124055},
\eqn{\label{eq:dgBC}
\delta{}_sG^T_{\ell} =-2i\sigma \frac{q^T(\sigma)}{|W^T|^2}{}_s\hat{R}^{in}_{\ell}(r,\omega){}_s\hat{R}^{in}_{\ell}(r';,\omega),
}
with $\sigma\equiv i\,\omega$, ${}_{s}\hat{R}^{in}_{\ell}\equiv {{}_{s}R}^\nt{in}_{\ell\omega}/{}_{s}{R}^\nt{in,tra}_{\ell\omega}$ and $q^T(\sigma)$ is the so-called ``branch cut strength".

The leading-order behaviour of ${}_{s}\hat{R}^{in}_{\ell}$ for small frequencies is $\order{\omega^0}$.
Thus, to leading order,  we can replace  ${}_{s}\hat{R}^{in}_{\ell}$ by ${}_{s}\hat{R}^{in,0}_{\ell}\equiv \lim_{\omega\to 0} {}_{s}\hat{R}^{in}_{\ell}$.
For ${}_{s}\hat{R}^{in,0}_{\ell}$, we use the small-frequency asymptotics for
${}_sR^{in}_{\ell}/{}_sR^{in,tra}_{\ell}$ given in (the second expression in)
Eq.~6.5~\cite{PhysRevD.92.124055} evaluated at $\bar\omega\equiv 2M\omega=0$.
That is,
\begin{align}\label{eq:Rin0}
{}_{s}\hat{R}^{in,0}_{\ell}(r)=  &
(2M)^{-2s}\lP\frac{r}{2M}-1\rP^{-s}\cdot
\nonumber\\&
{}_2F_1\lP-\ell,\ell+1,1-s;1-\frac{r}{2M}\rP,
\end{align}
where ${}_2F_1$ is a hypergeometric function.
In particular, for $s=-2=-\ell$, this yields
\eqn{\label{eq:Rin0,s=-2=-l}
{}_{-2}\hat{R}^{in,0}_{2}(r)=4M^2r^2
\lP\frac{r}{2M}-1\rP^{2}.
}
In its turn, the small-frequency asymptotics of the radius-independent factor in \eqref{eq:dgBC} for $s=-2$ and $\ell=2$ are given in
Eq.~(5.22)~\cite{PhysRevD.92.124055}:
\eqn{\label{eq:qW2}
4M^4\frac{q^T(\sigma)}{|W^T|^2}\sim\frac{\pi}{75}\bar\sigma^5,\quad \bar\sigma\to 0, \quad s=-2=-\ell,}
where $\bar\sigma\equiv 2M\sigma$.
Inserting \eqref{eq:qW2} into \eqref{eq:dgBC} and then into \eqref{eq:GTBC}, and replacing  ${}_{-2}\hat{R}^{in}_{2}$ by the leading-order, $\omega$-independent ${}_{-2}\hat{R}^{in,0}_{2}$, we obtain 
\begin{align}
\label{eq:GTBC,late-t,no2}
&{}_{-2}G_{2}^{T,BC}(r,r',\Dt)\sim
\nonumber \\&
-\frac{8M}{75}
{}_{-2}\hat{R}^{in,0}_{2}(r){}_{-2}\hat{R}^{in,0}_{2}(r')\int_0^{\infty} d\sigma \sigma^6 e^{-\sigma \Delta t}\sim
\nonumber \\&
-\frac{384M}{5}{}_{-2}\hat{R}^{in,0}_{2}(r){}_{-2}\hat{R}^{in,0}_{2}(r')\,\lP\Delta t\rP^{-7},\quad \Delta t\to\infty,
\end{align}
as the leading-order late-time asymptotics at fixed, finite radii $r, r'>2M$.
Due to using different normalizations, we find that we need to multiply by $2$ the final 
value in the right hand side of \eqref{eq:GTBC,late-t,no2}, so that we have 
\eqn{\label{eq:GTBC,late-t}
{}_{-2}G_{2}^{T,BC}(6M,6M,\Dt)\sim
\mathcal{A}_2\lP\Delta t\rP^{-7},\quad \Delta t\to\infty,}
with
\eqn{\label{eq:A-late-time}
\mathcal{A}_2
\equiv -    \frac{768M}{5}\lP{}_{-2}
\hat{R}^{in,0}_{2}(r=6M)\rP^2.
}

\bibliography{references.bib}

\bibliographystyle{apsrev}


\end{document}